\documentclass[12pt]{article}

\usepackage[vcentermath]{youngtab}
\usepackage{multirow}
\usepackage{graphicx}
\usepackage{epstopdf}
\usepackage{amsfonts}
\usepackage{amsbsy}
\usepackage{fullpage}
\usepackage{amsmath,amssymb}
\usepackage{latexsym}
\usepackage{array}

\makeindex

\def\Tr{{\mathrm{Tr}}}
\def\tr{{\mathrm{tr}}}
\def\talpha{\tilde{\alpha}}

\def\F{\mathbb{F}}
\def\P{\mathbb{P}}
\def\Z{\mathbb{Z}}
\def\C{\mathbb{C}}
\def\R{\mathbb{R}}
\def\T{{T}}
\newcommand{\eq}[1]{(\ref{#1})}

\begin{document}

\title{TASI Lectures on Supergravity and String Vacua\\ in Various Dimensions}

\author{Washington Taylor \\
Center for Theoretical Physics\\
Massachusetts Institute of Technology \\
Cambridge, MA 02139, USA\\
{\tt wati at mit.edu}}


\maketitle

\begin{abstract}
These lectures aim to provide a global picture of the spaces of
consistent quantum supergravity theories and string vacua in higher
dimensions.  The lectures focus on theories in the even dimensions 10,
8, and 6.  Supersymmetry, along with with anomaly cancellation and
other quantum constraints, places strong limitations on the set of
physical theories which can be consistently coupled to gravity in
higher-dimensional space-times.  As the dimensionality of space-time
decreases, the range of possible supergravity theories and the set of
known string vacuum constructions expand.  These lectures develop the
basic technology for describing a variety of string vacua, including
heterotic, intersecting brane, and F-theory compactifications.  In
particular, a systematic presentation is given of the basic elements
of F-theory.  In each dimension, we summarize the current state of
knowledge regarding the extent to which supergravity theories not
realized in string theory can be shown to be inconsistent.

\end{abstract}
\vspace*{0.2in}

\begin{center}
Based on lectures given at the TASI 2010 Summer School
\end{center}

\vspace*{0.4in}

\hfill MIT-CTP-4227

\newpage

\section{Introduction}
\label{sec:intro}

\subsection{Motivation}

Quantum field theory is an incredibly successful framework for
describing the fundamental processes underlying most observable
physics.  Quantities such as the anomalous magnetic dipole moment of
the electron can be computed from first principles in quantum
electrodynamics (QED), giving results that agree with experiment for up to 10
significant digits.  While in quantum chromodynamics (QCD) many physically
interesting questions cannot be addressed using perturbative
calculations, the framework of field theory itself is believed to be
adequate for describing most observed phenomena involving strong
nuclear interactions.  The difficulty in the case of QCD arises from
the strong coupling constant, which necessitates nonperturbative
treatment by methods such as lattice simulation.  The standard model
of particle physics, which underlies essentially all observed phenomena
outside gravity, is itself a quantum field theory.  And it is likely
that whatever phenomena are discovered at the LHC will also be
describable in terms of a quantum field theory, whether in terms of a
Higgs scalar field, supersymmetry, new strongly coupled physics, or
other more exotic possibilities.

Quantum field theory itself, however, does not place stringent
constraints on what kinds of theories can be realized in nature.
Field theories can be constructed with virtually any gauge symmetry
group, and with a wide range of possible matter content consisting of
particles transforming in various representations of the gauge group.
While the absence of quantum anomalies in gauge symmetries, as well as
other macroscopic consistency conditions, place some simple
constraints on what is possible within a quantum field theory, the
space of apparently consistent field theories is vast.

The one physically observed phenomenon that cannot be directly
described in terms of quantum field theory is gravity.  As discussed
in other lectures in this school, in some cases quantum gravity has a
dual description in terms of a quantum field theory in a lower
dimension.  Attempting to describe a diffeomorphism invariant theory
in space-time of dimension 4 or greater in terms of a field theory in
the same space-time, however, runs into well-known difficulties.
Quantum field theory is defined in a fixed background space-time, and
allowing the space-time metric and topology itself to fluctuate takes
us outside the regime of applicability of standard field theory
methods.

Constructing a consistent quantum theory of gravity has proven to be
substantially more difficult than identifying a quantum theory
describing the other forces in nature.  String theory unifies gravity
with quantum physics.  From the point of view of the low-energy field
theory, however, string theory requires an infinite sequence of
massive fields to smooth the divergences encountered when the theory
is coupled to gravity.  While there are many different ways in which
the extra dimensions of string theory can be ``compactified'' to give
different low-energy field theories coupled to gravity in four
dimensions, the space of such models still seems much smaller than the
full space of 4D quantum field theories.

This leads to a fundamental question, which can be taken as motivation
for these lectures:
\vspace*{0.05in}
\newpage

\noindent
{\bf Motivating question:} {\it Does the inclusion of gravity place
  substantial constraints on the set of consistent low-energy quantum
  field theories?}
\vspace*{0.05in}

By low-energy here, we  mean below the Planck scale.

To address this question we will attempt to give a global
characterization of two general classes of theories, one of which is
contained within the other.
\begin{eqnarray}
{\cal G} & = & \{\makebox{apparently consistent low-energy field
  theories coupled to gravity}\} \nonumber\\ {\cal V} & = &
\{\makebox{low-energy theories arising from known string
  constructions}\}
\end{eqnarray}
\begin{figure}
\begin{center}
\begin{picture}(200,130)(- 100,- 65)
\put(0,5){\makebox(0,0){\includegraphics[width=8cm]{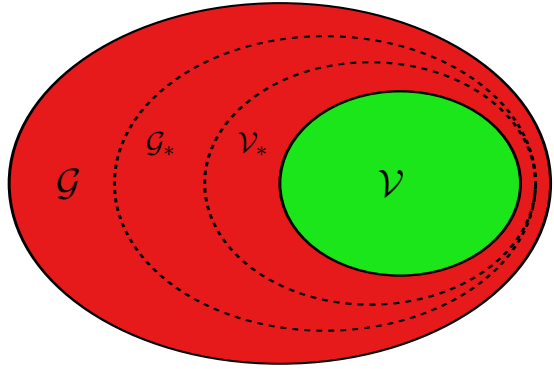}}}
\put(-45,20){\makebox(0,0){\small ${\cal G}_*$}}
\put(-80,5){\makebox(0,0){\large ${\cal G}$}}
\put(42,5){\makebox(0,0){\large ${\cal V}$}}
\put(-10, 20){\makebox(0,0){\small ${\cal V}_*$}}
\end{picture}
\end{center}
\caption[x]{\footnotesize Venn diagram of supergravity theories.
The set ${\cal G}$ 
of apparently consistent quantum gravity theories (largest set, red +
green regions) contains the set ${\cal V}$ of known string vacua in
any particular dimension (smallest set, green region).
Intermediate sets ${\cal G}_*$, ${\cal V}_*$ denote the mathematically
complete sets of consistent gravity theories and string vacua
respectively.  Set inclusions satisfy
${\cal G} \supseteq {\cal G}_* \supseteq {\cal V}_* \supseteq {\cal V}$.}
\label{f:Venn-diagram}
\end{figure}

By ``apparently consistent'', we  mean that there is no known
obstruction to constructing a consistent UV-complete quantum theory with the
desired properties.  By ``known string constructions'', we mean a
compactification of string theory in any of the regimes in which it is
currently understood ({\it i.e.}, heterotic, type I/II, M-theory,
F-theory, {\it etc.}).  While some of these vacua cannot yet be
described even perturbatively through a sigma model, and
no background-independent fundamental formulation of string theory yet
exists, we assume that there is a consistent quantum theory underlying
string constructions.  Pragmatically, this means that we assume that
a few high-dimensional supergravity theories have consistent UV
completions through string/M-theory, and consider the network of models
that can be constructed by compactifying and adding branes and other
features.  

Assuming that string theory is indeed a consistent theory of quantum
gravity, we have the inclusion
\begin{equation}
{\cal G} \supseteq {\cal V} \,.
\end{equation}
The sets ${\cal G}$ and ${\cal V}$ as defined above are dependent upon
our state of knowledge.  We denote by ${\cal G}_*$ the space of {\it
  actually} consistent gravity theories, and ${\cal V}_*$ the complete
space of possible string vacuum constructions (including those not yet
discovered).  The sets ${\cal G}_*$ and ${\cal V}_*$ are presumed to
admit a mathematically precise definition, which has not yet been
determined; these sets do not depend upon our state of knowledge.  We then
have the series of inclusions
\begin{equation}
{\cal G} \supseteq {\cal G}_* \supseteq {\cal V}_* \supseteq {\cal V} \,.
\end{equation}
By discovering new quantum consistency constraints, the space ${\cal
  G}$ can be reduced, and by discovering new string vacuum
constructions the space ${\cal V}$ can be expanded\footnote{It is of
  course possible that some string vacuum constructions currently
  viewed as plausible by some parts of the community may be
  inconsistent, even if string theory is a fine theory.  In this case
  we define ${\cal V}$ to be the subset of hypothesized string vacua
  that are actually in ${\cal V}_*$; in addition to the goals
  outlined in the text, the program described here can also be of
  assistance in determining which hypothetical string vacuum
  constructions are inconsistent (for example if a constraint on
  ${\cal G}$ can be identified that is violated by some proposed
  vacua).}.  As we discuss in these lectures, focusing on theories $x
\in{\cal G} \setminus{\cal V}$ can be a useful way of advancing
knowledge.  If $x \in{\cal V}_*$ then there is a new string
construction including $x$.  If $x\not\in{\cal G}_*$ there is a
physical constraint that makes $x$ an inconsistent theory.  And if $x
\in{\cal G}_*\setminus{\cal V}_*$ then there is a constraint
particular to string theory that is violated by $x$.  This discussion
is closely related to Vafa's notion of the ``swampland'' \cite{Vafa-swamp,
  Ooguri-Vafa}, which consists of gravity theories that appear to be
consistent but which are not realized in string theory.  In the
notation defined above, the swampland is the set ${\cal G}
\setminus{\cal V}_*$.  We will sometimes refer to the set ${\cal G}
\setminus{\cal V}$ as the ``apparent swampland''.  Part of the
motivation for the approach taken here is the notion that
we can learn a lot by trying to remove theories from the apparent
swampland.

When the dimension of space-time is sufficiently large, and we
restrict attention to gravity theories with supersymmetry, the
constraints on ${\cal G}$ are quite strong.  In particular, as we
discuss below in more detail, in 11 and 10 space-time dimensions the
constraints are sufficiently strong that all supergravity theories not
realized in string theory are known to be inconsistent.  Thus, for the
space of 11- and 10-dimensional supergravity theories we have ${\cal
  G} ={\cal V}$.  It follows that
\begin{equation}
{\cal G}_*={\cal V}_* \;\;\;\;\;\makebox{for 10D and 11D
  supergravity} \,.
\label{eq:universality}
\end{equation}
We say that {\it string universality} holds for supersymmetric
theories of gravity in these dimensions, meaning that every theory not
known to be inconsistent is realized in string theory \cite{universality}.  

In these lectures we focus attention on the discrete data
characterizing the structure of supergravity theories: namely the
field content and symmetries of the theory.  The statement of string
universality in \eq{eq:universality} has been demonstrated in 10 and
11 dimensions at the level of these discrete structures.  Any complete
supergravity theory will contain a range of additional detailed
structure encoded in continuous parameters such as coefficients of
higher-derivative terms, the metric on moduli space, etc.  Determining
the uniqueness of such additional structure would refine our
understanding of the relationship between the spaces ${\cal G}$ and
${\cal V}$.  While this represents a very interesting class of
questions and challenges, we do not discuss these issues here.

In dimensions 10 and 11, as we shall describe in more detail below,
there are only a handful of possible theories.  As the dimension of
space-time and the amount of supersymmetry decreases, the constraints
on possible theories become weaker, and the range of possible string
constructions increases.  We focus here on the classes of
gravity theories with minimal supersymmetry in even dimensions.  As
the dimension decreases, the range of interesting physical phenomena
also expands.  In 8 dimensions, there are supergravity theories
coupled to gauge theories with many different gauge groups.  In 6
dimensions there can also be matter fields living in many different
representations of the gauge group.  In 4 dimensions, couplings such
as Yukawa-type interactions arise, which increase the complexity of
the theories significantly.  These lectures focus primarily on
dimensions 10, 8, and 6.  By following possible string constructions
through the decreasing range of dimensions, we can systematically
introduce the new features and mechanisms giving rise to the phenomena
that arise in each dimension.  Thus, this approach gives a convenient
pedagogical framework for introducing many fundamental aspects of
different approaches to the construction of string vacua.  

We do not discuss odd-dimensional supergravity theories here.  While
there are also interesting questions in odd dimensions, the global
picture of theories in dimensions such as 9D is similar to the story
described here for 8D theories.  In even dimensions, there is also a
somewhat richer structure both from the point of view of supergravity,
where anomaly constraints can be stronger, and string theory, where
F-theory provides a powerful nonperturbative framework for describing
the space of string vacua.  A systematic analysis of the discrete
classes of string vacua with minimal supersymmetry in dimensions 9, 8,
and 7 can be found in the work of de Boer {\it et al.}
\cite{7-triples}.

Another aspect of the theme of global structure of the space of
theories is the connectivity of the space of theories.  While
different string theories, and different string vacua, may seem
physically distinct, most of these theories are connected in various
ways.  Perturbative and nonperturbative duality symmetries relate
different string constructions.  In eight dimensions and six
dimensions, most or all of the wide range of possible supersymmetric
string vacua are different branches of a single theory, living in a
continuous moduli space that is described in different regimes by
different string constructions.  In each of these dimensions we
describe how the connectivity of this set of spaces arises, and the
sense in which the diversity of string models fit into a single
overarching theory.

We conclude this introductory section with a brief outline of the
material presented in these lectures.  In each of the dimensions 10,
8, and 6, we characterize the supergravity theories and introduce
string vacuum constructions, then we compare what is known about the
spaces ${\cal G}$ and ${\cal V}$.  Ten dimensions provides a good
starting point for a systematic discussion of both supergravity
theories and string vacua.  After a brief summary of 11-dimensional
supergravity, we describe the basic supergravity theories in 10
dimensions from which the lower-dimensional theories descend.  On the
supergravity side, anomaly constraints and the Green-Schwarz anomaly
cancellation mechanism provide powerful tools for understanding the
space ${\cal G}$.  On the string side, basic objects such as strings
and branes arise naturally and can be most easily understood in 10
dimensions.  As mentioned above, in 10 dimensions ${\cal G} ={\cal
  V}$.  Going down to 8 dimensions, we introduce the heterotic and
F-theory approaches to string compactification.  We describe how
different gauge groups can be realized in these two types of
constructions, and how the constructions are related.  In each case, 8
dimensions provides a natural domain in which to introduce the
essential features of these classes of string vacua.  In 8 dimensions
we believe that we can identify ${\cal V}$ with ${\cal V}_*$, but many
theories still lie in ${\cal G} \setminus{\cal V}$.  In six
dimensions, as in 10 dimensions, anomaly constraints provide a
powerful tool for understanding the space ${\cal G}$ of gravity
theories without known inconsistencies.  We discuss additional
complexities in the heterotic and F-theory vacuum constructions in 6D,
and introduce intersecting brane models as another class of string
vacua.  F-theory vacua provide the largest set of 6D vacua
constructions.  Mathematical structure appearing in the anomaly
cancellation conditions for any six-dimensional supergravity theory
corresponds very closely to topological data for F-theory
constructions.  This 
enables us to identify a ``bottom-up'' map from low-energy theories to
candidate string vacua in six dimensions.  This map gives us a
characterization of the embedding ${\cal V} \subset {\cal G}$ and
allows us to identify general features of models in the ``apparent
swampland'' ${\cal G} \setminus {\cal V}$.  F-theory provides a
framework in which known supersymmetric 6D string vacua fit together
into a single theory, with a moduli space connected through continuous
deformations and phase transitions that can be understood in terms of
F-theory geometry.  We conclude the lectures with some comments on
four-dimensional theories.  While in four dimensions the spaces ${\cal
  G}$ and ${\cal V}$ are much larger and less well understood than in
higher dimensions, some lessons from higher-dimensional supergravity
and string constructions may be helpful in characterizing global
aspects of the space of possible theories.

\subsection{Background}

These lectures are intended for an audience with some knowledge and
experience with the basic principles and tools of quantum field
theory.  Some familiarity with elementary aspects of string theory is
also helpful, though little specific technical knowledge is assumed.
In the early part of the lectures, a number of concepts related to
supersymmetry, supergravity, and perturbative string theory are
reviewed briefly.  The reader interested in more background on these
topics should consult the textbooks of Green, Schwarz, Witten
\cite{gsw-1, gsw-2} and Polchinski \cite{Polchinski-I, Polchinski-II}.
The lectures generally follow the notation and conventions of
Polchinski.  A comprehensive overview of early work on supergravity
theories in various dimensions can be found in the two-volume
compilation by Salam and Sezgin \cite{Salam-Sezgin}.

The material covered in these lectures has evolved somewhat since the
lectures were given, as some new results have clarified parts of the
story.  These written lecture notes integrate developments up to the
time of writing (March 2011).

\subsection{Supersymmetry}

In these lectures we will restrict attention to theories with
supersymmetry.  Supersymmetry is a symmetry that relates bosons to
fermions.  We are interested here in theories where
supersymmetry can be described as an extension of the
Poincar\'{e} symmetry group that characterizes field theories in
Minkowski space.
At the level of the algebra of generators of the symmetry group,
supersymmetry extends the Poincar\'{e} algebra by a set of fermionic
generators $Q_\alpha, \bar{Q}_\alpha$ satisfying the anticommutation relations
\begin{equation}
\{Q_\alpha,  \bar{Q}_\beta\} = 2P_\mu \Gamma^{\mu}_{\alpha \beta}  \,,
\label{eq:SUSY}
\end{equation}
where $\alpha, \beta$ are spinor indices.  The way in which
bosonic and fermionic fields transform under an infinitesimal
supersymmetry transformation parameterized by a spinor $\epsilon$
takes the schematic form
\begin{equation}
\delta \phi \sim \bar{\epsilon}\psi,\;\;\;\;\; \;\;\;\;\;
\delta \psi \sim \Gamma^\mu \epsilon \partial_\mu \phi\,.
\end{equation}

Some theories have multiple supersymmetries, with generators 
$Q^A_\alpha,$
parameterized by an index $A = 1,\ldots {\cal N}$, where ${\cal N}$ is
the number of supersymmetries in the theory.  In this case
\eq{eq:SUSY} generalizes to
\begin{equation}
\{Q^A_\alpha, \bar{Q}^B_\beta\} = 2
  \delta^{AB} P_\mu \Gamma^\mu_{\alpha
\beta}
\label{eq:SUSY-multiple}
\end{equation}

The supersymmetry algebra \eq{eq:SUSY-multiple}
can be extended by central charges that identify topological charges
in the theory; for example,  \eq{eq:SUSY-multiple} can be extended to
\begin{equation}
\{Q^A_\alpha, \bar{Q}^B_\beta\} = 2 \delta^{AB} P_\mu \Gamma^\mu_{\alpha
\beta}+  Z^{A B}\delta_{\alpha \beta} \,,
\end{equation}
where the central charge $Z^{AB}$ commutes with all other generators.

A  description of supersymmetry and spinors in various
dimensions is given in Appendix B of Volume 2 of Polchinski's text on
string theory \cite{Polchinski-II}.  We will not use too many detailed
aspects of supersymmetry and spinors in these lectures, but will
assume some basic facts for which more detailed explanations can be
found in that reference.

For a quantum field theory in flat space-time, supersymmetry is
generally a global symmetry of the theory.
If a theory of gravity has supersymmetry, however, then the symmetry
becomes local.  Just as translation symmetry in flat space-time becomes
a symmetry under local diffeomorphisms in a generally covariant theory
of gravity,  in a supersymmetric generally covariant theory the spinor
parameter $\epsilon$ itself becomes a general space-time dependent
function.  A theory of gravity with local supersymmetry is called  a
{\em
  supergravity} theory.  Each supergravity theory contains ${\cal N}$
massless spin 3/2 {\em gravitino} fields $\psi^A_{\mu \alpha}$ that are
partners of the graviton $g_{\mu \nu}$.

A primary reason for considering only supersymmetric theories of
gravity in these lectures is that supersymmetry imposes 
additional structure that makes both  gravity  and string theory
easier to analyze and to understand.  There are, however, physical
reasons to be interested in supersymmetric theories, both from the
point of view of low-energy phenomenology, and from the top-down point
of view of string theory.

From the phenomenological point of view, supersymmetry has a number of
desirable features.  Supersymmetry modifies the renormalization group
equations so that the strong, weak and electromagnetic couplings
appear to unify at a high scale.  Supersymmetry protects the mass of
the scalar Higgs, giving a possible solution to the ``hierarchy''
problem.  Supersymmetry also provides a natural candidate for dark
matter.

From the string theory point of view, supersymmetry plays a crucial
role in removing tachyonic instabilities from the theory at the Planck
scale. While it is possible that some intrinsically non-supersymmetric
versions of string theory can be made mathematically consistent,
string theory is best understood as a supersymmetric theory of quantum
gravity in 10 dimensions.  

While understanding the space of supersymmetric theories is a
rewarding enterprising in its own right, which may also give new insights
into the structure of non-supersymmetric theories, the most optimistic
reason to study supersymmetric theories is the possibility that
supersymmetry is manifest in our world at an experimentally accessible
energy scale.  The simplest framework in which this occurs can be
analyzed following
the assumption that physics can be split into two different energy
scales.  Under this assumption, at and below some scale $\Lambda$, all
of the relevant physics describing our world can be characterized by a
supersymmetric quantum field theory coupled to (classical) gravity.
Above the scale $\Lambda$, supergravity and/or string theory are
needed to describe physics at energies up to the Planck scale.  If
this assumption is correct, it means that quantum gravity will play a
phenomenological role primarily in determining which supersymmetric
QFT's can arise at the intermediate scale $\Lambda$.  If, on the other
hand, this assumption is incorrect and supersymmetry is broken at the
Planck scale, making any progress in understanding the connection
between quantum gravitational consistency and low-energy physics will
be extremely challenging.  From a theoretical point of view, one of
the most exciting consequences of the discovery of supersymmetry at
the LHC would be the confirmation that this separation of scales
exists, guaranteeing that supersymmetry can be used in efforts to
understand physics up to the Planck scale.

\subsection{Supergravity in 11 dimensions}

We are interested, therefore, in understanding supersymmetric theories
of gravity (supergravity), which may also include gauge fields and
various kinds of matter.  In higher dimensions, supergravity theories
are quite constrained.  For dimensions $D > 11$, any representation of
the Clifford algebra associated with gamma matrices of the
relativistic symmetry group has dimension 64 or greater.  This leads
to massless particles related to the graviton by supersymmetry that
have spin greater than 2.  No interacting theories of this kind are
known in dimensions above 11, and it is believed (though perhaps not
rigorously proven) that the highest dimension in which a
supersymmetric theory of gravity can exist is 11 dimensions.

In 11 dimensions, there is a unique supersymmetric theory of gravity.
This theory has one supersymmetry (${\cal N} = 1$), and 32
supercharges $Q_\alpha$ carrying an index in the 32-dimensional spinor
representation of $SO(1, 10)$.  The massless fields in the theory
describe particles that are in the {\em supermultiplet} of the
graviton; {\it i.e.,} states that are related to the graviton by
acting with the supersymmetry generators.  These fields include:
\vspace*{0.05in}

\noindent
$g_{\mu \nu}$: the graviton (quantum of fluctuation in the space-time
metric)\\
The graviton is symmetric and traceless, with $(9 \times 10)/2 -1 = 44$
degrees of freedom.
\vspace*{0.05in}

\noindent
$C_{\mu \nu \lambda}$: an antisymmetric 3-form field (analogous to a
gauge field $A_\mu$ but with more indices)\\
The 3-form field has $(9 \times 8 \times 7)/6 = 84$ degrees of freedom.
\vspace*{0.05in}

\noindent
$\psi_{\mu \alpha}$: the gravitino, with 128 degrees of freedom.
\vspace*{0.05in}

In each case the number of degrees of freedom can be understood by
considering the appropriate representation of the $SO(9)$ little group for
massless states in $SO(1, 10)$, in a fashion directly analogous to the
standard analysis of states in 4D QFT.

The low-energy action for the bosonic fields of 11-dimensional
supergravity is given by
\begin{equation}
 S = \frac{1}{2\kappa^2_{11}} \left[
\int \sqrt{g} (R -\frac{1}{2} | F |^2) -\frac{1}{6} \int C \wedge F
\wedge F \right]\,,
\label{eq:11D}
\end{equation}
where $\kappa_{11}$ is the 11-dimensional Newton constant, and the
field strength $F$ is a 4-form given by
\begin{equation}
F^{(4)} = dC^{(3)}\,.
\end{equation}
This is a theory of pure supersymmetric gravity, with no conventional
gauge symmetries or matter content (though the 3-form field $C$ does
have a higher-index version of an abelian gauge symmetry).  For
further details and references on this theory, the reader is referred
to Polchinski's text \cite{Polchinski-II}.

From the point of view of these lectures, the significant feature of
supergravity in 11 dimensions is that there is only one possible
theory, at least at the discrete level of field content and
symmetries.  Thus,
\begin{equation}
{\cal G}^{(11)} =\{M_{11}\}
\end{equation}
where $M_{11}$ is the supergravity theory with bosonic action
(\ref{eq:11D}).  Although this theory does not itself contain strings,
it can be viewed as the strong coupling limit of a ten-dimensional
string theory \cite{Witten-dynamics}.  Thus, this theory is included in
the general space ${\cal V}$ of ``known string vacua'', and in 11
dimensions we have
\begin{equation}
{\cal G}^{(11)} = {\cal V}^{(11)}
\;\;\;\;\;\Rightarrow \;\;\;\;\;
{\cal G}^{(11)}_* = {\cal V}^{(11)}_*\,.
\end{equation}
So string universality holds in 11 dimensions.
The quantum theory of 11-dimensional supergravity is often referred to
as ``M-theory''.
As we discuss below,  M-theory can also be described in the light-cone
gauge through quantization of the membrane in the 11-dimensional
theory, or
alternatively in terms of pointlike branes in a 10D theory.

\section{Supergravity and String Vacua in Ten Dimensions}
\label{sec:10D}

We now consider supergravity theories in ten dimensions, and their UV
completions through string theory.  In ten dimensions, there are
theories with one or two supersymmetries.  We begin with a brief
summary of the gravity theories with two supersymmetries, before
considering the theories with one supersymmetry in more detail.  We
then discuss the string realization of these theories and the
relationship between ${\cal G}$ and ${\cal V}$.  Again, further
details on most of the material in this section can be found in
Polchinski \cite{Polchinski-II}.

In 10 dimensions we can define an 11th gamma matrix
\begin{equation}
\Gamma^{11}  = \prod_{\mu = 0}^{9} \Gamma^\mu
\end{equation}
analogous to $\gamma^5$ in four dimensions.  As in four dimensions,
$\Gamma^{11}$ has eigenvalues $\pm 1$ corresponding to Weyl spinors of
fixed chirality.  In dimensions of the form $D = 4k + 2$, the Weyl
representations are self-conjugate.  In ten dimensions it is also
possible to impose a Majorana (reality) condition.  The Lorentz group
in 10D thus has two distinct real 16-dimensional representations ${\bf
  16, 16'}$ corresponding to Majorana-Weyl chiral spinors.  The
different supergravity theories in 10D have different choices of
spinor representations for the supersymmetry generators $Q_\alpha^A$.

\subsection{${\cal  N} = 2$ supergravity in ten dimensions}

\subsubsection{Type IIA supergravity}

The type IIA ${\cal N} = 2$ supergravity theory in ten dimensions has
two supersymmetries of opposite chirality, generated by $Q^1 \in {\bf
  16}, Q^{2} \in {\bf 16'}$.  The theory has bosonic fields
\begin{equation}
g_{\mu \nu}, B_{\mu \nu}, \phi
\end{equation}
living in an ${\cal N} = 1$ supersymmetry multiplet.  The field
$B_{\mu \nu}$ is an antisymmetric two-form field, and the field $\phi$
is a scalar  (the {\em dilaton}).  The IIA theory has additional
bosonic fields described by a 1-form and 3-form
\begin{equation}
A_\mu, C_{\mu \nu \lambda} \,.
\end{equation}
From the nature of the stringy origin of these fields, they are
referred to as {\em Ramond-Ramond}, or {\em R-R} fields.

Counting degrees of freedom, the bosonic fields in the IIA theory have
128 components.  In accord with supersymmetry, this is the same as the
number of fermionic degrees of freedom, which are contained in a pair
of Majorana-Weyl gravitinos (${\bf 56}$ + ${\bf 56'}$) and a pair of
Majorana-Weyl spinors\footnote{Note that the Majorana-Weyl spinors
  characterizing the supersymmetry generators transform under the
  relativistic symmetry group $SO(1, 9)$, while the on-shell degrees
  of freedom in a massless spinor field transform under the
  little group $SO(8)$; this explains the discrepancy between the 16
  and 8 real degrees of freedom in these representations.} (${\bf 8}$ +
${\bf 8'}$).  Because there is one supersymmetry of each type, the
theory is non-chiral, with one spinor and one gravitino of each
chirality.

The IIA theory is directly related to 11-dimensional supergravity by
{\em compactification} on a circle $S^1$.  Wrapping the 11D theory described
by the action (\ref{eq:11D}) on a circle of radius $R$, as $R$ becomes
small the momentum (Kaluza-Klein) modes on the extra circle become
very massive, and in the low-energy limit the zero modes of the theory
combine into the fields of the IIA supergravity theory.
The dilaton, for example, comes from the component of the metric
tensor in 11D with both indices wrapped in the compact direction
$g_{11\; 11} \rightarrow \phi$.  The correspondence between the 11D
and 10D degrees of freedom for all the bosonic fields is given by
\begin{equation}
{ \rm IIA}: \;\;\;\;\;\begin{array}{ccccccc}
\makebox{11D SUGRA} &  \; \;& g^{(11)}_{\mu \nu}&
g^{(11)}_{\mu \;  11} & g^{(11)}_{11 \; 11}&
C^{(11)}_{\mu \nu \lambda} & C^{(11)}_{\mu \nu \;11}\\
\; \; \;\downarrow  {\tiny (S^1)} & &
\downarrow & \downarrow & \downarrow & \downarrow & \downarrow \\
\R^{1, 9} & &  g_{\mu \nu} &
  A_\mu & 
 \phi &   C_{\mu \nu \lambda} &
 B_{\mu \nu}
\end{array}
\end{equation}

\subsubsection{Type IIB supergravity}

Type IIB supergravity in ten dimensions is similar to the type IIA
theory in some ways, but is distinguished by having two
supersymmetries of the same chirality, $Q^1, Q^2 \in $ {\bf 16}.  The
type IIB theory again contains the $ {\cal N} = 1$ multiplet of fields
$g_{\mu \nu}, B_{\mu \nu}, \phi$, as well as a Ramond-Ramond axion
$\chi$, two-form $\tilde{B}_{\mu \nu}$, and self-dual 4-form field
$D^+_{\mu \nu \lambda \sigma}$ whose massless states transform in the
chiral {\bf 70} representation of the little group $SO(8)$.  In this
theory there are two gravitinos with the same chirality ($2 \times$
{\bf 56}), and two spinors of identical chirality ($2 \times {\bf
  8'}$).  The self-dual field $D^+$ makes it difficult to write a local
Lagrangian for this theory, but classical equations of motion for the
fields can be written in an unambiguous fashion.
The type IIB supergravity theory has a classical global symmetry under
$SL(2,\R)$, under which the two-form fields $B, \tilde{B}$ rotate into
one another as a doublet representation.

\vspace*{0.1in}

The type IIA and IIB supergravity theories are uniquely fixed by the
supersymmetries of the theory, at least in terms of the field content
and low-energy equations of motion.

\subsection{${\cal  N} = 1$ supergravity in ten dimensions}

We now turn to  ten-dimensional supergravity theories with only one
supersymmetry, $Q_\alpha \in$ {\bf 16}.  There are two kinds of
supersymmetry multiplet that can appear in such theories; in addition
to the gravity multiplet, there is a vector multiplet containing a
vector field and a spinor called the {\em gaugino} in the {\bf 8}
representation of SO(8).  The field content
of the multiplets, and the associated numbers of bosonic and fermionic
degrees of freedom, can be summarized as
\begin{equation}
{\rm 10D}\ {\cal N} = 1\ \makebox{multiplets}
\;\;\;
\left\{
\begin{array}{lll}
{\rm SUGRA} \;\;\;\;\; &  g_{\mu \nu}, B_{\mu \nu},
  \phi,{ \psi_{\mu \alpha}, \zeta_\alpha} &  \;\;\;\;
{\small [64 + 64 \;
{\rm DOF}]}\\[0.1in]
{\rm vector} &  A_{\mu},
  \lambda_{\alpha} & \;\;\;\;
{\small [8 + 8 \;
{\rm DOF}]}
\end{array} \right.
\end{equation}
Note that the spinor $\zeta_\alpha$ in the gravity multiplet is in the
${\bf 8'}$ of SO(8) and has the opposite chirality to any gauginos.

Classically,  ${\cal N} = 1$ supergravity can be coupled to a set of
vector fields $A_\mu^a$ realizing any abelian or nonabelian gauge group.  The
action for such a theory is
\begin{equation}
 S \sim \frac{1}{2 \kappa_{10}^2}  \int d^{10} x \sqrt{g}
\left[ e^{-2 \phi} 
\left(R + 4 \partial_\mu \phi \partial^\mu \phi
  -\frac{1}{2}| H |^2 \right) -\frac{e^{-\phi}}{g_{{\rm YM}}^2}
  F^a_{\mu \nu} F^{a \, \mu \nu} \right]
\end{equation}
where $g_{\rm YM}$ is the Yang-Mills coupling constant, $F$ is the
usual field strength of the abelian or nonabelian gauge group, and
the
three-form field strength $H$ is related to the two-form $B$ field through
\begin{eqnarray}
H   =   dB- \omega_Y
&\hspace*{0.2in} &d \omega_Y =  \sum_{a}  \; F^a \wedge F^a \,. \label{eq:h}
\end{eqnarray}
The additional contribution of the Chern-Simons term $\omega_Y$ to $H$
is a consequence of supersymmetry \cite{Bergshoeff, cm-cs}; this
term produces a coupling of the form $BF^2$ that is related through
supersymmetry to the gauge field kinetic term.
We now discuss the conditions for quantum consistency of supergravity
theories in 10D.

\subsection{Anomalies in ten-dimensional supergravity theories}

Not every theory that admits a classical Lagrangian formulation is
quantum mechanically consistent.  Classical symmetries can be broken
by quantum anomalies.  The most well-known example is the chiral
anomaly in four-dimensional gauge theories, in which the current
associated with a chiral symmetry develops a quantum correction at one
loop.  The non-vanishing of the divergence of the chiral current
$\partial_\mu j^{5\, \mu}\sim F \wedge F$ amounts to a quantum
breaking of the chiral symmetry.  The anomaly appears because the
quantum theory cannot be regulated or made UV complete without
breaking the symmetry.  Anomalies can be understood in terms of
one-loop corrections where the failure of a regulator to preserve the
symmetry is manifest.  Alternatively, anomalies can be understood in
terms of a failure of the measure in the path integral to respect the
classical symmetry of the theory
\begin{equation}
 \int d \psi d \bar{\psi}\neq \int d \psi' d \bar{\psi}'\, .
\end{equation}
In general, anomalies have a topological origin and can be related to
an appropriate mathematical index theorem.  For a good 
introduction to anomalies and more details on aspects relevant to
these lectures, see  Harvey's 2003 TASI lectures \cite{Harvey-anomalies}.

If a global symmetry in a theory becomes anomalous, such as the chiral
symmetry in some 4D gauge theories, then it simply means that the
symmetry is broken quantum-mechanically.  The theory can still be a
consistent quantum theory.  If, however, a local symmetry is
anomalous, it is generally impossible to find a consistent quantum
completion of the theory.  For example, if a theory contains a U(1)
gauge symmetry that is anomalous, then the theory cannot be fixed
unless an additional degree of freedom (such as a St\"uckleberg field)
combines with the $D-2$ degrees of freedom of the massless gauge field
to complete the necessary set of $D-1$ degrees of freedom for a
massive gauge field.

\begin{figure}
\begin{center}
\includegraphics[width=5cm]{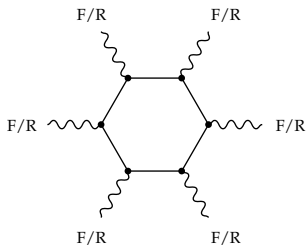}
\end{center}
\caption[x]{\footnotesize Hexagon diagrams give rise to gravitational,
gauge, and mixed gauge-gravitational anomalies in ten dimensions.}
\label{f:10D-anomalies}
\end{figure}

Similarly, if local diffeomorphism invariance is broken by a quantum
anomaly, there is no way to make the theory consistent as a quantum
theory of gravity.
As mentioned above, Weyl
representations are self-conjugate
in dimensions of the form $D = 4 k + 2$.  This means that particles and
anti-particles have the same chirality, so that chiral fermions can
contribute to gravitational and mixed gauge-gravitational anomalies as
well as to purely gauge anomalies.  The detailed form of such
anomalies in ten dimensions was worked out in a classic paper by
Alvarez-Gaume and Witten \cite{agw}.  In ten dimensions, anomalies
arise from hexagon diagrams with external gauge fields or gravitons
(see Figure~\ref{f:10D-anomalies}).  
The anomaly structure of any
theory in $D$ dimensions can be encoded in a
$(D +2)$-form {\em anomaly polynomial} 
$\hat{I}$.  Alvarez-Gaume and Witten showed that the respective contributions of $n$
gaugino spinor fields ({\bf 8}), the gravitino field ({\bf 56}), and a
self-dual 4-form field ({\bf 70}) to the 10D anomaly polynomial are
\begin{eqnarray}
\hat{I}_{(8)} & = & 
-\frac{\Tr (F^6)}{1440} 
+\frac{\Tr (F^4)\tr (R^2)}{2304} 
-\frac{\Tr (F^2)\tr (R^4)}{23040}
-\frac{\Tr (F^2)[\tr (R^2)]^2}{18432}  \nonumber\\
& &\hspace*{0.1in}
+\frac{n \;\tr (R^6)}{725760} 
+\frac{n\;\tr (R^4)\tr (R^2)}{552960} 
+\frac{n\;[\tr (R^2)]^3}{1327104} 
\label{eq:10D-8}
\\
\hat{I}_{(56)} & = & 
-495\frac{\tr (R^6)}{725760} 
+225\frac{\tr (R^4)\tr (R^2)}{552960} 
-63\frac{[\tr (R^2)]^3}{1327104} 
\label{eq:10D-56}
\\
\hat{I}_{(70)} & = & 
992\frac{\tr (R^6)}{725760} 
-448\frac{\tr (R^4)\tr (R^2)}{552960} 
+128\frac{[\tr (R^2)]^3}{1327104}
\label{eq:10D-70}
\end{eqnarray}
In the expression for $\hat{I}_{(8)}$, the terms containing $F$ arise
from the coupling of the gaugino spinor to the gauge field.  The
notation $\Tr$ indicates that these traces are evaluated in the
adjoint representation of the gauge group.  Note that for abelian
vector fields, these terms must vanish, as the adjoint representation
of $U(1)$ is trivial.

The anomaly contributions from eqs.\ (\ref{eq:10D-8}-\ref{eq:10D-70})
must be summed over all chiral fields in a theory to determine the
total anomaly polynomial.  The condition that the total anomaly must
cancel is a very strong condition on 10D gravity theories.

Let us first consider the anomaly contributions from the two ${\cal
  N}= 2$ theories.  The type IIA supergravity theory is non-chiral;
each chiral field has a counterpart with the opposite chirality.  So
in this theory all anomalies cancel.

The type IIB theory, on the other hand, has a more complicated set of
contributions.  The two gravitinos contribute two factors of
$\hat{I}_{56}$.  The self-dual four-form field $D^{+}$ contributes
one factor of $\hat{I}_{70}$.  The two spinors in the ${\bf 8'}$
contribute as two gauginos in \eq{eq:10D-8}, but with the opposite
sign and with no gauge contribution.  The total anomaly for the IIB
theory is then
\begin{equation}
-2\hat{I}_8 (F \rightarrow 0, n \rightarrow 1) + 2 \hat{I}_{56} +
\hat{I}_{70} = 0 \,.
\end{equation} 
Thus, the type IIB supergravity theory  can be a quantum consistent
theory through a rather intricate cancellation.

\subsection{The Green-Schwarz mechanism}

From the above analysis, it seems that all ${\cal N} = 1$ theories
must be anomalous.  Consider in particular the terms in
$\hat{I}_{(8)}, \hat{I}_{(56)}$ proportional to $\tr R^6$ and $[\tr
  (R^2)]^3$.  Since the spinor $\zeta_\alpha$ in the gravity multiplet
is in the ${\bf 8'}$ and contributes to the $R^6$ term with the opposite
sign of a gaugino, the $\tr R^6$ terms can only cancel if $n = 496$.
So the gauge group must have dimension 496.  But by a similar
argument, the $[\tr (R^2)]^3$ terms cannot cancel unless $n = 64$.  So
both terms cannot cancel simultaneously.

This seems to doom all ${\cal N} = 1$ theories of supergravity in ten
dimensions.  But, at the time of Alvarez-Gaume and Witten's analysis
it was already known that a string theory, known as the type I string,
exists and corresponds at low energies to an ${\cal N} = 1$
supergravity theory in ten dimensions with gauge group SO(32).  Green
and Schwarz analyzed this string theory carefully and showed that it
evades the apparent anomaly constraint through a feature now known as
the {\em Green-Schwarz mechanism} for anomaly cancellation
\cite{Green-Schwarz}.

The key to the Green-Schwarz mechanism is the realization that the
anomalous gauge variation associated with 1-loop diagrams can be
canceled by tree-level diagrams when higher-order terms that are
themselves  not gauge invariant are added to the action.  To
implement this mechanism in the case of a nonabelian gauge group , the
field strength (\ref{eq:h}) must be enhanced at higher orders in the
derivative expansion by a Chern-Simons term in the spin connection
\begin{equation}
H = dB- \omega_Y + k\omega_R, \;\;\;\;\;
d \omega_R =\tr R^2 \,,
\label{eq:h-complete}
\end{equation}
with $k$ a constant factor.
\begin{figure}
\begin{center}
\includegraphics[width=4cm]{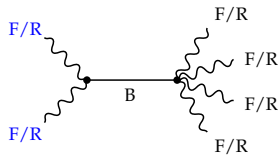}
\end{center}
\caption[x]{\footnotesize The Green-Schwarz mechanism:
A tree diagram describing exchange of a $B$
  field can cancel the anomalous part of the one-loop hexagon diagram
  in ten dimensions in special cases.}
\label{f:Green-Schwarz}
\end{figure}
The two-form $B$ must then transform as $\delta B = \Tr(\Lambda F)
-\tr (\Theta R)$ where $\delta A = d \Lambda$, and the connection
1-form transforms through $\delta \omega_1 = d \Theta$.  The anomaly
can now be cancelled by a tree diagram (see
Figure~\ref{f:Green-Schwarz}) in which a $B$-field is exchanged, when
a ``Green-Schwarz'' term
\begin{equation}
\Delta S \sim \int B \wedge X_8 (F, R)
\label{eq:Green-Schwarz}
\end{equation}
is added to the action such that the anomaly can be expressed in the factorized
form
\begin{equation}
\hat{I} = Y_4 (F, R) X_8 (F, R), \;\;\;\;\;Y_4 = {\rm tr} R^2 - \frac{1}{30}{\rm Tr} F^2   \,,
\end{equation}
where the constant $k$ has taken the value $k = 30$.

For a general nonabelian gauge group of dimension $n$, the total
anomaly can be rearranged to take the form
\begin{eqnarray}
\label{AnomPoly}
\hat{I}_{12} & = & {1 \over 1440} \left( - {\rm Tr} F^6+ {1 \over 48}
    {\rm Tr} F^4 {\rm Tr} F^2 - { ({\rm Tr} F^2)^3 \over 14400}
    \right) \\
& & \hspace*{0.1in}
+ (n - 496) \left(
{{\rm tr} R^6 \over 725760}
+\frac{\tr (R^4)\tr (R^2)}{552960} 
+\frac{[\tr (R^2)]^3}{1327104} \right)\\
& & \hspace*{0.1in} - {Y_4 X_8 \over 768} \,, \nonumber
\end{eqnarray}
where
\begin{equation}
X_8 = {\rm tr} R^4 + {({\rm tr} R^2)^2 \over 4}  -  { ({\rm Tr} F^2)({\rm tr} R^2) \over 30} +  {{\rm Tr} F^4 \over 3} -  {({\rm Tr} F^2)^2 \over 900} \,.
\end{equation}
Thus, the Green-Schwarz mechanism can cancel the anomaly for a
nonabelian group of dimension 496 precisely when the first term
vanishes.  There are exactly two nonabelian groups for which $\Tr F^6$
can be expressed as the necessary combination of $\Tr F^4 \Tr F^2$ and
$(\Tr F^2)^3$.  These two groups are $SO(32)$ and $E_8 \times E_8$.

Now consider a gauge group with abelian factors.  Because
supersymmetry requires \eq{eq:h} to contain a contribution from the
Chern-Simons term of each abelian factor, there is a coupling in the
action of the form $BF^2$ for each $U(1)$ factor in the theory.  For
the pure gravitational anomaly to vanish, the Green-Schwarz term
(\ref{eq:Green-Schwarz}) must be added to the action.  But the $R^4$
part of $X_8$ in this coupling combines with the $BF^2$ term required
by supersymmetry to give a tree diagram contribution of the form $F^2
R^4$.  And as mentioned above such terms do not appear in hexagon
diagrams for abelian gauge factors.  Thus, in any theory with an
abelian gauge group factor the Green-Schwarz mechanism cannot cancel
the anomaly in a way that is simultaneously compatible with
supersymmetry.  As a result, there are no consistent ${\cal N} = 1$
supergravity theories in ten dimensions with gauge groups having
abelian factors.  The details of this argument ruling out $U(1)$
factors were given in a paper with Adams and DeWolfe \cite{adt}.  This
shows, in particular, that the theories with gauge groups $U(1)^{496}$
and $E_8 \times U(1)^{248}$, which were for some time in the
``apparent swampland,'' cannot be consistent supersymmetric quantum
theories of gravity.

Summary: we have shown that there are only four possible consistent
ten-dimensional supergravity theories
\begin{equation}
{\cal G}^{(10)} =\{{\rm IIA,\ IIB,\ SO(32), E_8 \times E_8}\} \,.
\end{equation}

\subsection{String theories in ten dimensions}

In this section we give a short overview of how quantization of
strings and other extended objects can lead to consistent frameworks for
describing quantum gravity.  For a detailed pedagogical introduction
to string theory and branes, the reader should consult Polchinski
\cite{Polchinski-I, Polchinski-II}.

The fundamental excitations in any quantum field theory are pointlike
quanta associated with localized particles in space-time.  While
directly quantizing an interacting theory of gravity in more than
three dimensions using quantum field theory methods proves
problematic, greater success has been realized in quantizing gravity
using extended objects.  As we have seen, supersymmetric theories of
gravity in higher dimensions often contain antisymmetric $p$-form
fields.  For example, all supergravity theories in ten dimensions
contain an antisymmetric two-form field $B_{\mu \nu}$ in the gravity
multiplet, and 11-dimensional supergravity contains an antisymmetric
three-form field $C_{\mu \nu \lambda}$.  These antisymmetric tensor
fields can couple to extended charged objects in the same way that the
electromagnetic vector potential $A_\mu$ couples to charged particles.
The coupling of a pointlike object with charge $q$ under a vector
field $A_\mu$ is described by a contribution to the action given by an
integral along the particle world-line $x^\mu (\tau)$
\begin{equation}
S_0 =q \int A_\mu dx^\mu\,.
\label{eq:0-coupling}
\end{equation}
Similarly, the field $B_{\mu \nu}$ couples to charged stringlike
excitations with one direction of spatial extent through
\begin{equation}
S_1 = \int B_{\mu \nu} dx^\mu dx^\nu\,,
\label{eq:1-coupling}
\end{equation}
and a 3-form field couples to a membrane with two spatial dimensions
in a similar fashion, etc.
Just as gravity coupled to electromagnetism in four dimensions admits
classical solutions with mass and charge (Reissner-Nordstr\"om black
holes), gravitational theories in higher dimensions with $(p+1)$-form
fields admit extended charged ``black brane'' solutions.
In supersymmetric theories, the central charge in the supersymmetry
algebra places an upper bound on the charge/mass ratio possible for such brane
solutions.  Solutions which saturate this bound are known as {\em
  extremal} solutions.  Extremal brane solutions will be quantized in
a quantum theory of gravity.  By considering the quantum field theory
on the world-volume of the branes, a theory of quantum gravity in the
bulk space-time emerges in many cases.

The principle of {\em brane democracy} \cite{Townsend-democracy}
states that any of the branes appearing in a theory can in some sense
be treated as fundamental degrees of freedom.  Indeed, to the extent
it is possible to quantize branes in any supergravity theory, the
resulting quantum theory seems to give at least a limited description
of a consistent quantum gravity theory.  The $p$-form fields and
associated extended objects appearing in 11D and 10D supergravity
theories are listed in the following table
\vspace*{0.1in}

\begin{equation}
\begin{array}{ccl}
\underline{\rm Theory} & \underline{\rm Field} & \underline{\rm Brane}\\[0.1in]
{\rm 11D} &\hspace*{0.1in} C_{\mu \nu \lambda}\hspace*{0.1in} &\makebox{M2-brane (+
  dual M5)}\\[0.1in]
{\cal N} = 1, 2 \; {\rm 10D}\hspace*{0.1in} & {  B_{\mu
    \nu}} &
{ \makebox{(F) string (+
  NS5-brane)}}\\[0.1in]
\begin{array}{c}
{\rm IIA} \\{\rm IIB}
\end{array}
&
\hspace*{0.2in}
{ \left.\begin{array}{l}
{A}_\mu, {C}_{\mu \nu \lambda}\\
\tilde{B}_{\mu \nu}, {D}^+_{\mu \nu \lambda \sigma}
\end{array}\right\}}\hspace*{0.3in}&
{ \makebox{(D)-branes}}
\\[-0.03in]
&{\makebox{\small [RR-fields]}}\hspace*{0.15in} &
\end{array}\end{equation}
\vspace*{0.1in}
For each  $(p +1)$-form field, there is a corresponding $p$-brane
that is electrically charged, in the sense that the brane couples
to the field in a fashion analogous to \eq{eq:0-coupling} and
\eq{eq:1-coupling}.  There is also a dual $(D-p-4)$-brane that
couples magnetically, in the sense that there is a flux for the field strength
through a space-like $(p +2)$-sphere surrounding the magnetic charge.  For
example, there is a dual 5-brane charged magnetically under $B_{\mu
  \nu}$ in every 10D supergravity theory, so that a
3-sphere surrounding the 5-brane carries nonzero flux $\int_{S^3} H$.

The branes that can be quantized in the most well-understood fashion
are the ``fundamental'' strings appearing in all ten-dimensional
supergravity theories as objects charged under $B_{\mu \nu}$.  The
perturbative approach to string theory begins by considering the
theory on the world-sheet of these strings, described by a sigma model
mapping the world-sheet into space-time.  This approach leads to five
tachyon-free superstring theories in ten dimensions: the type IIA and
type IIB theories, whose low-energy limits are the type IIA and type
IIB supergravity theories; the heterotic $SO(32)$ and $E_8 \times E_8$
theories, with the associated ${\cal N} = 1$ supergravity theories as
low-energy limits; and type I string theory, which gives a different
description of the $SO(32)$ theory.  These theories give perturbative
descriptions of the quantum supergravity theory in each case, and can
be directly formulated in terms of a sigma model for backgrounds
solving the classical equations of motion with nontrivial geometry,
dilaton, and $B$ fields.  The perturbative description is less clear
when the background contains nonzero Ramond-Ramond fields, but recent
progress has been made in this direction \cite{Berkovits, Berkovits-review}.

Just as fundamental strings form a natural route to quantizing
ten-dimensional supergravity theories, the membrane carrying charge
under the three-form field $C$ is a natural candidate for quantizing
11-dimensional supergravity.  Although a generally covariant
quantization of the membrane has not been found, by restricting
to light-cone gauge the theory on the membrane can be regulated in
such a way that the resulting theory is a simple matrix quantum
mechanics theory.  This matrix quantum mechanics theory, known as
``M(atrix) theory'' gives a nonperturbative description of M-theory in
light-cone gauge \cite{dhn, BFSS, Taylor-matrix}.

A major breakthrough in understanding quantum gravity and string
theory was made when Polchinski observed that the supergravity brane
solutions charged under the R-R fields of type II string theory could
be identified with dynamical branes in space-time located on
hypersurfaces where open strings end with Dirichlet boundary
conditions (``D-branes'').  This insight provides two ways of
describing the dynamics of such branes: analyzing fluctuations around
the classical soliton solution, or quantizing open strings ending on
the brane.  Quantizing strings connecting $N$ coincident branes leads
to a world-volume $SU(N)$ gauge theory; in an appropriate near-horizon
limit the gauge theory becomes supersymmetric Yang-Mills theory.  The
celebrated AdS/CFT correspondence is the correspondence between this
Yang-Mills theory and the dual gravity theory in the vicinity of the
brane \cite{Maldacena, AdS-review}.  Again, quantizing fluctuations
around a brane has led to a theory of gravity.  In this case, the
gravity theory is in a space of higher dimensionality than the field
theory on the brane world-volume, and the correspondence gives a
nonperturbative definition of quantum gravity in terms of a field
theory.  The AdS/CFT correspondence underlies many of the talks at
this TASI  school.  One other situation where quantizing the
theory on branes leads to quantum gravity is realized by quantizing
the theory on $N$ D0-branes in type IIA string theory.  This gives an
alternate route to the M(atrix) model of M-theory mentioned above
\cite{BFSS}.

The upshot of this discussion is that theories of quantum gravity can
be studied by quantizing $p$-dimensional extended objects coupled
to dynamical $(p +1)$-form fields in supergravity theories.  String
theory, the AdS/CFT correspondence, and M(atrix) theory are all
examples of this general principle.  In later parts of these lectures
we will also use branes as a tool in constructing string vacua in
dimensions less than 10.

\subsection{Summary: string universality in ten dimensions}

We showed above that there are only four distinct theories of
supergravity in ten dimensions that do not suffer from
inconsistencies due to quantum anomalies.  Since each of these
theories can be realized in string theory, we have string universality
for supergravity theories
in ten dimensions
\begin{equation}
{\cal G}^{(10)} =
{\cal G}^{(10)}_* =
{\cal V}^{(10)}_* =
{\cal V}^{(10)} =
\{{\rm IIA,\ IIB,\ SO(32), E_8 \times E_8}\} \,.
\end{equation}
Of course, as mentioned above, we have only considered the discrete
field content and symmetries of the theory, and have not proven that
each of these quantum theories is perturbatively and nonperturbatively
unique.  Trying to prove, for example, that all higher-derivative
terms in the theory are uniquely determined by supersymmetry and
quantum consistency is a further interesting enterprise on which some
initial progress has been made \cite{Green-Sethi}.

Note that the strategy of analyzing theories in the ``apparent
swampland'' ${\cal G} \setminus {\cal V}$ has led to several steps of
progress enroute to this result.  After Green and Schwarz identified
their anomaly cancellation mechanism, the $E_8\times E_8$ theory was
in the apparent swampland.  Identifying a string theory underlying
this supergravity theory became a clear challenge, which was met when
the ``Princeton string quartet'' of Gross, Harvey, Martinec and Rohm
successfully constructed the $E_8\times E_8$ heterotic theory
\cite{Gross-heterotic}.  Thereafter, for some time the theories with
$U(1)^{496}$ and $E_8 \times U(1)^{248}$ gauge groups remained in the
apparent swampland, motivating the eventual demonstration that these
theories are inconsistent.

As discussed above, there are actually two string theory realizations
of the ${\cal N} = 1$ 10D supergravity theory with $SO(32)$ gauge
group, the type I and heterotic $SO(32)$ string theories.  While there
is no proof that these theories are equivalent, strong evidence
suggests that there is a nonperturbative {\em duality} symmetry which
identifies these two apparently distinct string theories
\cite{Witten-dynamics, Polchinski-Witten}.  From the discussion so far,
it seems that there are a number of disconnected theories in ten
dimensions.  These theories are all connected, however, through a
network of duality symmetries \cite{Hull-Townsend, Witten-dynamics}.
In particular, once the theories are compactified to lower dimensions,
it becomes clear that apparently different 10D
supergravity/superstring theories give rise to different descriptions
of the same continuous space of gravity theories in the lower
dimension.  We now turn to eight dimensions, where we see an explicit
example of such a duality symmetry for lower-dimensional theories.

\section{Supergravity and String Vacua in Eight Dimensions}
\label{sec:8D}

We begin our consideration of eight-dimensional theories with a brief
introduction to eight-dimensional supergravity.  We then discuss some
general aspects of the compactification of supergravity theories from
11D and 10D to lower dimensions.  This sets the stage for introducing
two approaches to constructing 8D superstring vacua: heterotic
compactifications on a two-torus, and F-theory compactifications on an
elliptically fibered K3 surface.  We show that these two rather
different string constructions give rise to the same set of 8D
theories, providing an example of a duality symmetry relating
ostensibly very different string constructions.

\subsection{Supergravity in eight dimensions}
\label{sec:8D-supergravity}

As in 10 dimensions, gravity theories can be constructed in 8
dimensions with either one or two supersymmetries.  The ${\cal N} = 2$
8D supergravity theory contains only the supergravity multiplet and
thus has a uniquely determined field content.  This supergravity
theory can be realized through dimensional reduction of 11D or 10D
supergravity on a torus, and thus has a natural string theory
realization.  In each dimension we focus on the supergravity theories
with minimal supersymmetry, where novel phenomena emerge.  In 8D
${\cal N} = 1$ theories, the new feature present is a wider range of
possible gauge groups relative to the highly constrained set of 10D
${\cal N} = 1$ supergravity theories.

Just as in ten dimensions,
the minimal ${\cal N} = 1$ supersymmetry in eight dimensions
has 16 supercharges, and the multiplets of interest consist of the
supergravity and vector multiplets
\begin{equation}
{\rm 8D}\ {\cal N} = 1\
\;\;\;
\left\{
\begin{array}{lll}
{\rm SUGRA} \;\;\;\;\; &  g_{\mu \nu}, B_{\mu \nu}, 2A_\mu, \sigma;
{ \psi_{\mu \alpha}, \chi_\alpha} &  \;\;\;\;
{\small [48 + 48 \;
{\rm DOF}]}\\[0.1in]
{\rm vector} &  A_{\mu}, 2 \phi,
  \lambda_{\alpha} & \;\;\;\;
{\small [8 + 8 \;
{\rm DOF}]}
\end{array} \right.
\end{equation}
In eight dimensions, the graviton multiplet contains two vector
fields, often referred to as {\em graviphotons}.  Spinors in eight
dimensions, as in four dimensions, can be Majorana or Weyl, but not
both.  Classically, the supergravity multiplet can be coupled to any
number of vector multiplets \cite{Salam-Sezgin-8D-2, Awada-Townsend}.
8D theories can in principle have pure gauge or mixed
gauge-gravitational anomalies from pentagon diagrams \cite{agw}.  The
gauge factor is always of the form $\tr_R F^3$ or $\tr_R F^5$ where
the trace is taken in the representation
$R$  under which the charged matter transforms; these terms vanish in
the adjoint representation, which is the only representation possible
for the charged fermions in a supersymmetric theory, so there are no
local anomalies in 8D supergravity theories.  In our current state of
knowledge regarding 8D theories, then, there are no restrictions from
anomalies on the gauge group and
\begin{equation}
{\cal G}^{(8D, {\cal N} = 1)} =
{\cal G} =\{ \makebox{SUGRA + YM for any\ }\ G\}.
\end{equation}
This is an infinite set of theories with distinct gauge groups.
We discuss at the end of this section how this set of theories may
be further constrained.

To physicists familiar with supersymmetry in 4 dimensions, or even 10
or 6 dimensions, it may be surprising that the supergravity multiplet
in 8 dimensions has a number of degrees of freedom (96) that is not a
power of 2.  To understand this it is helpful to briefly review the
construction of supermultiplets \cite{Wess-Bagger, Polchinski-II}.  In
any dimension $D$, a massless supermultiplet is formed by starting
with a representation of the corresponding little group $SO(D-2)$ and
acting with raising operators associated with the $\Gamma$ matrices
representing the supersymmetry algebra.  For example, the ${\cal N} =
1$ supergravity multiplet in 10 dimensions is formed by acting on the
8-dimensional vector representation ${\bf 8}_v$ of the little group
$SO(8)$ by the set of supersymmetry raising operators that combine
into the ${\bf 8}_v + {\bf 8}'$ representation of $SO(8)$.  The tensor
product of these representations gives
\begin{equation}
{\bf 8}_v \times ({\bf 8}_v + {\bf 8}') =
{\bf 35} + {\bf 28} + {\bf 1} + {\bf 56} + {\bf 8}'
\label{eq:10D-multiplet-structure}
\end{equation}
which corresponds to the set of degrees of freedom in the components
in the supergravity multiplet.  In eight dimensions, the little group
is $SO(6)$.  The 16 supersymmetry raising operators break up into
representations ${\bf 6} + 2 \times {\bf 1} + {\bf 4} + \bar{{\bf 4}}$
of $SO(6)$, and the supergravity multiplet is formed from the massless
${\bf 6}$ representation, giving
\begin{equation}
{\bf 6} \times ({\bf 6} + 2 \times {\bf 1} + {\bf 4} + \bar{{\bf
    4}})
= {\bf 20} + {\bf 15} + {\bf 1} + 2 \times {\bf 6} +
{\bf 20}_{\psi} + \bar{{\bf 20}}_\psi + {\bf 4} + \bar{{\bf 4}} \,,
\label{eq:8D-multiplet-structure}
\end{equation}
which are the degrees of freedom in the fields of the 8D supergravity
multiplet.

\subsection{Compactification of supergravity theories}
\label{sec:compactification}

We have described above the various supergravity theories in 11 and 10
dimensions.  As we have seen, the 10-dimensional type IIA theory can
be related to a compactification of the 11D supergravity theory
on a circle $S^1$.  More generally, a variety of supergravity theories
in lower dimensions can be constructed by compactifying the 11D and
10D supergravity theories on various geometries.  For example,
considering a $D$-dimensional supergravity theory on a manifold of the
form $X_k \times\R^{1, D-k -1}$, where $X_k$ is a $k$-dimensional
compact manifold, gives rise to a supergravity theory in $D-k$
dimensions.  To have a vacuum solution in the lower-dimensional
theory, the compactification must be constructed in such a way that
the equations of motion of the $D$-dimensional supergravity are
satisfied.  The massless spectrum in $D-k$ dimensions then comes from
the zero modes of the massless fields in the higher-dimensional
theory.  If the manifold $X_k$ is a torus $T^k$, then such a
compactification amounts to a periodic identification of all fields in
the theory (with some choice made for boundary conditions on fermions
and higher spin fields).  In general such a toroidal compactification
preserves all of the supersymmetries of the theory, giving a
supergravity theory in the smaller dimension with the same number of
supercharges.  The spectrum in the toroidally compactified
theory has the same number of degrees of
freedom as the higher-dimensional theory, organized according to the
relativistic symmetry group of the lower-dimensional theory.  More
complicated geometries can be chosen that break some of the
supersymmetry, giving a wider variety of theories in lower dimensions.

In addition to purely geometrical compactifications, additional
features such as branes and fluxes can be included that increase the
range of possibilities.  The inclusion of branes gives constructions
such as {\em intersecting brane models}, where branes wrapping cycles
on the compactification manifold break some of the supersymmetry and
give rise to gauge groups and matter.  {\em F-theory} models, as we
discuss later in this section, can be thought of as compactifications
of the type IIB theory with branes, although they also have a natural
interpretation in terms of pure geometry.  Including fluxes for the
various $p$-form fields in the supergravity theories can also break
some supersymmetry and increase the variety of possible low-energy
theories.  We do not discuss compactifications with fluxes much in
these lectures.  Although they are a rich source of structure in
lower-dimensional string vacua, there are no interesting classes of
flux compactifications known in dimensions 8 or 6.

For now, we focus on compactifications on purely geometrical spaces,
without branes or fluxes.
We assume then that space-time is of the form
\begin{equation}
{\cal M}_{10} =\R^{1, D-k-1} \times X_k \,,
\end{equation}
where $D = 10$ or $D = 11$.  We assume that we are expanding the
theory around a background where all fields other than the metric (and
dilaton when $D = 10$) vanish, with a constant value for the dilaton
when $D = 10$.  To preserve supersymmetry there must be a
supersymmetry transformation under which the variation of the gravitino
vanishes.  Thus, in particular, there must be a spinor $\eta$ on the
space-time so that $\delta \psi_\mu = D_\mu \eta = 0$.  Such a
covariantly constant spinor can exist on a manifold of dimension $k =
2 N$ only when parallel transport on the manifold, or {\em holonomy},
leaves a component of the spinor unchanged.  This is possible when the
holonomy around all curves lies in a subgroup $SU(N)$ of the structure
group $SO(2N)$.  A manifold of dimension $2 N$  has holonomy in $SU(N)$
if and only if it is a complex K\"ahler manifold 
with a nowhere vanishing holomorphic $N$-form.
Such manifolds are known as
{\em Calabi-Yau} manifolds.  We will not need to deal with many
technical aspects of Calabi-Yau manifolds in these lectures, but we
will learn a few things about some of the simpler Calabi-Yau spaces in
the context of various string compactifications.

The simplest example of a Calabi-Yau manifold is a torus of even
dimensionality $T^{2 N}$.  For a torus the holonomy is trivial around
any curve since the connection vanishes.  For compactification on a
torus, all supersymmetries of the higher-dimensional theory are
retained after compactification.  For complex dimension 1, the
two-torus is the only Calabi-Yau manifold.  For complex dimension 2,
besides the torus $T^4$, there is one other topological class of
Calabi-Yau manifold; this manifold is known as the {\em K3 surface}.
Compactification on a K3 surface breaks half of the supersymmetries of
the higher-dimensional theory.  We will describe this manifold in more
detail and study some of its features below in our discussion of
8-dimensional F-theory vacua.  The number of topologically distinct
Calabi-Yau manifolds in complex dimension three is much larger; it is
not even known if this number is finite.  Compactification on a
Calabi-Yau threefold reduces the number of supersymmetry generators by
a factor of 4.

There are a few other situations in which supersymmetry can be
preserved under purely geometrical compactifications.  If a 7-manifold
has holonomy in the group $G_2 \subset SO(7)$, then a supersymmetry is
preserved \cite{Joyce}.  For example, 11-dimensional supergravity
compactified on a $G_2$-holonomy 7-fold gives a theory with one
supersymmetry in four dimensions, reducing the number of supersymmetry
generators by a factor of 8.  Similarly, compactification on an 8-fold
with holonomy in Spin(7) or Sp(2) gives theories with supersymmetry
reduced by a factor of 1/16 or 3/16 \cite{Becker-Morrison,
  Morrison-TASI, Gauntlett-gp}.

A table of the primitive geometric compactifications that admit a
covariantly constant spinor, and hence supersymmetry in the
dimensionally reduced theory, is given below.  Products of these
spaces will, of course, also give a supersymmetric dimensionally
reduced theory; for example, compactification of a 10D theory on K3
$\times T^2$ gives a 4D theory with supersymmetry reduced by a factor
of 1/2.  In this table, CY3, CY4 refer to Calabi-Yau threefolds and
fourfolds with holonomy precisely equal to $SU(N)$; product spaces
have reduced holonomy and preserve more supersymmetry.  For reductions
on compact spaces of dimension up to 8, the manifolds in this table
and products thereof are the only smooth geometries which preserve
some supersymmetry.  A more detailed discussion of the mathematics
behind this assertion and further references can be found in the book
by Gross, Huybrechts and Joyce \cite{Gross-hj} and in
Morrison's TASI notes \cite{Morrison-TASI}.

\begin{center}
\begin{tabular}{|c|c|c|c|}
\hline
(real) dimension & manifold type & holonomy & SUSY\\
\hline
$k$ & $T^k$ &\{1\} & 1\\
\hline
4 & K3 (CY2) & SU(2) & 1/2\\
\hline
6 & CY3 & SU(3) & 1/4\\
\hline
7 & $G_2$&$G_2$& 1/8\\
\hline
8 & Sp(2) & hyper-K\"ahler & 3/16\\
\hline
8 & CY4 & SU(4) & 1/8\\
\hline
8 & Spin(7) & Spin(7) & 1/16\\
\hline
\end{tabular}
\end{center}

We can now ask the question: ``How can we get a supergravity theory in
eight dimensions by a geometric compactification?''.  Given what we
have learned so far, it would seem that this can only be done by
compactifying a 10-dimensional supergravity theory on a two-torus
$\T^2$.  Indeed, compactifying IIA and IIB supergravity/string theory
on $T^2$ gives rise to two (dual) descriptions of the same ${\cal N} =
2$ supergravity in eight dimensions, while compactifying ${\cal N} =
1$ supergravity on $T^2$ gives ${\cal N} = 1$ supergravity in 8D.  In
the following section we describe the compactification of the
heterotic theories to 8D in more detail.  We then describe another
approach to constructing eight-dimensional supergravity theories
through {\em F-theory}.  F-theory can be thought of as a
compactification of the type IIB theory on a curved manifold with
branes.  The branes give rise to an effective 12-dimensional geometric
picture, so that ${\cal N} = 1$ supergravity in eight dimensions can
be thought of as arising from F-theory on a K3 surface.  As we now
discuss in more detail, the two approaches of heterotic on $T^2$ and
F-theory on K3 give dual descriptions of the same set of 8D
supergravity theories.
\begin{center}
\begin{picture}(200,80)(- 80,- 40)
\put(-70,24){\makebox(0,0){het = I (10D) \small [16 Q]}}
\put(-70,-24){\makebox(0,0){8D  \small [16 Q]}}
\put(-70,15){\vector( 0, -1){30}}
\put(-60,2){\makebox(0,0){ $T^2$}}
\put(0,5){\makebox(0,0){dual}}
\put(0,-7){\makebox(0,0){$\Longleftrightarrow$}}
\put(75,24){\makebox(0,0){F-theory (``12D'')  \small [32 Q]}}
\put(75,-24){\makebox(0,0){8D  \small [16 Q]}}
\put(75,15){\vector( 0, -1){30}}
\put(85,2){\makebox(0,0){K3}}
\end{picture}
\end{center}

\subsection{Heterotic string vacua in eight dimensions}
\label{sec:heterotic-8}

\subsubsection{Compactification of 10D ${\cal N} = 1$ theories on $T^2$}

We now consider the theories that can be realized by compactifying
the $SO(32)$ and $E_8 \times E_8$ theories from ten
dimensions to eight dimensions on a torus $T^2$.  From the point of
view of the 10D ${\cal N} = 1$ supergravity theory, we can deduce the
field content of the 8D theory by considering how the representations
of the 10D little group $SO(8)$ for massless fields decompose when
reduced to the 8D little group $SO(6)$.  Operationally, we can perform
the reduction of fields by choosing for each 10D Lorentz index (0,
\ldots, 9) either an 8D Lorentz index ($\mu = 0, \ldots, 7$) or an
index in one of the compact directions ($c = 8, 9$).  For example, the
10D vector field (with massless states transforming in the ${\bf 8}_v$
of $SO(8)$) reduces to an 8D vector field $A_\mu$ (with states in the
{\bf 6} of $SO(6)$) and two scalar fields $\phi= A_8, \phi' = A_9$.
Thus, the 8D vector multiplet is simply the dimensional reduction of
the 10D vector multiplet.  The dimensional reduction of the 10D ${\cal
  N} = 1$ supergravity multiplet gives the 8D supergravity multiplet
and two vector multiplets ($128 = 96 + 2 \times 16$).  The fact that
the supersymmetry multiplet is reducible after dimensional reduction
comes from the decomposition of the ${\bf 8}_v$ multiplet on which the
SUSY generators act in \eq{eq:10D-multiplet-structure} through ${\bf
  8}_v \rightarrow {\bf 6} + 2 \times {\bf 1}$.  In the anomaly-free
10D ${\cal N} = 1$ supergravity theories, there are 496 gauge bosons.
The compactification of the bosonic fields of such a supergravity
theory on $T^2$, assuming all gauge fields vanish, gives an 8D theory
with one supergravity multiplet and 498 vector multiplets.
\begin{center}
\begin{picture}(200,60)(- 90,- 30)
\put(-75,24){\makebox(0,0){$g_{\mu c}, B_{\mu c}, A_\mu$}}
\put(-75,-24){\makebox(0,0){$500 \; A_\mu \; $   $[U(1)^4 \times G]$}}
\put(-75,15){\vector( 0, -1){30}}
\put(65,24){\makebox(0,0){$\phi, A_c, g_{c c'}, B_{c c'}$}}
\put(65,-24){\makebox(0,0){$\sigma +$ 996 $\phi$  (scalars/moduli)}}
\put(65,15){\vector( 0, -1){30}}
\end{picture}
\end{center}
The scalar $\sigma$ in the supergravity multiplet and the two scalars
in each vector multiplet combine to give 997 scalar fields in the 8D
theory.  

From the discussion so far, it would seem that the gauge group of the
dimensionally reduced theory must be given by $G \times U(1)^4$, where
$G = SO(32)$ or $E_8 \times E_8$ is the nonabelian gauge group of the
10D theory.  In either case, the rank of the 8D gauge group is 20.
The gauge group in the reduced theory is generically broken to a
smaller group, however, by {\em Wilson lines} around the circles of
the compactification space.  From the point of view of the low-energy
theory this corresponds to Higgsing the gauge group by turning on
expectation values for any of the 992 scalars arising from the
dimensional reduction of $A_8, A_9$, which transform in the adjoint
representation of the gauge group.  The gauge group can also be
increased to a larger group by a stringy gauge enhancement mechanism.
In either case the rank of the group stays fixed at 20.  We now
consider each of these possibilities in turn.

The possibility of nontrivial Wilson lines for the gauge fields
$A_\mu$ around each of the two directions in the torus can be
understood in a straightforward fashion.  A constant nonabelian gauge
field can be turned on in the directions $A_8$ and $A_9$ without
breaking supersymmetry, as long as the curvature $F = [A_8, A_9] = 0$
vanishes.  (Note that nonzero curvature gives an energy density $E
\sim F^2$, which is incompatible with the vanishing vacuum energy
required by supersymmetry in Minkowski space.)  In the presence of
Wilson loops, the 10D gauge group is broken to $H \subset G$ where $H
=\{h: [h, A_8] =[h, A_9] = 0\}$.  Since $A_8, A_9$ can always be
chosen as part of a maximal torus, the rank of the resulting 8D gauge
group is always 20.  This gives a large family of possible unbroken
gauge groups for the 8D theory.  For generic commuting Wilson lines
$A_8, A_9$, the nonabelian gauge group will be completely broken to
\begin{equation}
G_{\rm generic} = U(1)^{20} \,.
\end{equation}
In such a generic vacuum, two of the $U(1)$ factors are graviphotons.
Associated with the 16 unbroken generators of the original nonabelian
gauge group there are 32 scalar fields.  Four additional scalars are
associated with the shape and size of the compactification torus.
There is one more scalar coming from the 10D dilaton, for a total of
37 scalar fields.  These fields parameterize the space of
supersymmetric vacua of the theory.  There is no potential for these
massless scalar fields, which are known as {\em moduli} of the theory.
This gives us a picture of the 8D supergravity theory resulting from
the heterotic compactification as having a continuously connected
37-dimensional moduli space of vacua;  in lower-dimensional subspaces
of the moduli space the gauge group is enhanced.

From the preceding discussion it seems that the maximum enhancement of
the gauge group can be to  one of the original heterotic
groups.  As mentioned above, however, string theory also provides a
mechanism for enhancing the gauge group further beyond $U(1)^4 \times
SO(32)$ or $U(1)^4 \times E_8 \times E_8$.  To understand this
mechanism, it will be helpful to briefly review some basics of the
mathematics of lattices.  Lattices are relevant here because the
$k$-dimensional torus can be thought of as a quotient $\R^k/\Gamma$,
where $\Gamma$ is a $k$-dimensional lattice.  Lattices will also play
an important role in many other constructions in these notes.

\subsubsection{Interlude on lattices}

A $k$-dimensional lattice is defined to be the subset of $\R^k$ given
by integral linear combinations of a set of $k$ linearly independent
basis vectors $e_i$
\begin{equation}
\Gamma =\{n_i
e_i,n_i\in\Z\}\,.
\end{equation}
As in much of the mathematical literature, we are interested in
lattices that carry an integral symmetric bilinear inner product
\begin{equation}
v \cdot w \in\Z \; \; \forall v, w \in \Gamma\,.
\end{equation}
We assume that every lattice discussed in these notes carries such a structure.
We present here only some basic aspects of the theory of lattices.
For a more thorough introduction to the subject see the text by Conway
and Sloane
\cite{Conway-Sloane}.

For a given choice of basis, a lattice $\Gamma$ can be represented by
the integral matrix $e_i \cdot e_j$.  We will sometimes use $\Gamma$ to denote
this matrix as a convenient shorthand for describing a given lattice.
A lattice $\Gamma$ is {\em Euclidean} if
\begin{equation}
v \cdot v  > 0 \; \; \forall v \in \Gamma \,.
\end{equation}
This corresponds to the condition that the associated integral matrix
is positive definite.  More generally, we will be interested in
lattices of indefinite signature $(p, q)$.

A lattice is said to be {\em even} if
\begin{equation}
v \cdot v \in 2\Z \; \; \forall v \in \Gamma\,.
\end{equation}

The {\em dual} $\Gamma^*$ of a lattice $\Gamma$ is defined to be the
set of dual vectors whose inner product with all elements of $\Gamma$
is integral
\begin{equation}
\Gamma^*=\{v: v \cdot w \in\Z \; \; \forall w
\in \Gamma\}\,.
\end{equation}
A lattice is {\em self-dual} or {\em unimodular} if it is equal to its
dual
\begin{equation}
\Gamma= \Gamma^*\,.
\end{equation}
This is equivalent to the condition that the associated matrix has
unit determinant,
$\det \Gamma = \pm 1$.
Alternatively, the self-duality condition is equivalent to the
condition that the basis vectors $e_i$ span a unit cell of volume $\pm
1$.
An important theorem due to Milnor states that any unimodular lattice
has signature satisfying
\begin{equation}
p \equiv q \;\; ({\rm mod}\ 8) \;\;\;\;\; (\Gamma \;\makebox{unimodular})
\,.
\label{eq:Milnor}
\end{equation}
\vspace*{0.05in}

\noindent {\bf Example:}
As an example of  an even unimodular lattice consider the lattice
\begin{equation}
U = \left(\begin{array}{cc}
0 & 1\\1 & 0
\end{array}\right), \;\;\;\;\;
e_1^2 = e_2^2 = 0, \;\;\; e_1 \cdot e_2 = 1\,.
\label{eq:u}
\end{equation}
This lattice has signature $(1, 1)$.  A sublattice of $U$ is spanned
by the vectors with respect to which $U$ is diagonalized
\begin{equation}
(e_1 + e_2)^2 = 2, \;(e_1 - e_2)^2 = -2, \;
(e_1 + e_2)\cdot(e_1 -e_2) = 0 \,.
\end{equation}

A particularly noteworthy class of even lattices are the {\em root
  lattices} of the algebras associated with the
simply-laced Lie groups ($SU(N), SO(2N), E_6, E_7,
E_8$).  For such  an algebra the simple roots $r_i$ form a basis for the
lattice.  The
Cartan matrix is formed from the inner
products of the simple roots $a_{ij} = r_i \cdot r_j$, where the roots
are normalized such that $a_{ii} = 2$.
For simply-laced algebras, all off-diagonal entries in the Cartan matrix
are either 0 or -1.  These algebras and the associated matrices are
often conveniently described in terms of {\em Dynkin diagrams}. The
Dynkin diagram for a given algebra/Cartan matrix is given by drawing a
node for each simple root  $ r_i$, and connecting with a line each
pair of nodes $i, j$ satisfying $r_i \cdot r_j = -1$.
The number of nodes in the Dynkin diagram corresponds to the rank of
the associated Lie algebra, which
is the same as the dimension of the corresponding lattice.
The lattice and Dynkin diagram associated with the algebra of $SU(N)$
is denoted $A_{N-1}$, while that associated with the algebra of $SO(2N)$ is
denoted $D_N$.
\vspace*{0.05in}

\noindent {\bf Example:}
The lattice $A_2$ is described by the
Cartan matrix associated with the Lie algebra of $SU(3)$
\begin{equation}
(a_{ij}) = 
{\small \left(\begin{array}{cc}
2 & -1\\-1 & 2
\end{array}\right)} \,.
\end{equation}
This lattice is even but not unimodular.
The associated Dynkin diagram is
\begin{center}
\begin{picture}(40,15)(- 20,- 3)
\put(-10,0){\circle*{5}}
\put(10,0){\circle*{5}}
\put(-10,0){\line(1, 0){20}}
\end{picture}
\end{center}
\vspace*{0.05in}

\noindent {\bf Example:}
The lattice $E_8$ is described by the
Cartan matrix associated with the $E_8$ group
\begin{equation}
(a_{ij}) = 
{\small \left( \begin{array}{cccccccc}
2 & 0& -1& 0& 0& 0& 0& 0 \\
0 & 2& 0 &-1& 0& 0& 0& 0 \\
-1 & 0& 2&-1& 0& 0& 0& 0 \\
0  & -1& -1& 2& -1& 0& 0 &0 \\
0 & 0 &0 &-1& 2& -1& 0& 0 \\
0 & 0 &0 &0& -1& 2& -1& 0  \\
0 & 0 &0 &0& 0 &-1& 2 &-1 \\
0 & 0 &0 &0& 0 &0 &-1 &2 \\
\end{array} \right)} 
\end{equation}
This lattice is both even and unimodular.
The associated Dynkin diagram is
\begin{center}
\begin{picture}(80,18)(- 40,- 3)
\multiput(-45,0)(15,0){7}{\circle*{5}}
\multiput(-45,0)( 15,0){6}{\line(1, 0){15}}
\put(-15,15){\circle*{5}}
\put(-15,0){\line( 0,1){15}}
\end{picture}
\end{center}
\vspace*{0.05in}

As a consequence of \eq{eq:Milnor}, a Euclidean lattice can only be
self-dual if it has a dimension
$p \equiv 0$ (mod 8).
$E_8$ is the unique Euclidean self-dual lattice of dimension 8.
There are two even self-dual Euclidean lattices of dimension 16:
\begin{equation}
E_8 \oplus E_8, \; {\rm and} \;\Lambda_{16} ,
\label{eq:lattices-16}
\end{equation}
where $\Lambda_{16}$ is the {\em Barnes-Wall} lattice.  This lattice
is closely related to the root lattice $D_{16}$ of $SO(32)$.
$\Lambda_{16}$ includes $D_{16}$ as a sublattice, and contains
an additional
set of points corresponding to the weights of a spinor representation
of $SO(32)$ (essentially another copy of $D_{16}$, but shifted by an
offset moving each lattice point to one of the biggest ``holes'' in
the original lattice).

For lattices of indefinite signature $(p, q)$, 
there is a unique even self-dual lattice with any given signature when
$p \equiv q$ (mod 8).
This lattice is denoted $\Gamma^{p, q}$, and is given by
\begin{equation}
\Gamma^{p, q} = U\oplus \cdots
  \oplus U \oplus (\pm E_8)\oplus \cdots \oplus (\pm E_8)
\end{equation}
where the number of factors of $U$ is min ($p, q$) and the number of
factors of $E_8$ is $(p-q)/8$.

\subsubsection{Toroidal compactification and enhanced symmetries}
\label{sec:toroidal-compactification}

We can now describe enhanced symmetries of the toroidally compactified
heterotic string in terms of the language of lattices.
We begin by recalling the quantization of the simple bosonic string on
a compact circle of radius $R$, following the notation of
Polchinski \cite{Polchinski-I}.  In terms
of world-sheet coordinates $\sigma, \tau$, and setting $\alpha' = 1$,
the position of the string in the compact direction is given by
\begin{align}
X & = x + {w} \sigma +  p \tau + {\rm oscillators} (\alpha_n,
\tilde{\alpha}_n)\\
& = x +\frac{1}{\sqrt{2}} l_L (\tau + \sigma)
 +\frac{1}{\sqrt{2}}  l_R (\tau - \sigma)+ \cdots
\end{align}
where the winding number and momenta are quantized through
\begin{equation}
w = m R, \;\;\;\;\; p = n/R \,.
\end{equation}
T-duality is the symmetry that exchanges $w \leftrightarrow p$ 
through
$m \leftrightarrow n$
and $R
\leftrightarrow 1/R$ ($R \rightarrow \alpha'/R$ when $\alpha'$ is
reinstated).  The left- and right-moving momenta $l_{L, R}$ are
related to winding number and momenta through
\begin{equation}
l_L = \frac{p + w}{ \sqrt{2}}, \;\; l_R = \frac{p-w}{ \sqrt{2}} \,.
\end{equation}
These momenta live on the even lattice $U$ from \eq{eq:u}, with inner product
\begin{equation}
l \cdot l = l_L^2 -l_R^2 = 2pw = 2nm \in  2\Z \,.
\label{eq:u-momenta}
\end{equation}
Restoring units, physical momenta are given by
\begin{equation}
k = \sqrt{2/\alpha'} \;l \,,
\end{equation}
and the mass shell condition for string states is
\begin{equation}
M^2 = k_\mu k^\mu  =    \frac{2}{\alpha'}  \left[l_L^2  +  2 (N -1) \right]
=   \frac{2}{\alpha'}  \left[l_R^2  +  2 (N -1) \right] \,.
\end{equation}
There are then two different classes of massless vector fields arising
from closed string states.  There are states with both left and right
moving oscillator number
$N = \tilde{N} = 1$ and vanishing left and right momenta
\begin{equation}
 \alpha^\mu_{-1}
  \tilde{\alpha}^*_{-1}|l_L = 0; l_R = 0\rangle 
\pm
 \alpha^*_{-1}
  \tilde{\alpha}^\mu_{-1}|l_L = 0; l_R = 0\rangle  \,,
\end{equation}
where $*$ denotes the compact direction.  These states are the string
states associated with the  vector bosons $g^{\mu*}, B^{\mu*}$ in the
dimensionally reduced theory.  There are additional massless vector states
with only one nonzero oscillator number, namely
\begin{equation}
\tilde{\alpha}^{\mu}_{-1} | l_L; l_R  \rangle, \; l_R = 0, \;  l_L^2 =l \cdot l = 2\;
\end{equation}
and the analogous states with a single $\alpha_{-1}^\mu$.  These
massless states occur in the spectrum precisely when $R = \sqrt{\alpha'}$,
the self-dual radius for the circle.

The upshot of this analysis is that when there are points in the
lattice with $l \cdot l = 2$ and either $l_R = 0$ or $l_L = 0$ then
there are additional massless vectors in the theory.  In the case of
compactification on a single circle $S^1$, these massless vectors
appear at the self-dual radius, and enhance the generic $U(1)^2$ gauge
symmetry to $SU(2)^2$.  Note that the condition $l \cdot l = 2$ is
always satisfied by a set of elements of the momentum lattice, while
the existence of nonzero
lattice vectors satisfying the condition $l_R = 0$ or
$l_L = 0$ depends upon the values of the scalar moduli --- in this
case the compactification radius.

Generalizing to a compactification on a higher-dimensional torus
$T^D$, the momenta $l$ take values in a signature $(D, D)$ lattice
\begin{equation}
\Gamma^{D, D} = U \oplus \cdots \oplus U \,.
\label{eq:torus-lattice}
\end{equation}
Again, the generic gauge group arising from compactification is
$U(1)^{2D}$, but enhanced symmetries arise for states with $l
\cdot l = 2$ and either
$l_R = 0$ or $l_L = 0$.

The moduli space of theories associated with compactification on $T^D$
can be parameterized by the embeddings of the lattice $\Gamma^{D, D}$
in the momentum space $l_L, l_R$ of the compact directions with inner
product $l_L^2 -l_R^2$.  Given one embedding of the lattice in these
coordinates, any other embedding can be realized by acting with a
transformation in the group $SO(D, D)$ that preserves the inner
product.  Two embeddings that are related by a transformation in
$SO(D) \times SO(D)$ that
separately preserves $l_L^2$ and $l_R^2$ are physically equivalent, so
the moduli space is locally given by $SO(D, D)/SO(D) \times SO(D)$.
There is a further equivalence between embeddings related by a
transformation in $G_D =SO(D, D;\Z)$, the discrete (T-duality) group
of transformations that map the lattice $\Gamma^{D, D}$ to itself.
The global  space of toroidal
compactifications is thus given by the Narain  moduli space
\begin{equation}
SO(D, D;\Z)\backslash SO(D, D)/SO(D) \times SO(D) \,,
\label{eq:Narain-space}
\end{equation}
where the discrete duality group acts on the left, independently from
the $SO(D) \times SO(D)$ symmetry acting on the right.  As a simple
example, for the type IIA and IIB theories compactified on a single
circle ($D = 1$), the T-duality group is the $\Z_2$ group whose
nontrivial element exchanges momentum and winding, giving $n
\leftrightarrow m$ in \eq{eq:u-momenta}.  This T-duality
transformation gives a duality symmetry relating type IIA string
theory compactified on a circle of radius $R$ and the IIB theory
compactified on a circle of radius $\alpha'/R$.

For the toroidal compactifications just discussed, the unimodular form
of the lattice \eq{eq:torus-lattice} follows automatically from the
construction.  More generally, however, we can consider a world-sheet
theory with $p$ chiral left-moving bosons $\phi^a$ and $q$ chiral right-moving
bosons $\tilde{\phi}^b$.  The operators in such a theory are of the
form
\begin{equation}
{\cal O}_{l, \tilde{l}} =\exp \left( il_a \phi^a + i \tilde{l}_b
\tilde{\phi}^b \right) \,.
\end{equation}
The closure of the set of operators under operator products implies
that $(l, \tilde{l})$ lie in a lattice $\Gamma$ of dimension $p + q$.  The
single-valued nature of the operator product implies that
\begin{equation}
l_al_a-\tilde{l}_b \tilde{l}_b \in\Z \,,
\end{equation}
so that the lattice is integral and has signature $(p, q)$.  For a
consistent string theory, the world-volume theory must be modular
invariant.  As shown by Narain \cite{Narain, Narain-sw} (see also Polchinski
\cite{Polchinski-I} for
more details), modular invariance of the world-volume theory implies
that the lattice $\Gamma$ must be even and self-dual.  For $p = q$,
then, the lattice \eq{eq:torus-lattice} arising from toroidal
compactification is the only lattice  possible for a consistent string theory.

\subsubsection{Lattices and compactification of the heterotic string}

We can apply the preceding general discussion to the heterotic string.
The world-sheet degrees of freedom on the heterotic string are 10
chiral left-moving bosons and 26 chiral right-moving bosons.
(Note that we use the opposite convention from Polchinski for  the right-left
splitting of degrees of freedom to match the dominant negative
signature convention for the associated lattices.)  
In ten dimensions, ten of the right-moving bosons correspond to
space-time momenta, so the compactification lattice is  an even unimodular
lattice $\Gamma^{0, 16}$.  From \eq{eq:lattices-16} we know that there
are only two possibilities for 16-dimensional Euclidean even
unimodular lattices,
\begin{equation}
(-)\Gamma^{0, 16}=E_8 \oplus E_8, \; {\rm or} \;
(-)\Gamma^{0, 16}=\Lambda_{16}  \,.
\end{equation}
These are precisely the heterotic theories giving ${\cal N} = 1$ 10D
supergravity with the gauge groups $E_8 \times E_8$ and $SO(32)$.
Since $\Lambda_{16}$ is not exactly the root lattice of $SO(32)$, the
correct symmetry group for the latter theory is actually ${\rm
  Spin}(32)/\Z_2$, though it is generally referred to as the
``$SO(32)$'' theory.  The group Spin(32) is the simply connected cover
of $SO(32)$ just as $SU(2)$ is the simply connected cover of $SO(3)$.
Spin(32) has a center $\Z_2^2$, and the group SO(32) is given by the
quotient $SO(32) = {\rm Spin}(32)/\Z_2$ by a particular $\Z_2$
subgroup of the center.  The quotients by the other two $\Z_2$ factors
both give the same group, which is different from SO(32) and generally
denoted Spin(32)$/\Z_2$.  Further details related to the discrete
quotient are discussed in the references \cite{gsw-2, Witten-spin}.

Now let us consider compactification of the heterotic string to
dimensions below 10.
Consider first the compactification on $S^1$ to 9 dimensions.
In this case, one of the left-moving bosons and 17 of the right-moving
bosons correspond to internal degrees of freedom associated with the
compactification, and the resulting lattice is
\begin{equation}
\Gamma^{1, 17} = U \oplus (-E_8)
  \oplus (-E_8)\,.
\end{equation}
Note that this lattice is the unique even unimodular lattice of
signature $(1, 17)$.  It follows that compactification of both the
$E_8 \times E_8$ and Spin$(32)/\Z_2$ theories on a circle $S^1$ give
the same moduli space of theories in 9 dimensions.  

Compactifying to eight dimensions, the lattice is
\begin{equation}
\Gamma^{2, 18} = U \oplus U \oplus (-E_8) \oplus (-E_8)\,.
\end{equation}
The moduli space for this theory is the 36-dimensional space
\begin{equation}
SO(2, 18;\Z)\backslash SO(2, 18;\R)/SO(2)\times SO(18)\,,
\end{equation}
along with the single scalar in the gravity multiplet (the dilaton).
The generic gauge group is $U(1)^{20}$, as discussed above.  It is
possible to generate an enhanced gauge group by choosing moduli where
there are additional massless gauge bosons.  This can produce gauge
groups even larger than the original $E_8 \times E_8$ or $SO(32)$ of
the uncompactified theory.  In the heterotic string the ground state
for the left-moving oscillators is massless, like the type II string,
and only the ground state for right-moving oscillators is tachyonic.
So enhanced gauge symmetries only arise when $l_R^2 = 2, l_L^2 = 0$.
Since there are 18 right-moving bosons, the symmetry group is
determined by the sublattice of $\Gamma^{2, 18}$ lying in the
negative-signature 18-dimensional space spanned by the $l_R$.  The
possible enhanced nonabelian gauge symmetry groups of the theory are
therefore precisely the set of groups whose root lattices $G$ admit an
embedding into $\Gamma^{2, 18}$.
\begin{equation}
G: \; -\Lambda_G  \hookrightarrow \Gamma^{2, 18} \,.
\label{eq:8D-condition}
\end{equation}
Such groups can have rank up to 18.  For example, there is an
embedding
\begin{equation}
-D_{18} \hookrightarrow \Gamma^{2, 18}
\end{equation}
from which it follows that the group $SO(36)$ can be realized in the
8D heterotic theory \cite{Polchinski-II}.  Similarly, it was shown by
Ganor, Morrison and Seiberg \cite{Ganor-Morrison-Seiberg} that
$-A_{17}$ can be embedded in $\Gamma^{2, 18}$, associated with a
theory having gauge group $SU(18)/\Z_3$.

Determining which groups do or do not admit an embedding of the form
\eq{eq:8D-condition} is a nontrivial problem.  Fortunately, some
powerful theorems on lattice embeddings were proven by Nikulin
\cite{Nikulin} that state necessary and sufficient conditions for
such an embedding to be possible.  In many cases it is also possible
to prove that the embedding, when it exists, is unique up to lattice
automorphisms.  The precise statement of these theorems is somewhat
intricate; in many situations, however, the following simplified
theorem, which is a corollary of the stronger theorems, is sufficient.
\vspace*{0.05in}

\noindent
{\bf Theorem} [Nikulin]

Let $S$ be an even lattice of signature $(s_+,s_-)$ and let $T$ be an
even, unimodular lattice of signature $(t_+,t_-)$.  There exists an
embedding of $S$ into $T$
provided the following
conditions hold:
\begin{enumerate}
\item $t_+\geq  s_+ $ and $t_-\geq  s_-$
\item $t_++t_--s_+-s_-  > l (S^*/S)$
\end{enumerate}
where the lattice quotient $S^*/S$ is a finite abelian group, and $l
(S^*/S)$ is the minimum number of generators of this group $\Box$
\vspace*{0.05in}

It follows from this theorem, for example, that the rank 18 group
$SU(19)$ can be realized as the gauge group of a heterotic string
compactification in 8D.  The associated lattice $S =-A_{18}$ has $s_+=
0, s_{-}= 18$, so satisfies the first condition for $T = \Gamma^{2,
  18}$.  For $A_{n -1}$ the discrete group $S^*/S$ is $\Z_{n}$
\cite{Conway-Sloane} (an easy exercise for the reader is to check this
for $A_2$).  Thus, for $S =A_{18}$, $S^*/S=\Z_{19}$, which is a cyclic
group with one generator, so the second condition becomes $20-18 > 1$
and is satisfied.  The structure of the discrete group $S^*/S$ also
affects the global structure of the gauge group, for example leading
to the $\Z_3$ quotient in the case of $A_{17}$ mentioned above
\cite{Ganor-Morrison-Seiberg}.

The stronger theorems proven by Nikulin  \cite{Nikulin} use more
detailed aspects of the number-theoretic structure of $S^*/S$.  The
proofs are carried out using $p$-adic analysis.

This concludes our brief introduction to heterotic compactifications
to 8D.

\subsection{F-theory vacua in eight dimensions}
\label{sec:8D-F-theory}

We now turn to another approach to string compactifications, known as
{\em F-theory} \cite{Vafa-F-theory, Morrison-Vafa-I,
Morrison-Vafa-II}.  F-theory can be thought of in several ways: as a
limit of a class of type IIA or M-theory vacua, or as a framework for
characterizing nonperturbative type IIB vacua\footnote{Most of what I
know about F-theory I learned from David Morrison, my guru in the way
of F-theory.  Credit for any clear and correct insights in the
F-theory portions of these notes should go to him, while I of course
am responsible for all errors and confusion.}.  We will primarily
approach the subject from the latter perspective in these lectures.
Other pedagogical introductions to F-theory can be
found in the Les Houches notes of Denef \cite{Denef-F-theory} and in
the TASI notes by Morrison \cite{Morrison-TASI}.

\subsubsection{7-branes and geometry}
\label{sec:7-branes}

As described in Section \ref{sec:10D}, type IIB supergravity has a
classical $SL(2,\R)$ symmetry.  In the quantum theory, the spectrum of
string excitations charged under the $B$ fields is quantized, and this symmetry is broken to
$SL(2,\Z)$.  Just as the two-form fields $B, \tilde{B}$ transform
as a doublet under this symmetry, fundamental strings and
Dirichlet strings (D1-branes) also transform as a doublet under this
discrete group.  The group $SL(2,\Z)$ thus gives a group of
nonperturbative duality symmetries for the IIB string.  The axion and dilaton combine into a complex {\em
  axiodilaton}
\begin{equation}
\tau = \chi + i e^{-\phi}\,.
\end{equation}
The imaginary part
of the axiodilaton is positive,    so $\tau$ lives in the upper
half-plane of $\C$.  The axiodilaton transforms under $SL(2,\Z)$ as
\begin{equation}
 \tau \rightarrow \frac{a \tau + b}{c \tau + d} , \;\;\;\;\; ad-bc = 1\,.
\end{equation}
This group is
generated by the transformations
\begin{eqnarray}
T: &  &  \tau \rightarrow \tau + 1\\
S: &  &  \tau \rightarrow -1/\tau \,
\end{eqnarray}
and is precisely the group of transformations on the upper half-plane
corresponding to modular transformations on a two-torus parameterized
by the complex structure parameter $\tau$.  It is thus natural to
interpret the set of possible values for the axiodilaton as the
modular parameter for a two-torus.  In the language of algebraic
geometry, a two-torus equipped with a complex structure is an {\em
  elliptic curve}.

A D7-brane in the IIB theory is magnetically charged under the
axiodilaton. This means that the value of the axion field
undergoes a monodromy around a small circle surrounding the D7-brane
in $\R^9$, given by the $T$ modular transformation
\begin{equation}
\tau \rightarrow \tau + 1  \,.
\label{eq:t-monodromy}
\end{equation}
Acting on a D7-brane with an $SL(2,\Z)$ transformation produces a more
general class of $(p, q)$ 7-branes, with monodromies conjugate to
(\ref{eq:t-monodromy}).  A supergravity background containing a
configuration of $(p, q)$ 7-branes can be
geometrized by interpreting $\tau$ at each point in space-time as
parameterizing a torus locally fibered over space-time.  This leads to
the essential geometric picture of F-theory, a space that is {\em
  elliptically fibered} over space-time.
While this geometric picture suggests a 12-dimensional enhancement
of space-time, it is important to emphasize that there is no
underlying 12-dimensional supergravity theory associated with this
picture.  The presence of different kinds of 7-branes can produce an
axiodilaton configuration that may be strongly coupled in any duality
frame, and which is therefore essentially nonperturbative.
While F-theory captures many nonperturbative aspects of the moduli
space of supersymmetric string vacua, there are also limitations to
this approach as it is currently understood.  F-theory does not at
this time have an intrinsic definition including a dynamical
principle.  Thus, explicit computations, for example of the metric on
the compact space or the effect of internal fluxes, depend upon
finding a dual perturbative description in which the desired physics
can be formulated and computed more precisely.

An alternative to the type IIB picture of F-theory is through a limit
of compactification of M-theory.  Consider a compactification of the
11-dimensional M-theory on a torus with modular parameter $\tau$.
Treating one cycle of the torus as compactification of M-theory to
type IIA, we have the type IIA theory on a circle of radius $R$
parameterizing the other cycle of the original torus.  By T-duality,
as discussed in Section \ref{sec:toroidal-compactification}, this is
equivalent to type IIB on a circle of radius $\alpha'/R$.  In the
limit where the size of the original torus becomes small, $R
\rightarrow 0$ and this becomes the type IIB theory in an infinite
flat space with axiodilaton parameter $\tau$.  In principle, this
construction can be fibered over any base space, reproducing F-theory
as a limit of M-theory compactified on a torus shrunk to 0 size; for a
more detailed pedagogical description of this picture of F-theory, see
Denef's Les Houches notes \cite{Denef-F-theory}.

An F-theory
compactification is associated with an elliptic fibration over a
compact base space $B$ of complex dimension $d$, giving a space-time
theory in dimension $D = 10-2d$.  For the theory to be supersymmetric,
the total space of the elliptic fibration must be a Calabi-Yau
manifold.  Thus, for example, F-theory vacua in eight dimensions are
described by an elliptically fibered Calabi-Yau manifold with two
complex dimensions, F-theory vacua in 6 dimensions come from
elliptically fibered Calabi-Yau threefolds, and F-theory vacua in 4
dimensions come from elliptically fibered Calabi-Yau fourfolds.
We denote the total space of the elliptic fibration by $X$,
characterizing the fibration through the following diagram
\begin{center}
\begin{picture}(60,60)(- 30,- 28)
\put( 19, 19){\makebox(0,0){X}}
\put(19,-19){\makebox(0,0){B}}
\put(19,10){\vector( 0, -1){20}}
\put(-19, 19){\makebox(0,0){$T^2$}}
\put(-10,19){\vector(1, 0){20}}
\end{picture}
\end{center}

Away from 7-brane sources, the axiodilaton varies smoothly, describing
a smooth elliptic fibration over the base space.  At the position of a
7-brane, the elliptic fibration becomes singular.  This does not mean,
however, that the total space $X$ is necessarily singular.  As a rough
analogy, consider the two-sphere $S^2 =\{(x, y, z): x^2 + y^2 + z^2 =
1\}$ as a circle fibration over a line segment parameterized by $z \in
[0, 1]$.  At every $z \neq 0, 1$ there is a copy of the circle that
varies smoothly with $z$.  The fibration is singular at the endpoints,
where the circle degenerates to a point.  But the total space $S^2$ is
still smooth.  A similar thing happens in the vicinity of a $(p, q)$
7-brane in F-theory.  The total space $X$ is smooth, though the
structure of the fibration becomes singular.

We now focus specifically on compactifications of F-theory to 8
dimensions, the simplest class of F-theory vacua.  First we discuss
the structure of the base $B$ and elliptically fibered Calabi-Yau
complex two-fold $X$.  Like any codimension  two massive object, 7-branes
produce localized curvature in space-time that leads to a deficit
angle in the geometry away from the brane.  For a single 7-brane, the
metric at large distances becomes asymptotically flat, but the total
angle around the brane is reduced by $\pi/6$.  This can be seen from
the ratio between the circumference and diameter of a large circle,
which approaches $11 \pi/6$ (instead of $2 \pi$) as the circle
becomes large.  For F-theory to be defined on a compact space, we
need enough 7-branes to close the transverse 2D space into a compact
geometry.  With 12 7-branes, the deficit angle becomes $2 \pi$, so
that the asymptotic space becomes a cylinder (no change in
circumference at large radius).  Putting together two such
configurations, we can construct a topological sphere from a
space-time containing 24 7-branes.  This can also be understood from
the Gauss-Bonnet theorem, which states that the total curvature of a
sphere is $4 \pi$, where each 7-brane locally contributes $\pi/6$ to
the curvature.

\begin{figure}
\begin{center}
\centering
\begin{picture}(200,100)(- 100,- 50)
\put(0,0){\makebox(0,0){\includegraphics[width=7cm]{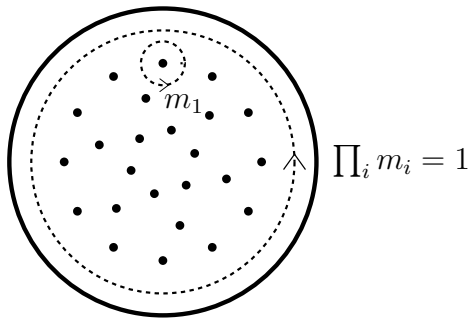}}}
\put(90,0){\makebox(0,0){$\prod_{i}m_i = 1$}}
\put(8, 22){\makebox(0,0){$m_1$}}
\end{picture}
\end{center}
\caption[x]{\footnotesize K3 as an elliptic fibration over the sphere
  $S^2$.  The fibration is singular at 24 points.  The product of the
  monodromies around the individual points must be the identity, as
  the associated curve is contractable.}
\label{f:K3-sphere}
\end{figure}
Thus, a compactification of F-theory to eight dimensions is associated
with a collection of 24 7-branes on a base space that is
topologically
a two-sphere
$B = S^2$.  Around each 7-brane there is a monodromy $m_i$ acting on
$\tau$.  The product of the monodromies must be trivial, $\prod_{i}m_i
= 1$, since a curve surrounding all 24 7-branes can be contracted to a
point on the sphere (see Figure~\ref{f:K3-sphere}).  Note that not all
the 7-branes can be D7-branes, since if each $m_i$ gives the modular
transformation \eq{eq:t-monodromy} then $\prod_{i}m_i$ would give the
modular transformation $T^{24}: \tau \rightarrow \tau + 24 \neq \tau$.
The total space $X$ of the elliptic fibration must be a Calabi-Yau
two-fold, and is not a torus if there is only one supersymmetry in 8
dimensions.  $X$ is therefore a K3 surface.  While the presentation of
K3 as an elliptic fibration over $S^2$ is the principal
characterization of K3 that we will use in studying F-theory, it is
useful to briefly digress on the structure of the K3 surface as seen
from other perspectives.

\subsubsection{The K3 surface}
\label{sec:K3}

We now introduce some basic features of the K3 surface.  For a much
more detailed introduction to the mathematics and physical
applications of K3 surfaces, see the  lecture notes by
Aspinwall on K3 \cite{Aspinwall-K3}.

The K3 surface is the only topological type of Calabi-Yau manifold of
complex dimension two besides the four-torus $T^4$, as mentioned
above.  K3 surfaces are simply connected, so they have no first
homotopy ($\pi_1$) or homology ($H_1$) structure.  K3 surfaces have a
22-dimensional second homology group $H_2 (K3,\Z)$.  The intersection
pairing on this homology group gives the lattice $\Gamma^{3, 19}$,
which as discussed above must take the form
\begin{equation}
H_2 (K3,\Z)= \Gamma^{3, 19}  = U \oplus U \oplus U \oplus (-E_8)\oplus (-E_8)\,.
\label{eq:K3-lattice}
\end{equation}

\begin{figure}
\begin{center}
\includegraphics[width=9cm]{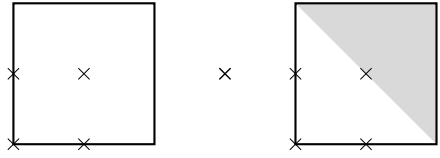}
\end{center}
\caption[x]{\footnotesize An orbifold limit of the K3 surface can be
  viewed as the space $T^4/\Z_2$, which is locally flat except at 16
  orbifold points (marked with ``x'''s).  The product of the regions
  on the two 2-tori
  that have not been marked in gray gives a fundamental domain for
  the orbifold space.}
\label{f:K3}
\end{figure}

One of the simplest ways to view the K3 surface is in a particular
limit where all the curvature is concentrated at points and the
surface can be viewed as an orbifold of a torus.  Consider the torus
$T^4$, considered as a product of two 2-tori with modular parameters
$\tau_1 = \tau_2 = i$, so that coordinates on the two 2-tori are
subject to the usual identifications on each 2-torus $z_i \sim z_i + 1
\sim z_i + \tau$.  We now take an orbifold of the space by imposing
the additional identification of points related through the $\Z_2$
action
\begin{equation}
\rho: (z_1, z_2)\rightarrow (-z_1, -z_2) \,.
\end{equation}
The resulting space is flat everywhere except at 16 orbifold points that are
locally of the form $\C^2/\Z_2$ (see Figure~\ref{f:K3}).
This singular limit of the K3 surface can be continuously deformed
into a smooth K3 surface by {\em blowing up} the 16 singular points.
The blowing-up procedure, which will be described in more detail
below, essentially involves replacing a singular point in a surface
with a complex projective space $\C\P^1$ consisting of the set of
limiting lines approaching the singular point (note that all
projective spaces discussed in these lectures are complex projective
spaces; from here on we denote $\P^n$ for $\C\P^n$.).  For a
singularity of the form $\C^2/\Z_2$, blowing up the singular point at
the origin produces a smooth space.  From this picture we can see
fairly clearly how the 22 homology cycles arise.  There are six
homology classes $ \bar{\pi}_{ij}$ in the original $T^4$ corresponding
to cycles wrapped around the $ij$ coordinate axes (where the real
coordinate axes are labeled by $i = 1, 2, 3, 4$).  In the $T^4$ these
have intersection numbers
\begin{equation}
\bar{\pi}_{ij} \circ \bar{\pi}_{kl} = -  \;\epsilon_{ijkl} \,.
\end{equation}
In the orbifold space, there are closely related cycles $\pi_{ij}$.
These lift in the covering space to two copies of $\bar{\pi}_{ij}$,
with 4 intersections in the covering space.  The intersection in the
orbifold space is then
\begin{equation}
\pi_{ij} \circ \pi_{kl} = -  2 \;\epsilon_{ijkl} \,.
\label{eq:orbifold-intersection}
\end{equation}
Another 16 cycles come from the $\P^1$'s
formed by blowing up the 16 singular points.  We denote these cycles
$e_{ij}$, where the values of the indices
$i, j \in \{1, \ldots,  4\}$
correspond to the points $(0, 0), (1/2, 0), (0, \tau/2), (1/2,
\tau/2)$ on the two toroidal factors $T^2$.
The inner product on these cycles
is $e_{ik} \cdot e_{jl} = -2 \;\delta_{ij} \delta_{kl}$.
The 22 cycles $\pi_{ij}, e_{ij}$ span a
22-dimensional lattice.  This is not quite the complete homology
lattice of K3, however.
This is a
sublattice, known as the Kummer lattice,
of the full homology lattice.
To complete
the full lattice additional fractional cycles must be added, such as
\begin{equation}
\gamma_{13} =\frac{1}{2}\pi_{13} +\frac{1}{2}(e_{11}+e_{12}+e_{21}+e_{22}) \,.
\end{equation}
These cycles are topologically spheres, which can be seen from
$\gamma_{13} \cdot \gamma_{13} = -2 = 2g-2$ where $g = 0$ is the genus
of the surface\footnote{This formula relating the intersection form
  of a curve with itself to the genus of the curve is the special case
  of a more general formula (\ref{eq:curve-genus}), which we derive
  later, for a curve in flat or toroidal space.}.  Including a set of
such fractional cycles gives a complete set of generators for the
homology lattice (\ref{eq:K3-lattice}) of K3, as reviewed in more
detail in several references \cite{Aspinwall-K3, Blumenhagen-6D,
  kt-K3}.

\subsubsection{Elliptically fibered K3 surfaces}

Now we return to our discussion of F-theory.  As described above,
F-theory on a K3 surface is described geometrically by an elliptic
fibration over the sphere $S^2$ with 24 singularities, where each
singularity has a monodromy conjugate to $\tau \rightarrow \tau + 1$.
Locally, an elliptic fibration can be described by a modular
parameter $\tau$ varying holomorphically over the base $\P^1$.
Generically, the total space of such a fibration is a smooth K3
surface.  

A convenient algebraic-geometric description of an elliptic curve is
as the set of points in the projective plane satisfying a cubic
equation.  More specifically, consider the projective space $\P^{2,
  3, 1}$ defined by the set of coordinates $(x, y, z)$ with the
equivalence
\begin{equation}
(x, y, z) \sim (\lambda^2 x, \lambda^3y, \lambda z), \;
\forall \lambda \in\C \setminus\{0\} \,.
\label{eq:p-equivalence}
\end{equation}
The equation
\begin{equation}
F = -y^2+ x^3 + fx z^4 +gz^6 = 0
\label{eq:elliptic-curve}
\end{equation}
defines a complex curve of genus one, {\it i.e.}, an elliptic curve,
for general complex values of $f$ and $g$.  We can take a local
coordinate chart leaving out the points where $z = 0$ by using eq.\
(\ref{eq:p-equivalence}) to set $z =1$ so that the curve in the local
$(x, y)$ coordinate chart is defined by
\begin{equation}
F = -y^2+ x^3 + fx  +g  = 0\,.
\label{eq:elliptic-2}
\end{equation}
This is known as the {\em Weierstrass} form of the elliptic curve.
One easy way to see that this equation should define a genus one curve
is to think of $x$ as giving a local coordinate chart on $\P^1$.  The
vanishing of $F$ then gives $y$ as a double-valued function of $x$,
with branch points at the points where the cubic $x^3+ fx + g$
vanishes.  Generically there are 3 such branch points, and another
branch point at $x = \infty$ ($z = 0$), so \eq{eq:elliptic-curve}
describes a branched cover of $\P^1$ with 4 branch points, which is
topologically a torus or genus one curve.  Note that while there are
two complex parameters $f, g$ defining the Weierstrass model, the
moduli space of elliptic curves is only one-dimensional; rescaling $f
\rightarrow \lambda^4f, g \rightarrow \lambda^6g, x \rightarrow
\lambda^2 x, y \rightarrow \lambda^3y$ gives equivalent elliptic
curves.

We can now give a more precise mathematical description of a K3
surface realized as an elliptic fibration over a complex
one-dimensional base space such as $B =\P^1$.  Consider $w$ as a local
coordinate on $B$.  An elliptic fibration is given locally by a
choice of elliptic curve at each point on $B$.  We can parameterize such
a family of elliptic curves in the Weierstrass form by taking $f, g$
to be functions of $w$
\begin{equation}
F = -y^2+ x^3 + f(w)x  +g (w)  = 0\,.
\label{eq:Weierstrass-w}
\end{equation}
This gives a local Weierstrass description of the elliptic fibration
as a hypersurface in $\C^3$.  To give the global description of the
surface, we can reinstate the projective variable $z$ in the fiber,
and a projective variable $v$ in the base.  The Weierstrass form for
an elliptic fibration selects a point at $\infty$ on each elliptic
curve (the point at $z = 0$).  This choice of a point on each fiber
defines a global section of the fibration.  Thus, F-theory is defined by
a compactification on an elliptically fibered Calabi-Yau {\em with
  section}.

A (complex) codimension one space defined as the vanishing locus in
$\C^n$ of a single function such as \eq{eq:elliptic-2} is singular
when all derivatives of the function simultaneously vanish.  Such
singularities are treated rigorously in the mathematical framework of
algebraic geometry, but we can understand this simply by noting that
level surfaces of the function define locally smooth complex analytic
(and algebraic) sets as long as the gradient of the function is
nonvanishing.  Thus, the elliptic curve defined by \eq{eq:elliptic-2}
is singular at a point $(x, y)$ when
\begin{equation}
F =\frac{\partial F}{\partial y}  =
\frac{\partial F}{\partial x}  =  -2y = x^3 + f x + g = 3x^2 + f = 0 \,,
\label{eq:singularity}
\end{equation}
Combining these equations, we have
$3x^3 + 3fx + 3g = 2fx + 3g = 0 \Rightarrow
x = -3g/2f$, so the curve is singular when the
{\em discriminant} $\Delta$ vanishes
\begin{equation}
\Delta = 4f^3 +27g^2 = 0 \,.
\label{eq:discriminant}
\end{equation}

Considering again the elliptic fibration over the base $\P^1$, we take
$f (w)$ and $g (w)$ to be functions on the base, and the K3 surface is
locally defined by \eq{eq:Weierstrass-w}.  At a singularity where the
discriminant \eq{eq:discriminant} vanishes the partial derivative of
$F$ with respect to $w$, $\partial F/\partial w = f' x + g'$, is
generically nonzero, so at such points the total space of the K3
surface is locally nonzero even though the fiber degenerates.  We
generically expect 24 singularities in the elliptic fibration as
discussed above.  Thus, the discriminant should be a degree 24
polynomial in the coordinate $w$ on the base.  We therefore expect
that $f (w)$ has degree 8 and $g (w)$ has degree 12.

It is helpful to view $f, g$, and $\Delta$ from a more global
perspective.  For this we need a little more mathematical machinery,
which will be extremely helpful in understanding F-theory
compactifications on Calabi-Yau manifolds of higher dimension.  While
these notes are intended to be relatively self-contained, the reader
interested in understanding the concepts outlined in the next few
paragraphs more thoroughly may find it useful to study some basic
aspects of algebraic geometry.  For the material presented here, the
text of Perrin \cite{Perrin} is a good starting point.  The book by
Barth, Hulek, Peters, and Van de Ven \cite{complex-surfaces} also
contains a great deal of useful material on compact complex surfaces,
including aspects of the structure of elliptic fibrations.

While locally $f$ and $g$ are functions of the  coordinate $w$
taking values in $\C$, when these functions are considered on the
complete base $B = \P^1$ they are actually sections of {\em line
  bundles} over $\P^1$.  Complex line bundles over any complex space $X$ are
characterized by their first Chern class $c_1 (X)$, and are thus
associated with an element of $H^2 (X, \Z)$.
In particular, for $X = \P^1$, the first Chern class is in
$H^2 (\P^1, \Z) =\Z$.  

We will freely move between the cohomology representative of a line
bundle on a space $X$ and the Poincar\'{e} dual homology class, which
is a {\em divisor} on $X$.  A divisor on $X$ is a linear combination
(over $\Z$) of irreducible algebraic hypersurfaces of the form
$\sum_{i}n_iH_i$.  There is a one-to-one correspondence between
divisors and line bundles over any complex space $X$.  The divisor
associated with a line bundle ${\cal L}$ can be thought of as the
linear combination of the homology class containing the zero locus of
a section of ${\cal L}$ and the negative of the homology class
containing the poles of the section.  We will refer to the line bundle
associated with a divisor $D$ as ${\cal O} (D)$.  On $\P^1$ the only
irreducible algebraic hypersurface is (the homology class of) a point
$P$, so every divisor is characterized as a multiple $nP$.  For
example, the function $z$ on $\P^1$ can be extended to a section of
the line bundle $T \P^1 ={\cal O} (2P)$, and vanishes at two points (0
and $z =\infty \rightarrow w = 1/z = 0$).  On the other hand, a
section $dz$ of the cotangent bundle goes to $dz= -dw/w^2$ on the
chart around $z = 1/w = \infty$, and has a double pole at $w = 0$, so
$T^*\P^1 ={\cal O} (-2P)$.  We can take the product of two line
bundles by multiplying the transition functions; this corresponds to
adding the corresponding divisors ${\cal O} (D) \otimes{\cal O} (F)
={\cal O} (D + F)$.

On any complex manifold there is a special class known as the {\em
  canonical class}, $K$.  On a complex manifold of dimension $d$, the
canonical class corresponds to the line bundle associated with the
$d$th power of the cotangent bundle, and is given locally by the
maximum antisymmetric power of the holomorphic differential, $dz_1
\wedge \cdots \wedge dz_d$.  The canonical class is essentially a
measure of the total curvature of a space.  On $\P^1$ the canonical
class is
\begin{equation}
K= c_1 (T^*) = -2P\,.
\end{equation}
A manifold in any complex dimension is Calabi-Yau if and only
if\footnote{Technically, manifolds with $K$ in a torsion class, $nK =
  0$, are sometimes classified as Calabi-Yau, but we do not worry
  about such issues here.} the manifold is K\"ahler and $K = 0$.

Given the machinery just defined, we can now understand a formula
that will be useful in studying several relevant aspects of
curves on surfaces.  For a smooth curve $C$ on a surface $S$ with
canonical class $K_S$, we have
\begin{equation}
(K_S + C) \cdot C = 2g-2 \,.
\label{eq:curve-genus}
\end{equation}
This statement follows from a fairly straightforward three-line
argument that uses some important theorems from elementary algebraic
geometry.  The first step is the {\em adjunction formula}, which
determines the canonical class on $C$ in terms of $K_S$, ${\cal O}
(K_S) \otimes{\cal O} (C) |_C ={\cal O} (K_C)$, where $|$ denotes the
restriction of the line bundle to $C$.  The second step is relating
this restriction to the intersection form using ${\cal O} (D) |_C = D \cdot
C$.  And the third step uses the {\em Riemann-Roch} theorem, which
essentially says that on a smooth curve $C$, ${\cal O} (K_C) = 2g-2$.
In flat space, or on a torus or K3 surface, $K_S = 0$ so (\ref{eq:curve-genus})
becomes $C \cdot C = 2g-2$, which we used in discussing the K3 surface
earlier.

As another simple application of the adjunction formula, consider a
hypersurface $D$ in projective space $\P^{d + 1}$ defined by a degree
$d + 2$ homogeneous polynomial.  There is a single irreducible
hypersurface $H$ (up to linear equivalence) on $\P^{d + 1}$,
associated with the vanishing of any coordinate function $z_i$.  The
canonical class of $\P^{d +1}$ is $K ={\cal O}(-(d + 2)H)$, while the
divisor $D$ associated with a degree $d + 2$ polynomial is $D = (d +
2)H$, so the canonical class on $D$ is ${\cal O} (K_D) ={\cal O} (K +
D) |_D = 0$.  Thus, any such $D$ is a Calabi-Yau.  This gives an
alternative proof that the cubic on $\P^2$ gives an elliptic curve.
Similarly, a quartic on $\P^3$ gives a K3 surface, and a quintic on
$\P^4$ gives a Calabi-Yau threefold.

Returning to our elliptically fibered K3 surface, in terms of the
canonical class, $f$ and $g$ are sections of the line bundles ${\cal
  O} (-4K)$ and ${\cal O} (-6K)$, and $\Delta$ is  a section
of the bundle ${\cal O} (-12K)$.  Characterizing $\Delta$ by the
curvature class of the associated bundle, the condition for an
elliptic fibration to describe an elliptically fibered Calabi-Yau
surface over $B$ can be written as
\begin{equation}
-12K = \Delta \,.
\end{equation}
We will refer to this condition as the ``Kodaira condition''.  It is a
special case of the relation proven by Kodaira between the canonical
class of the total space of an elliptic fibration and the canonical
class of the base.  In the case $B = \P^1$, the Kodaira condition states that
$\Delta = 24 P$, which is just the statement above that the
singularity locus consists of 24 points on the sphere $S^2$.

The space of elliptically fibered K3 surfaces is parameterized by the
coefficients of the polynomials (sections) $f (w), g (w)$.  The
coefficients are moduli that parameterize the complex structure of
the K3 surface just as $\tau$ parameterizes the complex structure of
$T^2$.  The number of coefficients is 9 + 13 = 22.  This
parameterization is, however, redundant.  Just as constants $f, g$
provide one extra parameter for a single elliptic curve with a
redundancy under scaling, the homogeneous functions $f (w, v), g (w,
v)$ have four redundant degrees of freedom through general linear
transformations on the homogeneous coordinates $w, v$.  Thus, the
number of complex degrees of freedom for the set of elliptically
fibered K3 surfaces (with section) is $22-4 = 18$, in agreement with
the number of degrees of freedom found in the heterotic construction.
One additional scalar field is associated with the volume of the base
$B = \P^1$.

\subsubsection{Gauge groups and singularities}

As discussed above, the generic elliptic K3 is smooth.  The resulting
F-theory vacuum is an 8D supergravity theory with an abelian gauge
group $U(1)^{20}$ (where as usual two of the $U(1)$'s lie in the
gravity multiplet.)  The derivation of the rank of the gauge group
from the F-theory picture is actually somewhat subtle.  The abelian
gauge group factors arise from global structure of the
compactification, unlike nonabelian gauge group factors, which, as we
discuss below, arise from the geometry in a simpler local fashion.
Naively it might seem that the gauge group should be $U(1)^{24}$, with
each 7-brane associated with a singularity giving a separate $U(1)$
factor.  Global constraints, however, reduce the rank of the gauge
group to 20.  This cannot be understood easily from a simple
supergravity picture, since the 7-branes cannot all be described
perturbatively in the same duality frame.  The constraint to rank 20
can be understood from the topology of the global K3, or from the
point of view of probe branes on the base, which interact with the
7-branes through open strings and string junctions
\cite{DeWolfe-Zwiebach, DeWolfe-hi-Zwiebach}; more geometrically this
corresponds to the structure of rational sections of the fibration
(Mordell-Weil group) \cite{Morrison-Vafa-II, Fukae-mw}, which encode
part of the topology of the global K3 space.

Now let us consider the non-generic situation in which several of the
7-branes on the spherical base $B$ coincide.  This leads to a more
complicated singularity in the elliptic fibration and often to a
singularity in the total space of the K3.  At such a singularity of
the K3, one or more two-cycles may shrink to a point.  When a
two-cycle shrinks to a point, the theory develops an extra massless
gauge boson.  This gives rise to an enhanced gauge group in the 8D
supergravity theory.  In the M-theory/IIA picture, the extra gauge
boson comes from a membrane/D2-brane wrapped on the vanishing cycle
\cite{Witten-dynamics}.  It was shown by Kodaira that the possible
types of singularity in a complex surface that can be resolved by a
succession of blow-ups can be systematically classified \cite{Kodaira,
  Kodaira-2}.  For each type of singularity there is a corresponding
Dynkin diagram, encoding the intersection form on the shrunk
two-cycles.  In the F-theory picture, these Dynkin diagrams precisely
characterize the nonabelian gauge group arising in that singular limit
of the K3 surface.
Note the chronic sign difference between the intersection form on the
two-cycles and the Cartan matrix; this arises from a difference in
conventions between the algebraic-geometric and Lie algebra frameworks.

In the following section we work through an explicit example of an
F-theory singularity of type $A_3$.  In general, the singularity type
depends upon the degree of vanishing of $f, g$ and $\Delta$ at the
singular point.  For example, when $f$ and $g$ are nonvanishing, and
$\Delta$ vanishes to order $n$, there is a singularity of the form
$A_{n -1}$, associated with nonabelian gauge group $SU(n)$.  The
complete list of Kodaira singularity types is given in
Table~\ref{t:Kodaira}.
\begin{table}
\begin{center}
\begin{tabular}{|c |c |c |c |c |}
\hline
ord ($f$) &
ord ($g$) &
ord ($\Delta$) &
singularity & nonabelian symmetry \\ \hline \hline
$\geq $ 0 & $\geq $ 0 & 0 & none & none \\
0 & 0 & $n$ & $A_{n-1}$ & $SU(n)$ \\
 $\geq 1$ & 1 & 2 & none & none \\
1 & $\geq 2$ &3 & $A_1$ & $SU(2)$ \\
 $\geq 2$ & 2 & 4 & $A_2$ & $SU(3)$ \\
2 & 3 & $n +6$ & $D_{n +4}$ & $SO(8 +2 n)$  \\
$\geq 2$ & $\geq 3$ & $6$ &$D_{4}$ & $SO(8)$  \\
 $\geq 3$ & 4 & 8 & $E_6$ & $E_6$ \\
3 & $\geq 5$ & 9 & $E_7$ & $E_7$ \\
 $\geq 4$ & 5 & 10 & $E_8$ & $E_8$ \\
\hline
\end{tabular}
\end{center}
\caption[x]{\footnotesize  Table of singularity types for elliptic
  surfaces and associated nonabelian symmetry groups.}
\label{t:Kodaira}
\end{table}
It is an educational exercise to work through the blow-up procedure
for some of the different singularities in the table, and to verify
the appearance of the stated Dynkin diagram in the intersection form
of the blown-up $\P^1$'s.  When a (complex) codimension one
singularity occurs which has degrees higher than any allowed in the
Kodaira table, the geometry cannot be resolved to a space which is
locally Calabi-Yau.  In Section \ref{sec:6D-F-theory}, we discuss
non-Kodaira singularities of higher codimension, which can lead to
physically interesting transitions in the space of theories.

We now have a general picture of how F-theory describes 8D
supergravities through compactification on elliptically fibered K3
surfaces.  The elliptically fibered K3 surface is described through a
Weierstrass equation of the form \eq{eq:Weierstrass-w}, where $f, g$
are degree 8 and degree 12 polynomials respectively on the base $\P^1$
(sections of ${\cal O} (-4K), {\cal O} (-6K)$ respectively in the
global picture).  Generically the gauge group is $U(1)^{20}$, though
as the moduli vary the singularities associated with vanishing of
$\Delta$ from \eq{eq:discriminant} can coincide, giving more
complicated singularity types and enhancing the 8D gauge group to
include nonabelian factors.

As a global example of an F-theory vacuum on an elliptically fibered
K3 with large symmetry group, we describe the theory with $E_8 \times
E_8$ gauge symmetry, following Morrison and
Vafa \cite{Morrison-Vafa-II}.  Choose
\begin{equation}
f = \alpha z^4, \;\;\;\;\;
g = z^5 + \beta z^6 + z^7\,.
\label{eq:8D-e8}
\end{equation}
This gives
\begin{equation}
\Delta = 27z^{10} + \cdots + 27z^{14} \,.
\end{equation}
This Weierstrass model has $E_8$ singularities at $z = 0, \infty$, as
can be verified from Table~\ref{t:Kodaira}.

We can now address the question of which gauge groups $G$ can be
realized in eight dimensions through an F-theory compactification on
an elliptically fibered K3.  The condition that the K3 be elliptically
fibered and have a section identifies two cycles $f, s$ on the total
space with intersection products $f \cdot f = 0, f \cdot s = 1, s
\cdot s = -2$.  The linear combinations $f, s + f$ then have
intersection products in K3 given by a copy of $U$.  The elliptic
fibration thus removes a factor of $U$ from $H_2 (K3;\Z) = \Gamma^{3,
  19}$, giving $\Gamma^{2, 18}$ \cite{Ganor-Morrison-Seiberg}.  In
this remaining lattice, we can shrink any combination of two-cycles
satisfying $c \cdot c = -2$ to get nonabelian gauge bosons.  The set
of nonabelian gauge groups that can be realized is then, just as in
the heterotic theory, the set of $G$ such that
\begin{equation}
G: \; -\Lambda_G  \hookrightarrow \Gamma^{2, 18} \,.
\label{eq:8D-condition-2}
\end{equation}
Again, as in the heterotic theory, the rank of the gauge group when
all $U(1)$ factors is included is always 20.  

\subsubsection{Example: $A_3$ singularity}

It is helpful to study the geometry of a particular case to understand
the general principles of the Kodaira classification.
Consider  an elliptic fibration over a local patch in $\C$
parameterized by $w$,
given by the Weierstrass form
(\ref{eq:Weierstrass-w}) with
\begin{eqnarray}
f & = &  -\frac{1}{3}-w^2 \label{eq:a3-fg}\\
g & = &  \frac{2}{27}   +\frac{1}{3}  w^2 \,.  \nonumber
\end{eqnarray}
While $f$ and $g$ are both nonvanishing at $w = 0$, a small
calculation shows that
the discriminant vanishes to order $w^4$
\begin{equation}
\Delta = 4 f^3 + 27 g^2  = -w^4-4w^6 \,.
\label{eq:discriminant-a3}
\end{equation}
At $w = 0$, $F = -y^2 + x^3-x/3 + 2/27$, which has a singularity at $x
= -1/3, y = 0$.
To simplify the analysis, it is convenient to change coordinates
\begin{equation}
x \rightarrow x + \frac{1}{3}
\end{equation}
to move the singularity to $x = 0$.  The Weierstrass equation then
becomes
\begin{equation}
F =
-y^2 + x^3 + x^2 -w^2 x = 0 \,.
\label{eq:a3}
\end{equation}
This gives a local equation for the complex surface described by the
elliptic fibration in coordinates $(x, y, w) \in \C^3$.  The surface
has a singularity at $x = y = w = 0$.
This singularity can be
resolved by blowing up the codimension one
singularity repeatedly until the space is smooth.  We can do this by
working in a sequence of coordinate
charts containing the various blow-ups.  
We refer to the local chart where the surface is defined through
\eq{eq:a3} as Chart 0.
\vspace*{0.05in}

\noindent
{\bf Chart 1}: 

To resolve the singularity in chart 0, we blow up the singular point.
We replace
the point $(0, 0, 0)$  with a $\P^2$ given by the set of limit points
described by homogeneous coordinates
$[x:y:w]$ along curves approaching $(0, 0, 0)$.  
The new space produced by the blow-up process can be described as a
subspace of $\P^2\times\C^3$, where homogeneous coordinates $[u:v:t]$ on the
$\P^2$ satisfy the relations
\begin{equation}
uy = vx, \; uw = tx, \; vw = ty \,.
\label{eq:blow-up-relations}
\end{equation}
When one or more coordinates $x, y, w$ are nonzero a unique point in
the $\P^2$ is determined, $[u:v:t] \sim [x:y:w]$. At the point $(x,
y, w)= (0, 0, 0)$, however, \eq{eq:blow-up-relations} imposes no
conditions on the coordinates $u, v, t$.

This blow-up process can be described in different coordinate charts
on the $\P^2$.
We choose, for example, a local
chart where $w \neq 0$.  
In this chart we cannot have $t = 0$ (or $u = v = 0$ also from
\eq{eq:blow-up-relations}), so we fix the homogeneous coordinates on $\P^2$ using $t = 1$.
We then have
\begin{equation}
x = uw, \; \; y = vw \,.
\label{eq:}
\end{equation}
Relabeling $u \rightarrow x_1, v \rightarrow y_1$,
this chart (Chart 1)
is realized by changing coordinates to
\begin{equation}
(x, y, w) = (x_1 w_1, y_1 w_1, w_1) \,.
\end{equation}
For $w_1 \neq 0$, there is a unique point $(x, y, w)$ in the original
chart (chart 0) corresponding to
each point $(x_1, y_1, w_1)$ in the new chart.  For $w_1 \rightarrow
0$, however, there is a set of points $(x_1, y_1, 0)$ in the new chart
that all correspond to $(0, 0, 0)$ in Chart 0.
This gives a local patch on the new $\P^2$ formed from the blow-up,
containing the points $[x_1:y_1:1]$.
In  the coordinates of Chart 1, the local equation (\ref{eq:a3})
becomes
\begin{equation}
F =
(-y_1^2 +  x_1^2 + x_1^3w_1-w_1x_1) w_1^2 = 0\,.
\label{eq:a3-t1}
\end{equation}
The $\P^2$ that is added through the blow-up process is known as the
{\it exceptional divisor} associated with the blow-up.  
The resulting equation \eq{eq:a3-t1} is reducible and contains two
copies of the exceptional divisor at $w_1 = 0$.  
Factoring out
the overall $w_1^2$ from (\ref{eq:a3-t1}) ({\it i.e.}, removing the
copies of the exceptional divisor), we have the equation for the
{\it proper transform} of the original space
\begin{equation}
F_w =
-y_1^2 +  x_1^2 + x_1^3w_1-w_1x_1  = 0\,.
\label{eq:a3-t}
\end{equation}
This equation describes the surface in chart 1 after the
original singularity at $(x, y, w) = (0, 0, 0)$ has been blown up.
(We use the subscript on $F$
to denote the coordinate chart used for the blow-up.)
The intersection of the space defined through (\ref{eq:a3-t}) with the
exceptional divisor at $w_1 = 0$ gives the exceptional divisor on the
surface, which is generally a curve or set of curves
associated with blowing up the point at $w_1 = 0$ within the surface
defined by $F$.  At $w_1 = 0$,
(\ref{eq:a3-t}) becomes $-y_1^2 +  x_1^2 = 0$
\begin{equation}
y_1 = \pm x_1 \,.
\label{eq:a3-c}
\end{equation}
This defines a pair of curves in $\P^2$, which we call $C_1^{\pm}$.
The equation (\ref{eq:a3-t}) still contains a singularity at the point
$(x_1, y_1, w_1)= (0, 0, 0)$, where the curves $C_1^\pm$ cross.  So we
must again blow up the singularity to produce a smooth space.
\vspace*{0.05in}

\noindent
{\bf Chart 2}: 

We replace the singular point in Chart 1 with another exceptional divisor
$\P^2$, this time using the local coordinates
\begin{equation}
(x_1, y_1, w_1) = (x_2, y_2  x_2, w_2 x_2)  = 0\,.
\end{equation}
After removing two copies of the exceptional divisor $x_2 = 0$
we get the new
local equation
\begin{equation}
F_{wx} =
-y_2^2 + 1 + x_2^2 w_2-w_2  = 0\,.
\label{eq:a3-tx}
\end{equation}
This gives another exceptional curve $C_2$, associated with the
intersection of (\ref{eq:a3-tx}) with the exceptional divisor $x_2 =
0$
\begin{equation}
C_2 =\{(x, y, w): x = 0,  w = 1 -y^2\} \,.
\label{eq:a3-c2}
\end{equation}
Since (\ref{eq:a3-tx}) has no further singularities, we have
completely resolved the local singularity and have a smooth space in
coordinate chart 2.

\begin{figure}
\begin{center}
\begin{picture}(200,100)(- 100,- 50)
\put(-100,0){\makebox(0,0){\includegraphics[width=7cm]{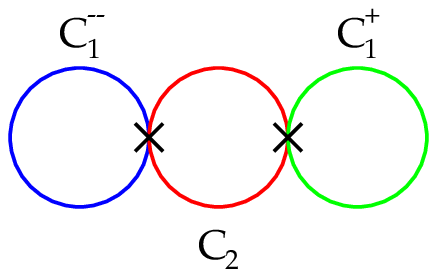}}}
\put(-100,-40){\makebox(0,0){(a)}}
\put(100,0){\makebox(0,0){\includegraphics[width=7cm]{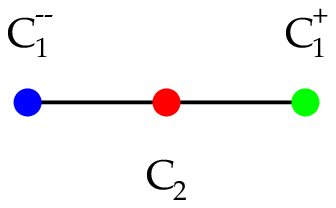}}}
\put(100,-40){\makebox(0,0){(b)}}
\end{picture}
\end{center}
\caption[x]{\footnotesize Exceptional curves resolving an $A_3$
  singularity, depicted as (a) geometry of the exceptional divisors, which take the
  form of $\P^1 = S^2$'s with associated intersections, (b) associated
  Dynkin diagram.  }
\label{f:a3}
\end{figure}

From the way in which the exceptional curves $C_1^\pm, C_2$ intersect,
we identify the $A_3$ form of the singularity found by Kodaira.  To
compute the intersections, we write (\ref{eq:a3-c}) in terms of
coordinates in chart 2
\begin{equation}
y_2x_2 = \pm x_2 \; \Rightarrow
\; y_2 = \pm  1\,,
\end{equation}
which combined with $w_1 = w_2x_2 = 0$ gives the points $[1:\pm 1:0]$ in homogeneous coordinates on $C_2 =\P^2$, showing that
$C_1^\pm$ each intersect $C_2$ at a single point but do not intersect
one another.  The intersections of these curves are shown graphically
in Figure~\ref{f:a3}, along with the associated Dynkin
diagram.
In principle, we would need to check all coordinate charts at each
step to be sure we have not missed any additional singularities.  Here
we have chosen charts that completely resolve the original
singularity.

The example just described is an $A_3$ singularity.  The shrinking
two-cycles in the K3 describe an extra set of massless gauge bosons
that extend the gauge group to a nonabelian group.  In particular,
the cycles $C_1^{\pm}, C_2$ form the simple roots of the Lie algebra
$SU(4)$.  By taking linear combinations of these cycles we can produce
all the nonzero roots of $SU(4)$, each of which corresponds to a curve
in the K3 with $c^2 = -2$.  When this type of singularity
occurs at a point in a global elliptically fibered K3 compactification
of F-theory, the resulting gauge group for the 8D theory has an
$SU(4)$ factor.  The rank of the group stays unchanged in such an
enhancement, so the total rank is still 20.

\subsection{The space of 8D supergravities}
\label{sec:8D-global}

We have now seen that two very different string-theoretic
constructions give rise to the same space of 8D supergravity theories.
For both the heterotic and F-theory constructions, there is a
low-energy 8-dimensional theory with one supergravity multiplet and
vector multiplets forming a rank 20 gauge group.  At a generic point
in this moduli space the gauge group is $U(1)^{20}$, but at special
loci in the moduli space the gauge group is enhanced.  The possible
nonabelian gauge groups that can be realized in each description are
precisely those whose root lattices can be embedded (with a factor of
$-1$) into $\Gamma^{2, 18}$.  The moduli space of theories has 37
dimensions, parameterized by the scalars in the gravity and vector
multiplets.  In both the heterotic and F-theory descriptions, this
moduli space of theories is a connected space.

The hypothesis that the heterotic string compactified on the torus
$T^2$ and F-theory compactified on K3 give rise not only to the same
low-energy supergravity theory, but to the same physical theory at the
nonperturbative level, asserts the existence of a duality symmetry
between these two classes of string vacua.  This duality symmetry,
first formulated by Vafa \cite{Vafa-F-theory},
underlies many duality symmetries relating lower-dimensional
theories.  While, as we have seen, this duality symmetry holds at the
level of the field content and symmetries of the theories, a complete
mathematical proof of this and all other string duality symmetries is
still lacking---in large part because there is as yet no
mathematically complete definition of string theory.  Nonetheless, the
duality between these two ostensibly very different realizations of 8D
supergravity theories has been explored in a number of ways beyond
what we have described here.  An explicit correspondence between the
moduli has been identified for some classes of heterotic
compactifications \cite{Morrison-Vafa-II, Barrozo}.  The duality
between F-theory on K3 and type I string theory on $T^2$ was explored
in detail by Sen \cite{Sen-F-theory-orientifolds}.  Protected one-loop
corrections to the effective action on the heterotic side have also
been matched from the geometry of F-theory \cite{Bachas-8D,
  Bachas-8D-2, Lerche-s, Lerche}, suggesting some deeper geometric
structure underlying these theories.  All evidence suggests that the
heterotic and F-theory compactifications are physically equivalent;
this suggests in turn that there may be a unique quantum supergravity
theory in 8 dimensions with minimal supersymmetry.

Classically, we can couple the 8D supergravity theory to any number of
vector multiplets, and the restriction to a rank 20 gauge group that
embeds through eq.\ (\ref{eq:8D-condition-2}) is not apparent.  As
for ten-dimensional ${\cal N} = 1$ supergravity, we would like to
understand whether quantum consistency restricts the gauge group in a
way that matches the range of models realized in string theory.
There is at this time no complete argument from the point of view of
the macroscopic supergravity theory that restricts the gauge group of
the theory in this way.  It does seem plausible, however, that quantum
consistency of the supergravity theory may impose additional
constraints that have not yet been fully elucidated.  One possibility
is that extra consistency conditions on the gauge group may arise from
supersymmetric constraints on the geometric structure of the moduli
space.  

While space-time anomalies in the 8D theories are not restrictive, we
can also consider anomalies on the world-volume of solitonic string
excitations of the theory; such anomalies may impose additional
constraints on the theory.  Every 8D supergravity theory has an
antisymmetric two-form field $B_{\mu \nu}$ in the spectrum.  There are
classical solutions describing stringlike black brane excitations of
the theory that couple to the $B$ field.  While there is no proof
that quantum excitations of this kind of string soliton must be
included in the complete quantum gravity theory for such an 8D
supergravity, it seems likely that this is the case.  Locally, a
classical solution describing a small loop of such string represents a
small deformation from the flat space-time background.  Some general
arguments for the conclusion that quanta of any  object carrying an
allowed conserved charge must be present in any supergravity theory were recently stated
by Banks and Seiberg \cite{banks-Seiberg}.

A hint for how world-volume anomaly conditions on a solitonic string
may place constraints on the set of allowed theories can be seen in 10
dimensions.  Anomaly cancellation in the world-volume
theory of $N$ D-strings in the type I description of the $SO(32)$
supergravity theory was analyzed by Banks, Seiberg and Silverstein
\cite{bss}.  This world-volume theory has $(0, 8)$ supersymmetry in 2
dimensions.  The theory is chiral and carries a $Spin(N)$ gauge
symmetry in its world-volume, which would be inconsistent without
cancellation of the gauge anomaly.  The theory contains a vector
multiplet with 8 left-moving chiral fermions in the adjoint
representation of the gauge group.  There is a matter multiplet
containing 8 bosons corresponding to transverse fluctuations of the
string and 8 right-moving fermions in the symmetric tensor
representation of the gauge group.  There are also 32 left-moving
fermions in the singlet representation of the gauge group, which can
be associated in the string theory picture with open strings
stretching from the D-string to the space-filling type I D9-branes that
generate the $SO(32)$ gauge group (these D9-branes and the type I
picture are discussed in more detail in the following section).
For the gauge anomaly to cancel we must have
\begin{equation}
\sum x_R A_R  = 0
\end{equation}
where $x_R$ denotes the number of chiral left-moving fermions in the
representation $R$ of the gauge group (right-moving fermions entering
with the opposite sign), and $A_R$ is the constant of
proportionality between the trace of $F^2$ in representation $R$ and
the fundamental representation
\begin{equation}
\tr_R F^2  = A_R  \tr F^2 \,.
\label{eq:a}
\end{equation}
Here, as previously, $\tr_R$ is the trace in representation $R$, with
the absence of index on the trace on the RHS indicating the
fundamental representation.  For $Spin (N)$, we have $A_{\rm adj} = N
-2$ and $A_{\rm sym} = N + 2$.  Denoting the number of left-moving
fermions in the singlet representation by $2r = 32$ where $r$ is the
rank of the space-time gauge group, the anomaly cancellation condition
is
\begin{equation}
8 (N -2) +2r- 8 (N + 2)  = 0 \,,
\end{equation}
which vanishes precisely when $r = 16$.  This confirms the consistency
of the $SO(32)$ ${\cal N} = 1$ theory in 10 dimensions, and can be
interpreted as a constraint on the rank of the space-time gauge group.
While this argument is formulated in the language of type I string
theory, it should be possible to reproduce the analysis from the point
of view of the low-energy theory on the D1-brane itself.  By analyzing
fluctuations around the solitonic string solution, the rank of the
space-time gauge group should thus be fixed to be 16 by cancellation
of the world-volume anomaly.  Note that this argument may become more
subtle for more general space-time gauge groups, in particular for the
$E_8 \times E_8$ theory.

It has been suggested by Uranga \cite{Uranga} that this kind of
argument can be applied in supergravity theories with 16 supercharges
in fewer dimensions.  If correct, this could lead to a demonstration
that the world-volume theory of the solitonic string would be
anomalous in any $D$-dimensional supergravity theory with 16
supercharges unless the rank of the space-time gauge group is fixed to
be $16 + 2 \times(10-D)$ = $36-2D$.  In particular, in 8D the gauge
group would need to be rank 20.  While plausible in schematic form,
the details of this argument have not been worked out.  In particular,
as we have seen, in eight dimensions, the 18 vector multiplets in the
Cartan algebra of the gauge group appearing in the theory are on an
equal footing, and should all play the same role in the world-volume
theory of the solitonic string solution.  Thus, we expect a
cancellation between the extra vector multiplets and the $U(1)$
factors in the supergravity multiplet.  Such a cancellation is
plausible since the graviphoton has different properties from the
other vector fields; for example, the couplings of the $BF^2$ terms
for the $U(1)$ factors in the different multiplets have opposite
signs.  But a detailed proof that this works out is still lacking, and
is left as a challenge for future work.  Proving by such an argument
that the rank of the gauge group in eight dimensions is constrained is
also not enough to prove that the set of consistent theories is
precisely those given by string theory.  There are groups, such as
$SU(2)^{18} \times U(1)^2$ that are of rank 20 and yet cannot be
embedded into $\Gamma^{2, 18}$ as in \eq {eq:8D-condition}.  A proof
of the stronger embedding constraint from the point of view of the
solitonic strings, combined with the known string constructions of
supergravity theories with 16 supercharges through toroidal
compactification of the 10D theory, would amount to a proof of string
universality for this class of supergravity theories.  This could lead
to similar conclusions for 4D theories with ${\cal N} = 4$
supersymmetry and 6D theories with ${\cal N} = (1, 1)$ supersymmetry
as well as the 8D theory with ${\cal N} = 1$ supersymmetry, and would
show that the rank and number of possible nonabelian gauge groups for
these theories is finite in each case.

The story just outlined must be incomplete in at least some respects.
One complication that must be addressed is the existence of orbifold
string compactifications that give rise to theories with 16
supercharges and gauge groups of lower rank.  For example, the CHL
string \cite{CHL, Witten-spin, Lerche-CHL} gives a theory in eight
dimensions with only 10 vector multiplets, and another class of
heterotic orbifolds gives rise to an 8D theory with 2 vector
multiplets \cite{7-triples}.  A complete argument for a bound on the
rank of the gauge group in 8D supergravity theories would need to be
compatible with the presence of these other discrete structures for
the spectrum.  Another interesting question is whether these discrete
families of string vacua can be smoothly connected in some way to the
moduli space of 8D supergravity theories with rank 20 gauge group.

\section{Supergravity and String Vacua in Six Dimensions}
\label{sec:6D}

We now turn to supergravity theories in six dimensions.  We again
focus on theories with the minimum amount of supersymmetry, where the
most interesting new phenomena arise.  In six dimensions the SUSY
generators are chiral, and there are theories with $(2, 2), (2, 0),
(1, 1),$ and $(1, 0)$ supersymmetry.  The (2, 2) theory with maximal
supersymmetry arises from compactification of 10D type II supergravity
on a torus, and the field content of the theory is uniquely
constrained by the supersymmetry structure.  The theories with $(1,
1)$ supersymmetry are in the class of theories with 16 supercharges
discussed at the end of the previous section.  Theories with $(2, 0)$
supersymmetry are strongly constrained by anomalies and correspond to
the theories realized through compactification of the type II theory
on a K3 surface \cite{Seiberg-16}.  Theories with $(1, 0)$
supersymmetry have the richest structure.  In particular, these
theories can contain matter fields that transform in a variety of
representations of the gauge group.  These are the 6D theories on
which we focus in these lectures.  These supersymmetric theories can all be
formulated in 6D Minkowski space.  While in four dimensions, there are
supersymmetric models in AdS space, and gauged supergravity theories
with stable supersymmetric backgrounds, such models do not occur in
six dimensions \cite{Nahm}, so all the supersymmetric 6D theories of
interest admit Minkowski vacua.

We begin by describing the constraints from supergravity and then
consider string constructions.  There is a much wider range of
possible string constructions for 6D ${\cal N} = 1$ supergravity
theories than for 8D supergravities.  We consider several approaches
here, adding intersecting brane models on a K3 compactification of
type IIB to to our repertoire, and explaining the additional
complications involved in heterotic and F-theory constructions beyond
those encountered in 8D compactifications.
\begin{center}
\begin{picture}(200,60)(- 100,- 30)
\put(-95,24){\makebox(0,0){type IIB}}
\put(-95,-24){\makebox(0,0){6D}}
\put(-95,15){\vector( 0, -1){30}}
\put(-125, 8){\makebox(0,0){K3 + D7}}
\put(-125, -8){\makebox(0,0){(IBM)}}
\put(-50,5){\makebox(0,0){dual}}
\put(-50,-7){\makebox(0,0){$\Longleftrightarrow$}}
\put(0,24){\makebox(0,0){het/I}}
\put(0,-24){\makebox(0,0){6D}}
\put(0,15){\vector( 0, -1){30}}
\put(15,0){\makebox(0,0){K3}}
\put(50,5){\makebox(0,0){dual}}
\put(50,-7){\makebox(0,0){$\Longleftrightarrow$}}
\put(95,24){\makebox(0,0){F-theory}}
\put(95,-24){\makebox(0,0){6D}}
\put(95,15){\vector( 0, -1){30}}
\put(133,0){\makebox(0,0){CY3 (elliptic)}}
\end{picture}
\end{center}

In Section \ref{sec:6D-SUGRA} we describe the basic structure of 6D
supergravity theories with minimal supersymmetry, and in Section
\ref{sec:6D-anomalies} we characterize the space of theories that
satisfy the 6D Green-Schwarz anomaly cancellation conditions.  We then
introduce some basic aspects of orientifolds in Section
\ref{sec:branes-orientifolds}, which we use to describe intersecting
brane models in Section \ref{sec:IBM} and magnetized brane models,
which are equivalent to compactifications of the heterotic string, in
Section \ref{sec:heterotic-6D}.  We briefly describe a variety of
additional approaches to heterotic and type I/II constructions that
have been used for 6D vacua in Section \ref{sec:6D-others}.  We
describe F-theory constructions in six dimensions in Section
\ref{sec:6D-F-theory}.  We show in Section \ref{sec:6D-map} that the
close relationship between a lattice determined by anomaly cancellation in the
low-energy theory and the mathematical
structure of F-theory
allows us to use the data from a low-energy theory to
characterize the topological structure of any corresponding F-theory
compactification.  This
is helpful in characterizing the global structure of the set of
possible theories.  We summarize the current state
of knowledge regarding 6D gravity theories with minimal supersymmetry
in Section \ref{sec:6D-global}.

\subsection{Six-dimensional gravity with ${\cal N} =  (1, 0)$ supersymmetry}
\label{sec:6D-SUGRA}

We focus on the massless spectrum of six-dimensional supergravity theories.
There are four massless
supersymmetry multiplets that appear in ${\cal N} =
(1, 0)$ theories with 8 supercharges.  These multiplets are summarized
in Table~\ref{t:6D-multiplets}.
\begin{table}
\begin{center}
\begin{tabular}{|c|c|}
\hline 
Multiplet & Matter Content \\
\hline
SUGRA & $( g_{\mu \nu}, B^+_{\mu \nu}, \psi^-_{\mu})$ \\
Tensor   (T) & $( B^-_{\mu \nu}, \phi, \chi^+)$\\
Vector  (V) & $(  A_\mu, \lambda^-)$ \\
Hyper  (H) & $(  4\varphi, \psi^+)$ \\
\hline
\end{tabular}
\end{center}
\caption[x]{\footnotesize 6D $(1, 0)$ supersymmetry multiplets}
\label{t:6D-multiplets}
\end{table}
The supergravity multiplet contains, in addition to the metric, a
bosonic self-dual two-form field $B^+_{\mu \nu}$.  There are also
tensor multiplets that contain anti-self-dual two-form fields
$B^-_{\mu \nu}$ as well as a single scalar field.  In general an
${\cal N} = 1$ supergravity theory can have any number $T$ of tensor
multiplets, although the theory only has a Lagrangian description when
$T = 1$.  The two-form fields $B^\pm$ transform under an $SO(1, T)$
action that also transforms the scalar fields in the tensor
multiplets.  These scalar fields parameterize a $T$-dimensional moduli
space $SO(1, T)/SO(T)$ that is closely analogous to the moduli space
for toroidal compactifications \eq{eq:Narain-space}.  (A further
discrete quotient by a duality symmetry group must be taken in the
quantum theory, as discussed further below.)  As in higher dimensions,
the vector multiplet contains the 6D gauge field and a chiral gaugino
field.  The gauge group of the theory in general takes the form
\begin{equation}
G = G_1 \times G_2 \times \cdots \times G_k\times U(1)^n/\Gamma
\end{equation}
where $G_i$ are simple nonabelian gauge group factors, and $\Gamma$ is
a discrete group.  The matter hypermultiplets in 6D supergravity
theories live in a manifold with a quaternionic K\"ahler structure.
These hypermultiplets can transform in an arbitrary representation
(generally reducible) of the gauge group.

To summarize,
the discrete data characterizing the field content
and symmetries  of a
6D  ${\cal N} = 1$ supergravity theory  consist of the following:
\vspace*{0.05in}

\noindent $T$: the (integer) number of tensor multiplets
\vspace*{0.05in}

\noindent $G$: the gauge group of the theory; we denote by $V$ the
number of vector multiplets in the theory.
\vspace*{0.05in}

\noindent ${\cal M}$: the representation of $G$ characterizing the
matter content of the theory.  We denote by $H$ the number of
hypermultiplets (including uncharged multiplets) in the theory.
\vspace*{0.05in}

A complete description of the theory would involve further information
such as the metric on the scalar moduli space and higher-derivative
terms in the action.  We do not address this more detailed structure
in these lectures.  Understanding the extent to which this structure
is uniquely determined by supersymmetry and quantum consistency is an
interesting direction for future research.
The Lagrangian for 6D supergravity theories with one tensor multiplet
($T = 1$) was originally described by Nishino and Sezgin
\cite{Nishino-Sezgin, Nishino-Sezgin-2}, and the field equations for
models with multiple tensors were developed by Romans \cite{Romans}.

The question we now want to address is: What combinations of $T, G,$
and ${\cal M}$ are allowed in a consistent 6D supergravity theory?
{\it i.e.}, what is the space ${\cal G}^{6D, {\cal N} = 1}$?
To begin to answer this question we consider the known quantum
constraints on this class of theories.

\subsection{Anomalies and constraints on supergravity in 6D}
\label{sec:6D-anomalies}

\begin{figure}
\begin{center}
\begin{picture}(200,60)(- 90,- 35)
\put(-75,0){\makebox(0,0){\includegraphics[width=4cm]{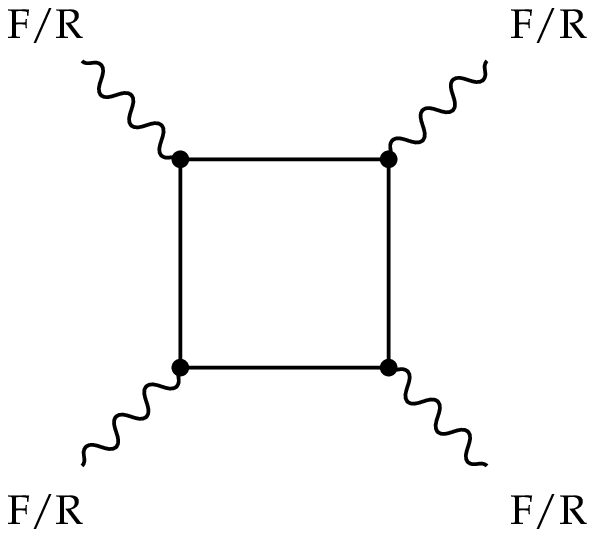}}}
\put(-20,0){\makebox(0,0){+}}
\put(55,0){\makebox(0,0){\includegraphics[width=4.5cm]{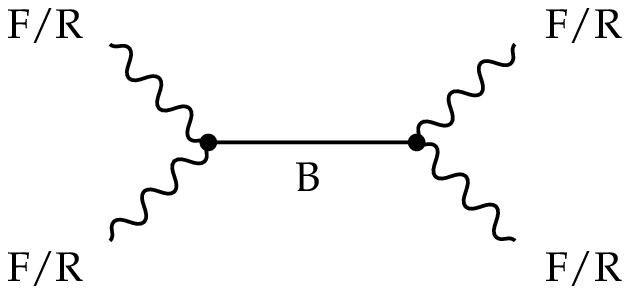}}}
\put(120, 0){\makebox(0,0){ = 0}}
\end{picture}
\end{center}
\caption[x]{\footnotesize Green-Schwarz anomaly cancellation in six
  dimensions cancels one-loop anomalies through tree-level diagrams
  mediated by self-dual and anti-self-dual $B$ fields.}
\label{f:gs-6}
\end{figure}

The structure of quantum anomalies in six dimensions is very similar
to that in ten dimensions.  The chiral fields of the theory that
contribute to anomalies are the self-dual and anti-self-dual two-form
fields $B^\pm_{\mu \nu}$, the gravitino $\psi_\mu^-$, gauginos
$\lambda^-$, and the chiral fermions $\chi^+$ and $\psi^+$ from the
tensor and hyper multiplets.  The anomaly is characterized by an
8-form anomaly polynomial $I_8 (R, F)$.  The anomaly arises from
one-loop ``box'' diagrams with 4 external gauge bosons or gravitons.
The Green-Schwarz mechanism again comes into play to cancel anomalies
through tree diagrams mediated by an exchange of $B$ fields
\cite{gs-west, Gates-Nishino} (See Figure~\ref{f:gs-6}).  The story in
six dimensions is complicated, however, by the presence of multiple
$B$ fields.
The generalization of the Green-Schwarz mechanism including multiple
$B$ fields was worked out by Sagnotti.
The 6D gravitational, nonabelian gauge, and mixed gauge-gravitational
anomalies cancel when the 8-form $I_8$ factorizes in the form
 \cite{Sagnotti, Sadov, KMT-II}
\begin{equation}
I_8 =  \frac{1}{2} \Omega_{\alpha\beta} X^\alpha_4 X^\beta_4
\label{eq:6D-cancellation}
\end{equation}
where
\begin{equation}
X^\alpha_4  =
\frac{1}{2} a^\alpha \tr R^2 +  \sum_i b_i^\alpha \ 
\left( \frac{2}{\lambda_i} \tr F_i^2 \right) \,.
\end{equation}
Here, $\Omega_{\alpha \beta}$ is a signature $(1, T)$ inner product,
$a^\alpha$ and $b_i^\alpha$ are vectors in $\R^{1, T}$, and
$\lambda_i$ are normalization constants for the simple group factors
$G_i$ appearing in $G$, where for example $\lambda_{SU(N)} = 1,
\lambda_{E_8} = 60, \ldots$.  We have not written the anomaly
conditions for $U(1)$ gauge factors.  These take a similar but
slightly more complicated form \cite{Erler, Honecker-1, Park-Taylor}.
We do not treat $U(1)$ factors systematically in these lectures; they
add some technical complications to the story but do not play an
important rule in the main points we wish to emphasize here.

It may be helpful to consider a special case of the anomaly conditions
where the factorization takes a particularly simple form.  When $T =
1$, there is always a basis for $\R^{1, T}$ where
\begin{equation}
\Omega_{\alpha\beta} = \begin{pmatrix}
0 & 1 \\
1 & 0
\end{pmatrix} 
\;\;\;\;\; a = (-2, -2), \;\;\;\; {\rm and}\;
b =\frac{1}{2} (\alpha, \tilde{\alpha})  \,.
\label{eq:t1}
\end{equation}
The anomaly factorization condition can then be written as
\begin{equation}
I_8 = X^1 X^2 =
(\tr R^2 - \sum_i \alpha_i \tr F_i^2)(\tr R^2 - \sum_i \talpha_i \tr F_i^2)
\end{equation}
This form of the anomaly cancellation condition appears in much of the
literature on 6D models with $T = 1$.

The anomaly cancellation conditions are not the only constraints that
a 6D supergravity theory must satisfy for consistency.  As observed by
Sagnotti \cite{Sagnotti}, there is also a constraint coming from the
condition that the gauge fields have kinetic terms with the proper
(negative) sign.  Wrong-sign gauge field kinetic terms would lead to
an instability in the theory.  As in 10D, the gauge kinetic term is
related by supersymmetry to the $BF^2$ coupling that plays a role in
the Green-Schwarz mechanism.  The gauge kinetic term is thus
proportional to $-j \cdot b \;\tr F^2$.  The sign constraint on this
gauge kinetic term gives the condition that $j \cdot b > 0$.  This
constraint plays an important role in restricting the set of possible
consistent supergravity theories in six dimensions.  We discuss other
quantum consistency constraints later in this section and in Section
\ref{sec:6D-global}.

Anomaly cancellation through \eq{eq:6D-cancellation} imposes a set of
conditions  that relate the field content of the theory and the
vectors $a, b_i$ appearing in the anomaly polynomial.  Each type of
term in the anomaly polynomial ($R^4$,  $F^2 R^2$, \ldots) must cancel
separately.  This gives the following conditions
\begin{eqnarray}
R^4:  &\hspace*{0.1in}& \; \;
H-V = 273- 29T \label{eq:hv}
\\[0.1in]
F^4:  &\hspace*{0.1in}& \; \;0
=B^{i}_{Adj} -   \sum_R x^i_{R} B^i_{R}  \label{eq:f4}
\\[0.1in]
(R^2)^2:  &\hspace*{0.1in}& \; \;
 a \cdot a = 9-T  \label{eq:aa}
\\[0.08in]
F^2 R^2:  &
\hspace*{0.1in}& \; \; a \cdot b_i  = \frac{1}{6} \lambda_i
 \left(A_{Adj}^i- \sum_R x^i_{R} A_{R}^i \right)  \label{eq:ab}
\\[0.1in]
(F^2)^2:  &
\hspace*{0.1in}& \; \; 
b_i \cdot b_i =\frac{1}{3} \lambda_i^2
\left( \sum_R x^i_{R} C_{R}^i -
C_{Adj}^i \right)\label{eq:bb}
\\[0.1in]
F_i^2 F_j^2:  &
\hspace*{0.1in}&\; \;  
 b_i \cdot b_j  = 2  \sum_{R,S} x^{ij}_{RS} A_{R}^i
A_{S}^j, \;\;\;\;\; i \neq j
\label{eq:bij}
\end{eqnarray}
In these expressions, $A_R$ is the set of group-theory coefficients
for each representation defined by \eq{eq:a}, and similarly $B_R$ and
$C_R$ are defined through
\begin{equation}
\tr_R F^4  = B_R \tr F^4+C_R (\tr F^2)^2 \,.
\end{equation}
The numbers $x^i_R$ represent the number of matter fields in the $R$
representation of $G_i$, and similarly $x^{ij}_{R S}$ is the number of
matter fields transforming in the $R \times S$ representation of
$G_i\times G_j$.  Values of the group theory coefficients $A_R, B_R,
C_R$ are straightforward to compute using elementary group theory
methods; a
table of values for a few representations of $SU(N)$ are given in 
Table~\ref{t:abc}, and more extensive lists and methods for deriving
these factors can be found in various
relevant papers \cite{Erler,  bound, 0}.
\begin{table}
\centering
\begin{tabular}{|c|c|c|c|c|}
\hline
 Rep. & Dimension & $A_R$ & $B_R$ & $C_R$\\
\hline
 ${\tiny\yng(1)}$ & $N$ & 1 & 1 & 0\\
 Adjoint & $N^2-1$ & $2N$ & $2N$ & 6\\
 ${\tiny\yng(1,1)}$ & $ \frac{N(N-1)}{2} $ & $ N-2 $ & $ N-8 $ & 3\\
 ${\tiny\yng(2)}$ & $ \frac{N(N+1)}{2} $ & $ N+2 $ & $ N+8 $ & 3\\
 ${\tiny\yng(1,1,1)}$ & $ \frac{N(N-1)(N-2)}{6} $& $
 \frac{N^2-5N+6}{2} $&$ \frac{N^2-17N+54}{2} $ & $ 3N-12 $ \\[0.07in]
\hline
\end{tabular}
\caption{Values of the group-theoretic coefficients $A_R, B_R, C_R$
and  dimension 
for some representations of $SU(N)$, $N \geq 4$.
For $SU(2)$ and $SU(3)$, $A_R$ is given in table, while
$B_R = 0$ and $C_R$ is computed by adding formulae for $C_R + B_R/2$
from table with $N = 2, 3$.}
\label{t:abc}
\end{table}

The relations (\ref{eq:aa}-\ref{eq:bij}) determine the inner products
of the vectors $a, b_i$ in terms of the matter content of the theory.
It can be proven using group theory identities \cite{KMT-II} that
these anomaly cancellation conditions\footnote{for gauge groups such
  as $SU(2)$ and $SU(3)$ with no fourth order invariant, cancellation
  of global anomalies \cite{Witten-global, Bershadsky-Vafa} is also
  needed \cite{KMT-II}} automatically lead to integral values for the
inner products $a \cdot a, a \cdot b_i, b_i \cdot b_j$.  Thus, the
vectors $a, b_i$ define an integral lattice
\begin{equation}
\Lambda \subset\R^{1, T} \,.
\end{equation}
We refer to $\Lambda$ as the {\em anomaly lattice} of the 6D theory.
(Note that the basis chosen in (\ref{eq:t1}) for $T = 1$ is not always
the basis in which the basis vectors for the lattice are integral.)
\vspace*{0.05in}

\noindent
{\bf Example:}

Consider a theory with one tensor multiplet ($T = 1$),
nonabelian gauge group 
\begin{equation}
G =SU(N)
\end{equation}
and charged matter content
\begin{equation}
{\rm matter} = 2 \times {\tiny\yng(1,1)} ({\bf  N (N -1)/2})
+ 16 \times {\bf N}
 \label{eq:example-matter-0}
\end{equation}
For each representation ${\bf R}$ of $G$ listed, the matter
content contains one complex scalar field in representation ${\bf R}$
and a corresponding field in the conjugate representation ${\bf
  \bar{R}}$.  Together these fields form the quaternionic structure
needed for the scalar moduli space.  (Note that for special
representations like the ${\bf 2}$ of $SU(2)$, the representation is
itself quaternionic, so that the conjugate need not be included.  In
cases like this the field is often referred to as a ``half
hypermultiplet'').  The total number of charged hypermultiplets in the theory
is $H = N^2 + 15 N$ and  the number
of vector multiplets is $V =  N^2 -1$, so from (\ref{eq:hv}) we have
\begin{equation}
H-V = H_{\rm neutral} +H_{\rm charged} -V = H_{\rm neutral} + 15 N + 1
= 244 \,.
\end{equation}
It follows that $N \leq 16$.
From Table~\ref{t:abc}, it is easy to verify
that \eq{eq:f4} is also satisfied for any $N$, and the matrix of inner products
defining the anomaly lattice for the theory is given by
\begin{equation}
\Lambda =
\left(\begin{array}{ccc}
a \cdot a & -a \cdot b\\
-a \cdot b &  b \cdot b
\end{array} \right)
= \left(\begin{array}{ccc}
8 & 2\\
2 & 0
\end{array} \right) \,.
\label{eq:example-matrix-0}
\end{equation}
\vspace*{0.05in}

\noindent
{\bf Example:}

Now
consider a theory with one tensor multiplet ($T = 1$),
gauge group 
\begin{equation}
G =SU(5) \times SU(6)
\end{equation}
and matter content
\begin{equation}
{\rm matter} = 2 \times ({\bf 10} \left({\tiny\yng(1,1)}\right),  {\bf 1})
+ 1 \times ({\bf 5},  {\bf  \bar{6}})
+ 10 \times ({\bf 5},  {\bf  1})
+ 7 \times ({\bf 1},  {\bf  6}) +161 \times ({\bf 1}, {\bf 1}) \,.
 \label{eq:example-matter}
\end{equation}
The total number of hypermultiplets  is $H = 303$ and the number
of vector multiplets is $V = 59$, so $H-V = 244 = 273-29T$, and
\eq{eq:hv} is satisfied.  From Table~\ref{t:abc}, it is easy to verify
that \eq{eq:f4} is also satisfied, and the matrix of inner products
defining the anomaly lattice for the theory is given by
\begin{equation}
\Lambda =
\left(\begin{array}{ccc}
a \cdot a & -a \cdot b_1 & -a \cdot b_2\\
-a \cdot b_1 &  b_1 \cdot b_1 & b_1 \cdot b_2\\
-a \cdot b_2 &  b_1 \cdot b_2 & b_2 \cdot b_2
\end{array} \right)
= \left(\begin{array}{ccc}
8 & 0 & 2 \\
0 &  -2 & 1\\
2 & 1 &  0
\end{array} \right) \,.
\label{eq:example-matrix}
\end{equation}
The lattice $\Lambda \subset\R^{1, T}$ is two-dimensional; we see from
\eq{eq:example-matrix} that the matrix of inner products is
degenerate, and there is a linear relation between the vectors $-a =
2b_1+4b_2$.  In the basis \eq{eq:t1}, we have $a = (-2, -2)$, $b_1 =
(1, -1)$ and $ b_2 =(0, 1)$, reproducing the desired inner products.
\vspace*{0.05in}

The anomaly cancellation conditions for 6D supergravity theories place
strong constraints on the range of possible consistent theories.  It
has been proven that when $T < 9$, the set of possible gauge groups
$G$ and matter representations ${\cal M}$ charged under the nonabelian
gauge group factors is finite \cite{bound, KMT-II}.  This bound is
valid for theories with nonabelian and abelian gauge group factors,
although for theories with abelian factors the charges of the matter
fields under the $U(1)$ factors are not constrained to a finite range
of possible values by anomaly cancellation \cite{Park-Taylor}.  The
proof of the finiteness result is slightly technical, but relies at
its core on the relation \eq{eq:hv}, which bounds the number of matter
multiplets for a given number of tensor fields and gauge content.
Roughly, if the gauge group factors are all of limited dimension
(dim$(G_i) < D \; \forall i$), then it can be shown that almost all
pairs of gauge group factors share hypermultiplets with charge under
both factors.  This leads to a situation where $H \sim{\cal O} (k^2)$
and $V \sim{\cal O} (k)$, where $k$ is the number of distinct
nonabelian factors in the gauge group.  As $k \rightarrow \infty$, the
constraint \eq{eq:hv} must be violated.  For the details of this
argument the reader is referred to the original papers \cite{bound,
  KMT-II}.  Anomaly cancellation alone does not rule out the
possibility of a finite number of gauge group factors with unbounded
dimension.  It was shown by Schwarz \cite{Schwarz} that there are
infinite families of theories with $T = 1$ that satisfy the anomaly
factorization conditions.  For example, the theories with gauge group
$G = SU(N) \times SU(N)$ and matter content $2 \times ({\bf N},
\bar{{\bf N}})$ satisfy the anomaly cancellation conditions and have
$H-V = 2$ for any $N$.  There are 5  infinite families of models
that satisfy the factorization conditions.  For each such family,
however, it is possible to prove that everywhere in the moduli space
at least one gauge group factor must have
the wrong sign on the kinetic term \cite{KMT-II}.  For example, for
the theories just mentioned that were found by Schwarz, it is easy to
check that the anomaly conditions give
\begin{equation}
 a \cdot b_1 = a \cdot b_2 = 0, \;\;\;\;\;
b_1^2 = b_2^2 = -2, \;\;\;\;\; b_1 \cdot b_2 = 2\,.
\end{equation}
From this it follows that, when $a^2 > 0$,
\begin{equation}
a \cdot (b_1 + b_2) = 0 \; \;
\& \; \; (b_1 + b_2)^2= 0 \; \; \Rightarrow
  \; \; b_1 + b_2 = 0\; \;
\Rightarrow \; \; j \cdot b_1 = -j \cdot b_2  \,.
\end{equation}
Thus, at least one of the gauge group factors has the wrong sign on
the kinetic term for any vector $j$.  This rules out families of
theories with gauge group factors of unbounded dimension, as long as $T
< 9$.  When $T \geq 9$, however, the norm of the vector $a$ is no
longer positive definite, (recall $a^2 = 9-T$), so the above argument
no longer works.  In fact, there are infinite families of models with
$T = 9$ and greater that satisfy the anomaly cancellation equations
and have proper-sign kinetic terms \cite{KMT-II}.

This gives a basic outline of the set of 6D ${\cal N} = 1$
supergravity theories that satisfy the anomaly cancellation and gauge
kinetic term sign constraints.  It was recently shown that a further
constraint can be placed on the set of 6D supergravity theories that
can give consistent quantum theories.  The set of allowed charges for
objects that couple to the $B$ fields of the theory form a lattice
$\Gamma$ of signature $(1, T)$.  This lattice must be integral, from
the Dirac quantization condition \cite{Deser-dyons}.  The consistency
of the dimensional reduction of the theory to 2D or 4D requires that
the lattice is furthermore self-dual \cite{Seiberg-Taylor}.  This
conclusion is compatible with the general mathematical framework for
treating (anti)self-dual $p$-form fields that has recently been under
development \cite{Hopkins-Singer, dmw, Moore-Witten,
  Freed-Moore-Segal}.  The charge lattice is invariant under a
discrete duality group $G^{1, T} \subset SO(1, T)$.  This reduces the
part of the moduli space of the theory parameterized by the scalars
$\phi$ in the tensor multiplets to
\begin{equation}
G^{1, T}\backslash
SO(1, T)/SO(T)\,.
\end{equation}

The vectors $b_i$ in the anomaly lattice are associated with gauge
dyonic strings \cite{Duff-gauge} associated with instantons in each
gauge group factor, and thus represent vectors in the charge lattice
$\Gamma$.  There must therefore be an embedding of the lattice spanned
by the vectors $b_i$ into the lattice $\Gamma$.  The vector $a$ should
also give a vector in the charge lattice $\Gamma$, though this is not
rigorously proven.  The condition that an embedding of the anomaly
lattice $\Lambda$ into $\Gamma$ is possible imposes further
constraints on the set of allowed 6D theories.  For example, consider
the theory with $T = 2$, gauge group
\begin{equation}
G = SU(N) \times SU(N) \,,
 \end{equation}
and charged matter content
\begin{equation}
2 N \times ({\bf N},  {\bf 1}) +
2 N \times ({\bf 1},  {\bf N})
\,.
\end{equation}
The anomaly
lattice for this model is spanned by vectors $-a, b_1, b_2$ with inner products
\begin{equation}
\Lambda= \left(\begin{array}{ccc}
7 & 0 &0\\
0 & -2 & 0\\
0 &  0 & -2
\end{array} \right) \,.
\label{eq:sick-lambda-3}
\end{equation}
This lattice does not admit an embedding into any unimodular lattice
$\Gamma$.  If it did, the unit cell of $\Lambda$ would have a volume
given by an integer times the volume of the unit cell of $\Gamma$.
The determinant of the matrix $\Lambda$ would then be a perfect
square.  But the determinant is 28, which is not a perfect square, so
the lattice does not admit a unimodular embedding and the theory is
not a consistent theory.  This criterion cuts down further the space
of allowed 6D supergravity theories, though it does not eliminate some known
infinite families of models that obey all known consistency
constraints.

Thus, we have characterized the set of theories ${\cal G}^{6D, {\cal
    N} = 1}$ with no currently known inconsistencies.  We now
turn to string constructions of 6D theories to see what subset of the
apparently consistent supergravity theories can be realized in string
theory.

\subsection{Branes and orientifolds}
\label{sec:branes-orientifolds}

Six dimensions is an excellent playground for systematically
developing the methods of string compactification.  The range of
possibilities for string constructions in 6D is still much simpler
than in 4D, but provides for a rich range of possible gauge groups and
matter structure.  The strong constraints of anomaly cancellation
provide a useful mechanism for checking the internal consistency of
different approaches to string vacuum constructions, as well as
providing a natural framework for relating different constructions.

We describe in some detail three distinct ways of constructing 6D
string vacua.  We expand upon the development in Section~\ref{sec:8D}
of the heterotic and F-theory approaches, illustrating some of the new
features that must be included for compactifications of these types
to 6 dimensions.  We also introduce the basic structures needed for
another class of compactification, the {\em intersecting brane model}
(IBM) construction.  We consider intersecting brane models  associated with
compactifications of type IIB on an orbifold limit of the K3 surface,
where D7-branes added to the geometry realize the gauge groups and
matter content of the theory.  A more detailed introduction to
the physics of intersecting brane model constructions, with a focus on
the phenomenology of four-dimensional vacua, can be found in the
lecture notes by Cvetic and Halverson from this school
\cite{Cvetic-Halverson}.  We give here a much more rudimentary
introduction to the subject of IBM's, focusing on how the models
generated through this construction fit into the broader set of 6D
string vacua.

To understand intersecting brane models and heterotic/type I
compactifications in 6D, we must first review some
basic aspects of D-branes and orientifolds.  These objects are
explained clearly and pedagogically in Polchinski's text
\cite{Polchinski-I, Polchinski-II}, but we summarize the basic
structures involved in order for these lectures to be somewhat self-contained.

The essential feature of a D-brane is that it is a locus in space-time
where open strings end.  A D$p$-brane in the flat ten-dimensional
space-time of type IIA or IIB string theory is a fluctuating
hypersurface of spatial dimension $p$, whose motion is described by
the quantized open strings ending on the brane.  When $N$ D$p$-branes
are coincident, the D-branes carry a world-volume  gauge
group $U(N)$; the dynamics of this gauge group and the scalars
describing transverse fluctuations of the brane are captured by the
physics of the open strings.  One approach to describing theories in
fewer than 10 dimensions with chiral matter is to use systems of
intersecting branes.   When  two different branes intersect, the strings
stretching between the branes give rise to chiral matter fields in the
dimensionally reduced theory.

\begin{figure}
\begin{center}
\includegraphics[width=6cm]{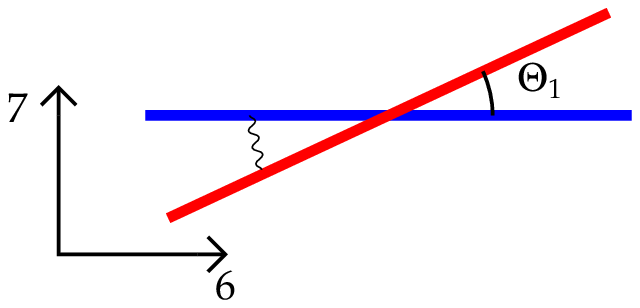}
\hspace*{0.8in}
\includegraphics[width=6cm]{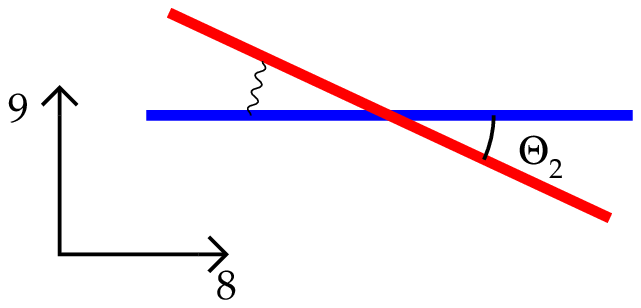}
\end{center}
\caption[x]{\footnotesize Brane stacks intersecting in two planes preserve
  some supersymmetry if the intersection angles are equal and opposite
in the two planes, $\theta_1 = -\theta_2$.  Strings stretching between
the branes live in the bifundamental representation $(N, \bar{M}) +
(\bar{N}, M)$ of the gauge groups on the two brane stacks.}
\label{f:D-brane-intersection}
\end{figure}

Consider a pair of flat D-branes in $\R^{1, 9}$ that are parallel and
coincident in all but 4 spatial dimensions.  For example, two
D7-branes that are both extended in dimensions 0-5, but each extended
in a different two-plane in the four-dimensional space 6789.  We can
view these branes as intersecting in two different coordinate planes;
for example the branes may each extend on different lines in the 67
plane and the 89 plane (See Figure~\ref{f:D-brane-intersection}).  The
branes preserve a common supersymmetry if the angles between the
branes in the two planes are equal and opposite  \cite{bdl}
\begin{equation}
\theta_1 + \theta_2
= 0 \,.
\label{eq:intersection-SUSY}
\end{equation}
If there are stacks of $N, M$ coincident branes intersecting,
then the open strings connecting the branes produce  fields living
on the brane intersection locus that transform in the bifundamental
representation $(N, \bar{M}) + (\bar{N}, M)$.  If we 
compactify the theory down to six dimensions, this will give matter
in the low-energy 6D theory transforming in the same representation.

The obstacle to considering compactifications with D-branes is that
the D-branes act as sources for the Ramond-Ramond fields.  A D-brane
that extends in all noncompact dimensions of space-time will produce a
flux in the transverse compact directions.  D-branes that preserve the
same supersymmetry will generally carry the same kind of charge.  With
net D-brane charge, the flux lines will have nowhere to end on the
compact space, and the geometry will not admit a solution of the
equations of motion.  The situation is like a set of positive charges
in a compact space; the electric field lines point away from the
positive charges, and without negative charges to collect the field
lines, a compact solution is impossible.  Thus, we need something to
cancel the R-R charges of the D-branes in order to use the branes to
generate the low-energy gauge group and matter content.  One solution
to this problem is the {\em orientifold}.  Orientifolds are similar to
D-branes, but carry negative tension and negative charge.  When
orientifolds are included, compactifications with D-branes can
preserve SUSY and give rise to interesting low-energy
theories\footnote{Note that the 7-branes on $S^2$ in the F-theory
  description of 8D vacua described in Section \ref{sec:7-branes}
  evade this problem in a different way; the condition that the
  product of monodromies cancels on the F-theory base is equivalent to
  the condition that the charges cancel, which can be done in a
  supersymmetric way when the total space of the elliptic fibration
  defining the F-theory model is a Calabi-Yau. Like everything else in
string theory, these apparently different constructions are related
through duality \cite{Sen-F-theory-orientifolds}.}.

To understand orientifolds, it is easiest to begin with the
space-filling orientifold that relates the type IIB theory to the
type I string theory.  Consider the transformation $\Omega$ that acts
by a reflection on the internal coordinate of a string
\begin{equation}
\Omega: \sigma \rightarrow - \sigma \, .
\end{equation}
This transformation exchanges the right- and left-moving string degrees
of freedom $\Omega: x_L  \leftrightarrow  x_R$,
$\Omega: \alpha \leftrightarrow  \tilde{\alpha}$.  This is a symmetry
of the type IIB string theory.
If we consider only
string states that are invariant under this symmetry (essentially
gauging the symmetry $\Omega$
by taking the orbifold of the theory by $\Omega$), then the string
spectrum will be simplified.  
For example, the $B$
field of the type IIB theory is associated with the first-quantized string
states
\begin{equation}
\left[\alpha^\mu_{-1} \tilde{\alpha}^\nu_{-1} 
-\tilde{\alpha}^\mu_{-1} \alpha^\nu_{-1} \right] | 0 \rangle \,.
\end{equation}
This set of states is projected out by the orientifold action.  The
theory realized by the orientifold of the type IIB theory is the type
I theory, which has no $B$ field, only the R-R field $\tilde{B}$.  The
type I theory has only one supersymmetry in 10 dimensions.  As we
know, an ${\cal N} = 1$ theory in 10D must have a gauge group of
dimension 496 to be anomaly-free.  Thus, some additional structure
must provide a gauge group for the type I theory.  The projection
under $\Omega$ produces an unoriented string theory.  String diagrams
with unoriented topology, such as the M\"obius strip and Klein bottle,
must be included.  From the analysis of such diagrams, the type I
orientifold background has been shown to be inconsistent without the
addition of 16 space-filling D9-branes, whose world-volume theory
(after including the orientifold projection) is precisely the gauge
group $SO(32)$ needed for an anomaly-free theory \cite{Polchinski-II}.
This can be understood in terms of the presence of a space filling
``orientifold 9-plane,'' with negative tension and D9-brane charge
-16, which cancels the charge of the 16 D9-branes.

By combining the world-sheet symmetry $\Omega$ with a space-time
reflection $\rho$, orientifold planes of lower dimension can be
produced.  One-loop string diagrams show that such an orientifold
plane of codimension $d$ carries D-brane charge $16/2^d$.  One easy way
to understand this result is by analyzing toroidal
compactification and T-duality for the type I theory.
Recall that for a string compactified on a circle $S^1$ of radius $R$,
the spectrum of momentum and winding modes is given by
\begin{equation}
M^2 = \frac{n^2}{R^2}  + \frac{m^2 R^2}{\alpha'}  \,.
\end{equation}
As discussed in Section \ref{sec:toroidal-compactification},
T-duality is the symmetry that exchanges 
\begin{equation}
{\rm T-duality}: \;n
\leftrightarrow m \,.
\end{equation}
T-duality also exchanges $x_R \leftrightarrow -x_R$.  It follows that
T-duality exchanges a D-brane wrapped in the compact direction with one
that is unwrapped ($D_p \leftrightarrow D_{p\mp 1}$); the gauge field
component $A_\mu$ on the D$p$-brane wrapped in the compact direction
$\mu$ becomes the transverse scalar field $X^\mu$ of the unwrapped
brane
\begin{equation}
T: \; \; A_\mu \leftrightarrow X^\mu \,.
\label{eq:T-duality-a}
\end{equation}
The relation \eq{eq:T-duality-a} can be understood from the point of
view of the string world-sheet \cite{Polchinski-I}, or from the
world-volume theory of a brane on a compact space \cite{WT-compact}.
Just as T-duality can reduce or increase the dimension of a D-brane,
in a similar fashion T-duality reduces the dimension of a
space-filling orientifold plane.  The action of $\Omega: x \rightarrow
x$ on $x = x_R + x_L$ before T-duality becomes the reflection $\Omega:
x \rightarrow -x$ on $x = -x_R + x_L$ after T-duality.  On the dual
circle, this action has two fixed points, at $x = 0, x = \pi R$.
Thus, under T-duality of the type I theory on $d$ circles in the torus
$T^d$ the 16 D-branes are distributed across $2^d$ orientifold
$(9-d)$-planes.  This confirms the statement above that each
orientifold $(9-d)$-plane carries D-brane charge $-16/2^d$.

This brief summary of the physics of D-branes and orientifold planes
gives us enough background to construct intersecting brane models on
compactifications of the type IIB theory to six dimensions.

\subsection{Intersecting brane models in 6D}
\label{sec:IBM}

Intersecting brane models (IBM's)
provide a rich range of examples of string
compactifications with a variety of gauge groups and chiral matter.
Semi-realistic  IBM's in four dimensions have been the subject
of much study \cite{IBM-review, Cvetic-Halverson}.  Here we will just
introduce the basics of the construction in six dimensions to show how
some 6D supergravity theories can be realized from this approach.  A
more detailed analysis of 6D intersecting brane models was given by
Blumenhagen, Braun, K\"ors, and L\"ust \cite{Blumenhagen-6D};
the presentation here follows the analysis and notation of that work.
Additional features of these models have been developed in work with
Nagaoka \cite{Nagaoka-Taylor}.

While in principle, intersecting brane models on smooth Calabi-Yau
manifolds give rise to a very general class of string
compactifications, the analysis is greatly simplified by working on a
toroidal orbifold.  We will use here the orbifold limit of K3
described in Section~\ref{sec:K3}.  Thus, we consider the orbifold of
$T^4$ 
by the symmetry
\begin{equation}
\rho: z_i \rightarrow -z_i\,.
\end{equation}
Here, as before, $z_1, z_2$ are complex coordinates on two $T^2$
factors of $T^4$.  We assume that both $T^2$ factors are rectangular,
and have moduli $\tau_j =  i R_j, \;  j = 1, 2$.

In order to include D-branes we must also include orientifolds.  We
include an orientifold 7-plane (O7-plane) defined through the orientifold action
$\Omega \sigma$, where $\sigma$ gives the reflection
\begin{equation}
\sigma: z_i
  \rightarrow \bar{z}_i\,.
\end{equation}
This gives an O7-plane in the 68 directions.  Since we are taking the
orbifold separately by $\rho$ and $\Omega \sigma$, we must also
include the combined orientifold action $\Omega \rho \sigma$; this
gives another O7-plane in the 79 directions.  As discussed above,
there are really 4 copies of each O7-plane in the covering space
$T^4$, at antipodal points on the perpendicular tori.  After taking
the orbifold action,  the combined set of cycles where
the orientifold is wrapped  becomes
\begin{equation}
\pi_{O7}=2(\pi_{68}-\pi_{79}) \,.
\end{equation}

\begin{figure}
\begin{center}
\includegraphics[width=10cm]{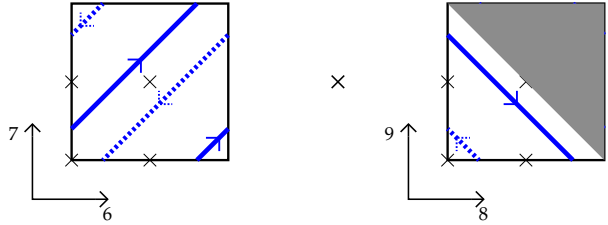}
\end{center}
\caption[x]{\footnotesize A diagonal D7-brane wrapped on the cycle  $(1,
  1)$ on the first $T^2$ and the cycle $(1, -1)$ on the second $T^2$.
The dotted part of the brane is associated with the piece that lifts
under the orbifold action $\rho:z_i \rightarrow -z_i$
to the shaded part of the covering space outside the fundamental domain.}
\label{f:IBM-brane}
\end{figure}

Now consider D7-branes that are wrapped on the $T^4$ as a product of
one-cycles on the two $T^2$ factors with winding numbers
\begin{equation}
(n_1, m_1; n_2, m_2)\,.
\label{eq:brane-winding}
\end{equation}
For example, a D7-brane wrapped on the cycle $(1, 1; 1, -1)$ is
depicted in Figure~\ref{f:IBM-brane}.
The homology class of a brane with winding numbers \eq{eq:brane-winding}
is
\begin{equation}
\pi  =  n_1 n_2\,\pi_{68}+ n_1 m_2\,\pi_{69}+ m_1 n_2\,\pi_{78}
+ m_1 m_2\,\pi_{79}\,.
\end{equation}
Associated with each such brane there is an orientifold image with
winding numbers (reflected across the horizontal axes) $(n_1, -m_1;
n_2, -m_2)$ and associated homology class $\pi'$.  We consider a
general D7-brane configuration to be composed of stacks of $N_a$
coincident branes with winding numbers $(n_1^a, m_1^a; n_2^a, m_2^a)$.
The total D-brane charge for all branes extended in space-time and wrapped
around any particular cycle of the compact space must vanish as discussed above.
Each O7-plane carries -4 units of D7-brane charge.  Combining the
branes and their orientifold images, the condition for the charges to
vanish on the compact space is then
\begin{equation}
\sum_{a}N_an_1^an_2^a = 8, \;\; \;\;\;\;\;
\sum_{a} (-N_a m_1^am_2^a) = 8\,.
\label{eq:IBM-tadpole}
\end{equation}
These equations are often referred to as ``tadpole cancellation''
conditions since they are needed for cancellation of tadpoles in the
R-R fields of the theory; similar conditions arise in most brane model
constructions.  For 6D models, these conditions can be related to the
anomaly cancellation conditions of the associated supergravity theory.

If we assume that the branes all preserve a common supersymmetry, then
the SUSY condition \eq{eq:intersection-SUSY} implies that for all
branes in the system
\begin{equation}
\frac{m_1}{n_1} = -\alpha  \frac{m_2}{n_2}
\label{eq:IBM-SUSY}
\end{equation}
for a common value of $\alpha$ that parameterizes the moduli of the
two $T^2$ factors.

A stack of $N$ diagonal branes that do not coincide with their
orbifold counterpart ({\it i.e.}, branes that are not parallel to the
orientifold plane and that do not pass through the orbifold points)
carry a $U(N)$ gauge group.  Each such brane is wrapped on a cycle
that is topologically a torus and will have a single matter field in the
adjoint representation,  which describes transverse motions of the
brane.  As discussed in the previous section, open strings between
intersecting branes produce bifundamental fields transforming under
the gauge groups on each brane.  Symmetric and antisymmetric
representations of the group $SU(N)$ for a particular brane stack will
also be produced by intersections of the brane with its orientifold
image.  A table of the multiplicity of matter representations
associated with a pair of branes $\pi_a, \pi_b$ follows; in each case
the representation given, $R$, signifies a hypermultiplet with matter
in the $R+ \bar{R}$ representation.
\[
\begin{array}{|c |c |}
\hline
{\rm representation} & 
{\rm multiple}\\
\hline
{\rm Adj} &  1\\
({\tiny\yng(1)}_a, \bar{{\tiny\yng(1)}}_b) & \pi_a \cdot \pi_b\\
({\tiny\yng(1)}_a, {\tiny\yng(1)}_b) & \pi_a \cdot \pi'_b\\
{\tiny\yng(1,1)}_a &\frac{1}{2} (\pi_a \cdot \pi'_a + \pi_a \cdot o_7)\\
{\tiny\yng(2)}_a &\frac{1}{2} (\pi_a \cdot \pi'_a - \pi_a \cdot o_7)\\[0.05in]
\hline
\end{array}
\]
This completes the specification of the simplest intersecting brane
models on K3.  Other types of branes can be included, such as branes
parallel to the orientifold planes, fractional branes, and branes
intersecting the orbifold points \cite{Blumenhagen-6D,
  Nagaoka-Taylor}; these brane types can include other gauge group
factors such as $Sp(N)$ and introduce additional interesting features
that we do not explore here.  Another complication that we do not
treat in detail here is associated with the $U(1)$ factors in the
gauge group.  In general some of the $U(1)$ factors in an intersecting
brane model will be anomalous, and will acquire a mass through the
St\"uckleberg mechanism.  As above, we primarily focus here on the
nonabelian part of the gauge group, and do not treat the $U(1)$
factors carefully.

To summarize, in the basic class of 6D IBM models just described, we
have a set of stacks of $N_a$ branes, with each stack carrying a gauge
group factor $U(N_a)$.  The winding numbers for each stack satisfy
\eq{eq:IBM-SUSY}, and the tadpole constraint 
associated with the total charge of the branes and orientifold plane
(on each homology class) is satisfied through
the conditions \eq{eq:IBM-tadpole}.  The set of models that satisfy
these conditions is fairly limited.  We mention here only the simplest
example.
\vspace*{0.05in}

\noindent {\bf Example:} 6D IBM on K3

The simplest example is to take a stack of 8 branes with winding
numbers $(1, 1; 1, -1)$ as depicted in Figure~\ref{f:IBM-brane}.  This
model clearly satisfies the SUSY and tadpole cancellation conditions.
The supergravity theory associated with this model has gauge group
\begin{equation}
G = U(8) \,.
 \label{eq:IBM-example-1}
\end{equation}
From \eq{eq:orbifold-intersection} it follows that $\pi \cdot \pi'
= \pi \cdot o_7 = 8$, so the matter content of the theory is
\begin{equation}
{\rm matter} = 1 \times ({\bf 63}) + 8 \times ({\bf 28}) \,.
\end{equation}
Considering only charged hypermultiplets, we have
$H_{\rm charged} -V = 224$.  This leaves 20 uncharged hypermultiplets
to saturate the gravitational anomaly constraint, which is precisely
the number of scalar hypermultiplets that come from the closed string
sector.  The $F^4$ anomaly equation \eq{eq:bb} for an $SU(N)$ gauge
group factor under which there are $f$ fundamental matter
representations, $D$ adjoint representations, and  $A$ antisymmetric
representations is
\begin{equation}
f = 2 N -2 N D-A (N -8)  \,,
\end{equation}
which is satisfied for this matter content.
In all 6D intersecting brane models of this type, there is
a single tensor multiplet in the 6D theory ($T = 1$); the
anti-self-dual two-form from the tensor and the self-dual two-form
from the gravity multiplet combine to form the  two-form field
that descends from the 10D $B_{\mu \nu}$.
Connecting to the general framework for 6D supergravities, we
can compute the anomaly lattice of this theory, which is spanned by
vectors $-a, b$ with inner product matrix
\begin{equation}
\Lambda= \left(\begin{array}{cc}
 8 & 8\\
8 & 8
\end{array} \right)
 \label{eq:IBM-example-matrix}
\end{equation}
This lattice is degenerate, and represents a 1D lattice spanned by the
generator $b = -a$ \,.
We show in Section \ref{sec:6D-map} how this data can be used to
construct the equivalent F-theory model.

\subsection{Magnetized brane models and heterotic bundles in 6D vacua}
\label{sec:heterotic-6D}

We now consider 6D theories associated with compactifying ${\cal N} =
1$ supergravity on a K3 surface.  The general compactification of this
type is described by a nonabelian instanton configuration on K3.  We
primarily consider a simple subclass of these models, where the
instantons are characterized by $U(1)$ fluxes on cycles in the K3
surface.  The models in this class were the first 6D compactifications
of string theory \cite{gs-west}.  We can think of such
compactifications from the string point of view either as heterotic or
type I string compactifications.  We begin from the supergravity point
of view and then connect to the type I picture, which in spirit is
like a T-dual of the intersecting brane models just considered.  As
for the intersecting brane models on the orbifold limit of K3
discussed above, all models constructed by compactification on a
smooth K3 have $T = 1$, with the anti-self-dual and self-dual
two-forms combining to form the dimensionally reduced $B$ field from
10D.

\subsubsection{D-brane charges on branes and orientifolds}

We begin from the point of view of supergravity.  In compactifying an
${\cal N} = 1$ 10D theory to eight dimensions on a torus, the only
additional structure available was the possibility of nontrivial
Wilson lines around the cycles of the torus.  Compactifying on a
non-toroidal Calabi-Yau requires the further introduction of a gauge
bundle carrying instantons.  To see why this is the case, recall from
\eq{eq:h-complete} that the 3-form $H$ contains Chern-Simons
contributions from the Yang-Mills and space-time curvatures.  The
Bianchi identity in a general background is then generalized from $dH
= 0$ to
\begin{equation}
dH =\tr R \wedge R -\Tr F \wedge F\,.
\end{equation}
Since the left-hand side is an exact form, the right-hand side must
vanish in cohomology, implying a nonzero instanton number for a
compactification on K3
\begin{equation}
\int_S c_2(F)= \frac{1}{8\pi^2}\int_S \Tr F \wedge F = \frac{1}{16\pi^2}\int_S \Tr R\wedge R = -\frac{1}{2}\int_S p_1(R) = 24\,,
\end{equation}
where $c_2$ and $p_1$ are the second Chern class and first Pontryagin
class, respectively\footnote{Note the factor of 2 in moving between
  the vector and adjoint representations of $SO(1, 3)$, denoted
  respectively by tr and Tr.  Note also that we have changed
  normalization relative to \eq{eq:h-complete} so that the instanton
  number is an integer when expressed in conventional form.}.  For
compactification of an ${\cal N} = 1$ 10D theory on K3, then, we must
have a background gauge field configuration with instanton number 24.

An alternative perspective on the conclusion that the gauge bundle on
K3 must have 24 instantons comes from the type I picture, which can be
used when the 10D gauge group is $SO(32)$ (really Spin$(32)/\Z_2$).
Recall that in the type I picture the gauge group comes from 16
D9-branes superimposed on an orientifold O9-plane.  An important
aspect of D-brane physics is the presence of Chern-Simons couplings in
the world-volume action between powers of the field strength $F$ and
the Ramond-Ramond $p$-form potentials \cite{Douglas-branes}.  For D9-branes, there is a coupling to the
6-form potential of the IIB and I theories proportional to
\begin{equation}
\int F \wedge F \wedge \hat{C}_6\,.
\label{eq:Chern-Simons-6}
\end{equation}
($\hat{C}_6$ is the dual to the R-R two-form $\tilde{B}$.)  This means
that an instanton on a D9-brane carries the charge of a D5-brane
\cite{Witten-instantons}. 
This statement, as well as the more general statement that a system of
D$p$-branes can carry charge associated with a D$(p\pm 2k)$-brane can be
understood easily from T-duality on the D-brane world-volume
\cite{Taylor-lectures}.  By taking the world-volume derivative of
\eq{eq:T-duality-a} we have
\begin{equation}
T: \; \; \partial A_\mu \leftrightarrow \partial X^\mu \,.
\end{equation}
Thus, T-duality can relate, for example, flux $F = \partial A$ on a
D9-brane to the slope $\partial X$ of a tilted D8-brane in a T-dual
picture.  The T-dual of a diagonal D8-brane on a torus has both
D9-brane and D7-brane charge; the D7-brane charge can be associated
with the flux encoded in $F$ on the D7-brane.  Generalizing this
picture to multiple T-dualities shows that $F \wedge F$ on a
D$p$-brane carries D$(p-4)$-brane brane charge, etc.  (A further
generalization to multiple D$p$-branes, replacing $\partial
\rightarrow \partial + iA_\mu$, leads to the T-dual statement that
$[X, X]^k$ on a stack of multiple D$p$-branes carries charge
associated with higher-dimensional D$(p + 2k)$-branes \cite{tv-0,
  tv-p, Myers-dielectric}.)  Through the geometric picture of
T-duality as relating fluxes to tilting on branes, we see that
intersecting diagonal D7-branes on a torus $T^4$ can be T-dual to a
system of D9-branes carrying D7-brane and D5-brane charges through
world-volume fluxes.  This relates the intersecting brane models
discussed in the previous section to the type I models discussed here.
There are some subtleties, however,  in this correspondence,
since we are really working on K3
and not a torus.  Also, both the IBM and magnetized brane models are
described in the supergravity approximation where the volume of the
compactification space is large, and T-duality maps a large
compactification torus to a small dual torus, where the supergravity
approximation is no longer valid.

In the type I picture, the need for a gauge field configuration with
instanton number 24 follows from structure of the D-brane world-volume
theory related to the Chern-Simons couplings \eq{eq:Chern-Simons-6}.
From supersymmetry, there are similar terms of the form $R \wedge R
\wedge \hat{C}_6$ in the D9-brane world-volume action
\cite{Bershadsky-vs}, and analogous couplings occur for the
orientifold plane \cite{Morales-ss, Stefanski}.  The quantization of
the coupling is such that each D9-brane wrapped on a K3 carries -1
units of D5-brane charge, while the O9-plane wrapped on K3 carries -8
units of D5-brane charge.  Combining one O9-plane and 16 D9-branes
gives a total deficit of 24 units of D5-brane charge, or instanton
number 24, needed to cancel the $\hat{C}_6$ tadpole.  The term
``magnetized brane'' refers to $U(1)$ fluxes $F$ on the D9-brane
world-volume that can be used to realize the needed instanton.

\subsubsection{Abelian instantons on K3}

To understand a general 6D compactification of the 10D ${\cal N} = 1$
theory, then, we must consider the moduli space of instantons for a
nonabelian gauge theory on K3. This leads into a rich and interesting
mathematical story, which leads beyond the scope of these lectures.
Here we will focus on the simple class of ``magnetized brane'' models
where the instanton structure is completely encoded in fluxes within a
commuting set of $U(1)$'s in the SO(32) gauge group.  This is the
class of 6D models originally studied by Green, Schwarz, and West
\cite{gs-west}.  By describing the structure of these theories in
terms of the homology lattice of K3 with associated intersection
product, following work with Kumar \cite{kt-K3}, we make contact with
the general theme of lattice embeddings that we have already
encountered in the 8D context, and put these models into a framework
that connects with the other classes of 6D vacua.

We begin the discussion with a review of some basic facts about K3.
Like any complex surface, K3 has a choice of complex structure
\begin{equation}
\Omega \in H^{2, 0} (S) \subset H^2
  (S)
\end{equation}
that is fixed up to a scale factor.  The complex structure satisfies
\begin{equation}
\int \Omega \wedge \Omega = 0, \quad \int \Omega \wedge \bar{\Omega}
\propto \mbox{Vol}(S) > 0 \,.
\end{equation}
Writing $\Omega = x + iy$, it follows that
\begin{equation}
x \cdot y = 0, \; x \cdot x = y \cdot y > 0\,,
\end{equation}
where in this section we will use $\cdot$ to denote the wedge product
in cohomology or the equivalent intersection form in homology, freely
moving back and  forth using Poincar\'{e} duality.
There is a K\"ahler structure associated with a K\"ahler
form $J \in H^{1, 1} (S)$ that satisfies
\begin{eqnarray}
\int J\wedge \Omega & = & 0 \; \;\Rightarrow \; \;
{J\cdot x = J \cdot y =
0} \\
\int J \wedge J & \propto & \mbox{Vol}(S) > 0 
\; \;\Rightarrow  \; \;{J\cdot J > 0}
\end{eqnarray}
It follows that $(x, y, J)$ define a positive-definite 3-plane in
$H^2(S;\R) =\R^{3, 19}$.

Now, let us consider fluxes of an $SO(32)$ gauge theory on K3.  Fluxes
wrapped around nontrivial two-cycles are quantized, and can be
normalized to $F = 2 \pi i f$ where
\begin{equation}
f \cdot f \in 2\Z, \;
f \in H^2 (K3,\Z)
= \Gamma^{3, 19}\,.
\end{equation}
We thus naturally identify such a flux with an element of the lattice
$\Gamma^{3, 19}$.  

The group $SO(32)$ contains 16 mutually commuting
$U(1)$ factors (associated with $SO(2)$ transformations on mutually
disjoint pairs of the 32 indices).  In the type I picture these 16
$U(1)$'s are the world-volume gauge fields on each of the 16 D9-branes
in the theory.  A flux $f$ on one D9-brane is accompanied by a flux
$-f$ on the orientifold image of that D9-brane.  (This is essentially
the T-dual of the relationship between diagonal branes and their
orientifold images in the intersecting brane models described in
Section \ref{sec:IBM}.) We consider a configuration of
the $SO(32)$ theory where we have ``stacks'' of $N_a$ D9-branes with
flux $f_a$, $a = 1, \ldots, K$
\begin{equation}
{\tiny F = \left(\begin{array}{lll}
\left.\begin{array}{ccc}
 f_1 & &\\
& \ddots &\\
& & { f_1}
\end{array} \right\}& \hspace*{-0.1in}N_1 &
{\LARGE 0}\\
&\hspace*{-0.25in} \left.\begin{array}{ccc}
{ f_2} & &\\
& \ddots &\\
& & { f_2}
\end{array} \right\}& \hspace*{-0.1in}N_2\\
{\LARGE 0} & & \hspace*{-0.2in}\ddots
\end{array} \right)}
\end{equation}

The constraint that the total instanton number is 24 becomes
\begin{equation}
\frac{1}{8\pi^2}\int_s \Tr F \wedge  F = 24
\; \; \Rightarrow \; \; \sum_{a}N_af_a \cdot f_a = -24 \,.
\label{eq:instanton-condition}
\end{equation}
The constraints from supersymmetry (analogous to \eq{eq:IBM-SUSY})
give
\begin{equation}
\int f^a\wedge \Omega = 0 \quad \int f^a\wedge \bar{\Omega} = 0 \quad
\int f^a\wedge J=0\,
\end{equation}
\begin{equation}
\Rightarrow \; \; f^a \cdot x = f^a \cdot y = f^a \cdot J = 0 \,.
\label{eq:magnetic-SUSY}
\end{equation}

The set of abelian instanton configurations on K3 is thus
parameterized by the set of fluxes $\{f_a\}$.  These fluxes generate
an even lattice
\begin{equation}
L \subset \Gamma^{3, 19} \,,
\end{equation}
which characterizes the theory.  In fact, the structure of the gauge
group and matter content of any 6D theory constructed in this way are
dependent only on the integers $N_a$ and the matrix of inner
products $m_{ab} =f_a \cdot f_b$.  (Note that the matrix $m$ may be
degenerate, in which case the dimensionality of $L$ is smaller than
that of $m$.)  As long as the lattice $L$ is
negative-definite, there always exists a perpendicular
positive-definite 3-plane in $\Gamma^{3, 19}$, so that there are some
moduli for which (\ref{eq:magnetic-SUSY}) holds and
the theory is supersymmetric.  The gauge group of the
6D theory is given by
\begin{equation}
G =U(N_1) \times U(N_2) \times \cdots \times
U(N_K) \times SO(M), \;\;\;\;\;
M =32 -2 \sum_{a} N_a \,.
\end{equation}
The matter content can be determined from a simple index theory
calculation and is given in the following table
\begin{center}
\begin{tabular}{|c|c|}
\hline
Rep. (+ c.c.) & \# hypermultiplets \\
\hline
$(N_a,N_b)$ & $(-2-(f_a+f_b)^2)$ \\
$(N_a,\bar{N}_b)$ & $(-2-(f_a-f_b)^2)$ \\
Antisym.  $U(N_a)$ & $(-2-4f_a^2)$ \\
$(N_a,2M)$ & $(-2-f_a^2)$ \\
Neutral &     20 \\
\hline
\end{tabular}
\end{center}
It is easy to verify that anomalies cancel for any theory with this
spectrum; for example, a small calculation shows that
$H-V+29T=273$.  

There are a few technical subtleties that we have glossed over here.
Some of the $U(1)$ factors may be anomalous and removed by a
St\"uckelberg mechanism, as in the intersecting brane models discussed
above \cite{Honecker-1}.  Furthermore, there is a
possibility of further massless states that may enhance the gauge
spectrum when $J \cdot f = 0, f^2 = -2$ corresponding to a rational
curve on the K3 shrinking to a point \cite{kt-K3}.

Given that the spectrum and symmetry group of a 6D model constructed
in this fashion is determined uniquely by the integers $N_a$ and inner
products of vectors $f_a$ in the lattice $L$, it is natural to
ask for what $N_a, L$ models can be constructed.  Given $N_a,
m_{ab}$ satisfying $\sum_{a} N_am_{aa}= -24$, the criterion for
existence of a model of this type is that there exist a lattice
embedding
\begin{equation}
L (m)
  \hookrightarrow \Gamma^{3, 19}\,.
\end{equation}
It can be shown \cite{kt-K3} using Nikulin's lattice
embedding theorems that such an embedding is always possible for stack
sizes and matrix $m$ compatible with the 24-instanton condition
\eq{eq:instanton-condition}.  Furthermore, such an embedding is often
unique, with degeneracies sometimes associated with possible
overlattice embeddings that would give rise to additional discrete
structure in the 6D theory.
\vspace*{0.05in}

\noindent {\bf Example:}

As a simple example,
consider a model with  gauge group 
\begin{equation}
G = U(4) \times U(4) \times SO(16) \,.
\label{eq:K3-example}
\end{equation}
We have $N_1 = N_2 = 4$, so 
\eq{eq:instanton-condition} becomes 
\begin{equation}
4m_{11} + 4 m_{22} = -24 \,,
\end{equation}
with $m_{aa}$ an even integer for each $a$.  The only solution
(up to exchange of stacks) is
\begin{equation}
f_1 \cdot f_1 = -2, \; f_2 \cdot f_2 = -4 \,.
\end{equation}
There are 5 distinct choices for $m_{12}$ giving different theories;
we consider the case where $m_{12} = 0$.  From the table above we can
compute the matter spectrum
\begin{eqnarray}
{\rm matter} & = &  4 \times ({\bf 4}, {\bf \bar{4}}, {\bf 1})
+4 \times ({\bf 4}, {\bf 4}, {\bf 1})
+6 \times ({\bf  6}, {\bf 1}, {\bf 1})
+14 \times ({\bf  1}, {\bf  6}, {\bf 1})
\label{eq:K3-example-matter}
\\
 & & 
\hspace*{0.1in}
+2 \times ({\bf  1}, {\bf   4}, {\bf 16})
+20 \times ({\bf  1}, {\bf  1}, {\bf 1}) \,. \nonumber
\end{eqnarray}
It is straightforward to check that this ($T = 1$) model satisfies the anomaly
equations and has anomaly lattice  
\begin{equation}
\Lambda = \left(\begin{array}{cccc}
8 & 6 & 14 & -2\\
6 & 4 & 8 & 0\\
14 & 8 & 12 & 4\\
-2 & 0 & 4 & -4
\end{array} \right) \,.
\label{eq:K3-example-1}
\end{equation}
Note that the two $U(1)$ factors must be included in the number of
vector multiplets for the condition \eq{eq:hv} to be satisfied, and
that both kinds of bifundamental hypermultiplets $({\bf 4}, {\bf 4})$
and $({\bf 4}, {\bf \bar{4}})$
contribute to $b_1
\cdot b_2$.
The matrix defining the anomaly lattice \eq{eq:K3-example-1} is
degenerate, which is expected since the dimension of this lattice must
be $T + 1 = 2$.  Indeed, the anomaly vectors can be chosen in the
coordinate system \eq{eq:t1} to be
\begin{eqnarray}
-a= (2, 2) &  \hspace*{0.1in}&  b_1 = (1, 2)\\
b_2 = (1, 6) &  &  b_3 = (1, -2) \,.
\end{eqnarray}
These vectors correctly reproduce the matrix of inner products in
\eq{eq:K3-example-1}.  We will see in Section \ref{sec:6D-map} how
this structure can be used to easily map this model to the topological
data for an F-theory compactification.
\vspace*{0.1in}

The general approach described here for compactification of the
$SO(32)$ theory on K3 can be repeated for the gauge group $E_8 \times
E_8$ with analogous results \cite{Honecker-t}.  

The models described here represent only
a small slice of the full space of type I/heterotic compactifications
on K3.  More generally, the instanton structure can be nonabelian;
the moduli space of vacua in this class is given by the moduli space
of instantons on K3 with instanton number 24.  
Tools for understanding instanton moduli spaces of this type were
developed by Friedman, Morgan, and Witten \cite{fmw}.
By embedding nonabelian instantons with total instanton number 24 in a
subgroup of the heterotic group, many other theories can be
constructed \cite{Kachru-Vafa, Duff-Minasian-Witten, Aldazabal-fiu}.
For example, as pointed out by Kachru and Vafa, putting all the
instantons in an $SU(2)$ subgroup of one of the factors of $E_8 \times
E_8$  gives a theory with gauge group
\begin{equation}
G = E_7 \times E_8
\label{eq:78-model}
\end{equation}
and charged matter content
\begin{equation}
{\rm matter} = 10 \times ({\bf 56}, 1) \,.
\end{equation}
Using the group theory coefficients 
\begin{eqnarray}
E_7:&  & A_{56} = 1, C_{56} = 1/24, A_{\rm adj} = 3, C_{\rm adj} = 1/6\\
E_8:&  & A_{\rm adj} = 1, C_{\rm adj} = 1/100
\end{eqnarray}
the anomaly lattice for this theory is
\begin{equation}
\Lambda= \left(\begin{array}{ccc}
8 &  -14  & 10\\
-14 &  12 & 0\\
10 &  0 & -12
\end{array} \right) \,.
\end{equation}
In the  $T = 1$ basis (\ref{eq:t1}) we have
\begin{eqnarray}
b_7=(1, 6), & &  b_8 = (1, -6)\,. \label{eq:78-b}
\end{eqnarray}

The continuous moduli space of theories including nonabelian
instantons connects many of the models that appear with distinct
gauge groups in the preceding analysis.  Generically, the bundle
structure will break the $SO(32)$ gauge group to $SO(8)$.  The
different models described above are connected by Higgsing to a
generic model, moving in moduli space, and un-Higgsing to restore a
different gauge group.  This connects part of the moduli space of 6D
theories; we describe a more extensive connectivity of this moduli
space in the context of F-theory in Section \ref{sec:6D-F-theory}.

Note that while the magnetized brane models just described are morally
``T-dual'' to the intersecting models described in the previous
section, the specific models arising from intersecting brane models of
IIB on K3 do not appear in the magnetized brane picture.  For example,
the  model from \eq{eq:IBM-example-1} with gauge group $SU(8)$
does not arise as a magnetized brane model.  This is because the
orbifold quotient needed to realize K3 breaks the naive T-duality
associated with $T^4$.  Further structure is needed to make the
connection between these classes of models more precise, some of which
we describe briefly in the following section.

\subsection{Other constructions of 6D type I/heterotic vacua}
\label{sec:6D-others}

A wide range of other 6D models have been constructed from string
theory, using a variety of techniques related to the heterotic/type I
constructions described in the previous section.  We mention briefly
some of these approaches and models to give a sense of the range of
possibilities.  A more comprehensive review of the range of models
known can be found in the review by Ib\'a\~nez and Uranga
\cite{Ibanez-Uranga}.  As we discuss in the following section,
F-theory provides a powerful unifying framework that seems to
encompass all of the vacua constructed from these other approaches.
The F-theory description of each of these models can be determined
directly from the anomaly structure, as we discuss in Section
\ref{sec:6D-map}.  There are other types of 6D models
related to the type I compactification on K3 that are constructed by
generalizing the geometry of K3 to a ``non-geometric space''
\cite{Hellerman-mw}.  These models also can be associated with
F-theory constructions.  

One class of models was constructed by Bianchi and Sagnotti based on
combining the $\Z_2$ orientifold action in the IIB theory with a
space-time orbifold action giving $T^4/\Z_2$ \cite{Bianchi-Sagnotti,
  Bianchi-Sagnotti-2, Gimon-Polchinski, Dabholkar-park-1}.  One model
of this type that was considered in more detail by Gimon and
Polchinski gives a gauge group
\begin{equation}
G = SU(16) \times SU(16) \,,
\label{eq:Gimon-Polchinski}
\end{equation}
with matter content
\begin{equation}
{\rm matter} =
2 \times ({\bf 120}, {\bf 1}) +
2 \times ({\bf 1}, {\bf 120}) +
 1 \times ({\bf 1}, {\bf 1}) \,.
\end{equation}
For each gauge group factor the anomaly cancellation in this model
follows as for the example with matter (\ref{eq:example-matter-0}).
This kind of type I model on K3 orbifolds can also give rise to
theories with additional tensor multiplets \cite{Gimon-Johnson,
  Dabholkar-park-2, Polchinski-tensors}.

A variety of
heterotic compactifications on various toroidal orbifold limits of K3
of the form $T^4/\Z_k$ were constructed by Erler and others
\cite{Erler, afiq}.
Tables of some of these models can be found in the work of Aldazabal
{\it et al.} \cite{afiuv}, and include models such as one with
\begin{equation}
G = SU(16), \;\;\;\;\; {\rm matter} = 2 \times {\bf 120} + 16 \times
{\bf 16} \,.
\end{equation}
Note that this is the model considered in (\ref{eq:example-matrix-0})
with the maximum possible value of $N$.
Many heterotic orbifold models were considered by Honecker and
Trapletti \cite{Honecker-1, Honecker-t} and connected to smooth K3
constructions.

From the heterotic point of view, more exotic theories with additional
tensors can arise when instantons shrink to a point.  This corresponds
to a limit where the 5-brane charge encoded in the instanton congeals
into a localized 5-brane and can separate from the ambient gauge
theory \cite{Witten-instantons}.  The 5-brane world-volume theory
carries a self-dual two-form field that corresponds to an additional
tensor in the 6D theory.  By shrinking $k$ coincident instantons, a
group $Sp(k)$ arises on the 5-brane world-volume.  The gauge groups
can become even more exotic when the instantons shrink to a point with
an orbifold singularity.  A variety of models of this type have been
constructed \cite{Dabholkar-Park-3, Gopakumar-Mukhi, Aspinwall-Gross}.
An extreme case of the kind of large gauge group that can arise was
identified by Aspinwall, Gross and Morrison \cite{Aspinwall-Morrison} in a
theory with 192 vector multiplets and
\begin{equation}
G =E_8^{17} \times F_4^{16} \times G_2^{32} \times SU(2)^{32} \,.
 \label{eq:big-group}
\end{equation}
A further complication that can arise in these heterotic
constructions is the appearance of bundles without vector structure
\cite{Berkooz-all, Aspinwall-point}.  This refers to a situation where
an instanton attached to an orbifold singularity describes a
Spin(32)$/\Z_2$ bundle but not an $SO(32)$ bundle.

From the heterotic/type I point of view, the
wide range of possible models and underlying physical mechanisms,
including all possible numbers of tensor fields, seems difficult to
connect into a single systematic framework.  We now turn to the point
of view of F-theory, which provides a unifying perspective on the
complicated network of possible 6D theories.

\subsection{Six-dimensional vacua of F-theory}
\label{sec:6D-F-theory}

We now consider F-theory compactifications to six dimensions.  The
story is very similar in spirit to the description of
eight-dimensional F-theory vacua in Section \ref{sec:8D-F-theory},
though the details are more complicated.  We begin by developing the
structure of F-theory further to incorporate compactification on
elliptically fibered Calabi-Yau threefolds.  We then show how the
anomaly data from any 6D theory can be used to identify the
topological data for an F-theory realization of the theory, if one
exists.  This brings us to a point where we can discuss the global
structure of the space of 6D ${\cal N} = 1$ supergravity theories.

For a compactification of F-theory to six dimensions, we consider a
Calabi-Yau threefold $X$ that is elliptically fibered over a complex
surface $B$.  The canonical class of the base is $K = c_1 (T^*)\in
H^2 (B)$, with a description in local complex coordinates $\sim dz_1
\wedge dz_2$.  As in the case of a surface elliptically
fibered over $\P^1$, there is a Weierstrass description of the
elliptic fibration in terms of local functions $f, g$ on the base
\begin{equation}
 -y^2+ x^3 + f (s, t)x  +g (s, t)  = 0 \,,
\end{equation}
where $s, t$ are local coordinates on the base.
The functions $f, g$ are given globally by
sections of ${\cal O} (-4K)$ and ${\cal O} (-6K)$,
as in the 8D story.  Again, codimension one singularities in the fibration
give rise to the gauge group of the 6D theory.  These are understood
through the Kodaira classification as in 8D.  In 6D theories, however,
there are also codimension two singularities.  These give rise to
matter in the 6D theory.  While some of the simplest kinds of
codimension two singularities, giving rise to the simplest types of
matter, are well understood, the mathematical description of
codimension two singularities is as yet incomplete.

One approach to systematically constructing many elliptically-fibered
Calabi-Yau geometries for  F-theory compactifications is through
the use of toric geometry \cite{Candelas-Font, Candelas-pr,
  Candelas-pr2, Candelas-pr3}.  We do not discuss this approach in
detail here, but it provides a concrete set of tools for explicitly
constructing and analyzing a broad subset of the space of
elliptically-fibered spaces used in F-theory constructions.

\subsubsection{Codimension one singularities and gauge groups}

The discriminant locus where the fibration is singular is given by
the set of points where $\Delta = 4f^3 +27g^2 = 0 $.  This is
generically a codimension one locus on the base, which can be
characterized as a divisor
\begin{equation}
\sum_{i}n_iH_i, \;\;\;\;\; n_i \in\Z
\label{eq:divisor}
\end{equation}
where $H_i$ are irreducible algebraic
hypersurfaces on the base $B$.
The divisor class of the discriminant locus is again
given through the Kodaira condition  for the total space to be Calabi-Yau
\begin{equation}
-12K =\Delta \,.
\end{equation}
This divisor is {\em effective}, meaning that the coefficients $n_i$
in an expansion of the form \eq{eq:divisor} satisfy $n_i \geq 0$.  The
irreducible components of $\Delta$ can give rise to nonabelian gauge
group factors, through the Kodaira classification in
Table~\ref{t:Kodaira}.  We denote by $\xi_i$ the irreducible
components associated with nonabelian factors and by $\nu_i$ the
corresponding multiplicity ({\it i.e.} the degree of $\Delta$ along
$\xi_i$).  
For example, an $A_{N -1}$ singularity on a divisor class $\xi$ has
multiplicity $\nu = N$, while an $E_8$ singularity has multiplicity
$\nu = 10$.
Each $\xi_i$ is associated with an algebraic curve on the
surface $B$.  In the IIB picture, multiple 7-branes are wrapped on the
curves $\xi_i$.  There is a residual part of the discriminant locus,
which we denote by
$Y$, that is not sufficiently singular to give rise to nonabelian
gauge factors.  The total discriminant locus can thus be written
\begin{equation}
-12K =\Delta = \sum_{i}\nu_i \xi_i + Y \,,
\label{eq:Kodaira}
\end{equation}
where $\xi_i$ gives rise to the nonabelian gauge group factor
\begin{equation}
\xi_i \rightarrow G_i \,.
\end{equation}
The semisimple part of the gauge group for the 6D theory is then given by
\begin{equation}
G = G_1 \times \cdots \times G_k /\Gamma
\end{equation}
where $\Gamma$ is a discrete group.  The abelian part of the gauge
group arises in a more subtle fashion from the global structure of the
elliptic fibration, as in 8D.  More precisely, the rank of the abelian
part of the gauge group is given by the dimension of the {\em
  Mordell-Weil} group of rational sections of the fibration
\cite{Fukae-mw, Hulek-Kloosterman}.  The structure of the discrete
group $\Gamma$ is determined by the torsion part of the Mordell-Weil
group \cite{Aspinwall-Morrison-mw}.  In terms of the
total space $X$ of the Calabi-Yau threefold, the total rank $r$ of the
gauge group, including both abelian and nonabelian factors, is given
by \cite{Morrison-Vafa-II}
\begin{equation}
r = h_{1, 1} (X) -h_{1, 1} (B) -1 \,.
\end{equation}
While this is easy to compute when the geometry of the threefold is
known, it is less straightforward to compute the abelian part of the
gauge group given only the local characterization of the fibration,
and requires either the explicit Weierstrass model or a detailed
characterization of the codimension one and two components of the
discriminant locus.

A further complication that can arise in 6D F-theory compactifications
is the appearance of non-simply-laced groups outside the A-D-E
classification \cite{Bershadsky-all, Berglund-kmt,
  Aspinwall-Katz-Morrison}.  This can occur when the set of singular
curves associated with a codimension one singularity undergoes a
monodromy around a nontrivial cycle in the base, so that the
associated Dynkin diagram is mapped into itself in a nontrivial
fashion ({\it i.e.}, through an outer automorphism).  For example, an
$A_7$ singularity that undergoes a monodromy reflecting the Dynkin
diagram from one end to the other contracts the Dynkin diagram through
the reflection, giving rise to a group $Sp(4)$ associated with a $C_4$
singularity.  Treating such groups in F-theory requires a slightly
more careful analysis of the singularity structure on the dimension
one discriminant locus, for which a systematic procedure known as the
``Tate algorithm'' is helpful \cite{Tate}.

\subsubsection{Codimension two singularities and matter}

Matter in the 6D theories is produced from codimension two
singularities in the elliptic fibration.  In general, at codimension
two points in the base $B$, the singularity structure of the fibration
can worsen.  This can lead to additional points in the base where the
total space of the fibration is singular and must
be blown up into two-cycles to give a smooth Calabi-Yau threefold.
These two-cycles represent additional matter fields in the theory.  In
general, the codimension two singularity at a point $x$ in the
base gives an enhancement of the Kodaira type of the singularity along
one or more codimension one curves $\xi_i$ passing through the point $x$.  In
the simplest kind of codimension two singularity, as described by Katz
and Vafa \cite{Katz-Vafa}, the singularity at $x$ is resolved by
blowing up a Kodaira type associated with a group of one rank higher
than the codimension one singularity/singularities intersecting that
point.  In this case the matter content is found by decomposing the
adjoint of the larger group into the gauge factor(s) associated with
the codimension one locus.  In other cases, the codimension two
singularity can give rise to matter in a more complicated fashion
\cite{Grassi-Morrison, Morrison-Taylor}.  One general class of codimension two
singularities has been analyzed by Miranda \cite{Miranda}, though more
general classes of singularities are possible that have not yet been
completely classified.
\vspace*{0.05in}

\noindent
{\bf Example:}

Consider a local singularity of type $A_{n-1}$ along a curve $\xi_1$,
which we can take to be on the line $t = 0$, with $\Delta \sim t^n$.
If there is also a codimension one singularity of type $A_{m-1}$ along
the curve $\xi_2$ given by the line $s = 0$, with $\Delta \sim s^m$,
then at the origin where the curves intersect, the total singularity
type is $A_{n + m - 1}$, since ord ($\Delta$) $= n + m$.
The two components of the codimension one singularity locus along $\xi_1,
\xi_2$ give a group $SU(n) \times SU(m)$.  The enhanced singularity at
the origin is associated with the adjoint of $SU(n + m)$ (although
this is not a part of the gauge symmetry).  Decomposing the adjoint
representation of $SU(n + m)$ gives a bifundamental field in the $(n,
\bar{m}) + (\bar{n}, m)$ representation of the two gauge group
factors.  This is the F-theory realization of the bifundamental strings
arising from intersecting branes.
\vspace*{0.05in}

The mechanism by which matter arises in F-theory is thus related to
the standard mechanism realizing matter from strings connecting
intersecting branes.  But in F-theory there is a much wider range of
possible matter structures \cite{Katz-Vafa, Grassi-Morrison-2,
  Morrison-Taylor}.  For example, a trifundamental matter field can be
produced by a local enhancement of an $A_1 \times A_2 \times A_4$
singularity to $E_8$
\begin{center}
\begin{picture}(40,10)(- 20,0)
\multiput(-22.5,0)(7.5,0){7}{\circle*{2.5}}
\multiput(-22.5,0)( 7.5,0){6}{\line(1, 0){7.5}}
\put(- 7.5,0){\makebox(0,0){${\times}$}}
\put(-7.5,7.5){\circle*{2.5}}
\put(-7.5,0){\line( 0,1){7.5}}
\end{picture}\;\;\;\;\;$\Rightarrow 
({\tiny\yng(1)}_3,{\tiny\yng(1)}_2,{\tiny\yng(1)}_4)$
\end{center}
In addition to matter from localized singularities, there are global
contributions to matter in F-theory.  In particular, a codimension one
singularity locus associated with a gauge group factor $G_i$ that is
wrapped on a curve of genus $g$ in the base gives rise to $g$
adjoint representations of $G_i$.  

\subsubsection{Bases for elliptically fibered Calabi-Yau threefolds}
\label{sec:6D-bases}

We now consider the question of which bases $B$ can be used for an
F-theory compactification on an elliptically fibered Calabi-Yau
threefold.  A full discussion of the mathematics underlying the answer
to this question is beyond the scope of these lectures, but we
summarize the main points here.

The only bases that can support an elliptically fibered Calabi-Yau
threefold are given by the following set of spaces \cite{Grassi-1,
  Grassi-2, Morrison-Vafa-II}:
\vspace*{0.05in}

\noindent $T^4$, K3: These spaces both have $K = 0$.  So there is no
discriminant locus and the fibration is trivial.  Thus, these are 
simple IIB compactifications, and the general structure of F-theory is
not needed.  
These models give theories with enhanced supersymmetry.  Models with enhanced
supersymmetry can also arise from compactification on hyperelliptic
surfaces or surfaces of the form $(T^2 \times \P^1)/G$ with $G$ a
discrete group; we do not consider any of these models   further here.
\vspace*{0.05in}

\noindent Enriques surface: This is an orbifold of K3 with no fixed
points.  The space has $-12K = 0$, with $K$ a torsion class.  So there
are no nonabelian gauge group factors on the discriminant locus. 
F-theory can nonetheless be compactified on the Enriques surface
\cite{Blum-Zaffaroni}.
The
dyon charge lattice for the theory is \cite{fhsv} $\Gamma^{1, 9}$.
\vspace*{0.05in}

\noindent $\P^2$: The two-dimensional complex projective space $\P^2$
has $H_2 (\P^2)=\Z$, with $K = -3H$, where $H$ is
the hyperplane divisor generating $H_2 (\P^2)$.
\vspace*{0.05in}

\noindent $\F_m$: Hirzebruch surfaces with $m \leq 12$ (described in
more detail below).
\vspace*{0.05in}

\noindent  Blow-ups of $\P^2,\F_m$ at one or more points: Blowing up one or more
points on these spaces gives surfaces with increasingly large $H_2
(B,\Z)$.
\vspace*{0.05in}

It is also possible that there may be consistent F-theory
compactifications on bases with orbifold singularities, whose resolution gives
one of the spaces above \cite{Morrison-Vafa-II}; understanding such
compactifications is an interesting open problem.

The fact that no other surfaces besides those listed above can be used
as bases for an elliptically fibered Calabi-Yau manifold follows from
results in {\em minimal surface theory}.  Basically, the idea of
minimal surface theory is that on any surface which contains a divisor
class describing a curve $C$ satisfying $C \cdot K = C \cdot C = -1$,
the curve $C$ can be blown down to give a simpler smooth surface
\cite{Friedman}.  Minimal surfaces are those which admit no further
curves which can be blown down in this fashion.  The only minimal
surfaces which can be bases for elliptically fibered Calabi-Yau
manifolds are those listed above.

It will be helpful to have on hand some more details of the structure
of the Hirzebruch surfaces $\F_m.$
These surfaces can be described as $\P^1$ bundles over $\P^1$; the
surface $\F_m$ is essentially equivalent to the line bundle over
$\P^1$ with first Chern class $c_1 = -m$
with all fibers compactified to $\P^1$.  For example,
\begin{equation}
\F_0 = \P^1 \times \P^1 \,.
\end{equation}
The surface $\F_1$ is also described by blowing up $\P^2$ at a point.
The surface $\F_2$ is given by compactifying $T^*\P^1$ on each fiber.

A basis for the linear space of divisors of $\F_m$ is given by the
divisors $D_v, D_s$, where $D_v$ is a section of the $\P^1$ fibration
with $D_v^2 = -m$, and $D_s$ is a fiber of the fibration.  The
intersection matrix in this basis is given by
\begin{equation}
\left(\begin{array}{cc}
D_v \cdot D_v & D_v \cdot D_s\\
D_s \cdot D_v & D_s \cdot D_s
\end{array}\right)
=
\left(\begin{array}{cc}
-m & 1\\
1 & 0
\end{array}\right) \,.
 \label{eq:fm}
\end{equation}
Geometrically we can picture $D_v$ as a horizontal line depicting the
section and $D_s$ as a vertical line depicting the fiber
\begin{center}
\begin{picture}(80,35)(-40,-10)
\put(-40,0){ {
 \line( 1,0){80}}}
\put(20,-8){\makebox(0,0){ { \small
  $D_v$}}}
\put(  -5,15){\makebox(0,0){\small 
  $D_s$}}
\put(-15,- 10){
 \line( 0,1){35}}
\end{picture}
\end{center}
The irreducible effective divisor classes in $\F_m$ are given by
\begin{equation}
D_v, \; aD_v + bD_s, \;\;\;\;\;b \geq ma \,.
\label{eq:fm-irreducible}
\end{equation}
The canonical class on $\F_m$ is
\begin{equation}
-K = 2D_v + (2 + m)D_s\,.
\end{equation}
It is easy to check that for any $m$, $K^2 = 8$ on $\F_m$.

From the divisor structure of $\F_m$ we can see that $\F_m$ can only
be the base for an elliptically fibered Calabi-Yau threefold with
singularities that are in the Kodaira table when $m \leq 12$.  From
$f  = -4K$, we see that for $m > 2$, $f$ must have an irreducible component
proportional to $D_v$, since from eq.\ (\ref{eq:fm-irreducible}) any
irreducible component other than $D_v$ must have a coefficient of
$D_s$ at least $m$ times the coefficient of $D_v$.  Writing
\begin{equation}
f = x_f D_v + y = x_f D_v + (A D_v + B D_s) = 8 D_v + (8 + 4m) D_s
\end{equation}
where $y$ is a sum of irreducible components other than $D_v$, we have
$B = 8 + 4m \geq mA$, so $A \leq 4 + 8/m$.  It follows that 
\begin{equation}
x_f \geq 4 - 8/m \,.
\end{equation}
A similar analysis for $g = -6K$ and $\Delta = -12K$ gives
\begin{equation}
x_g \geq 6-12/m, \;\;\;\;\; x_{\Delta} \geq 12-24/m \,.
\end{equation}
For $m > 12$, the divisor class $D_v$ thus carries a singularity of
degrees deg$ (f) \geq 4$, deg$ (g) > 5$, deg$ (\Delta) > 10$.  Since
$D_v^2$ is negative, there are no deformations of this divisor
class, so there is a singularity worse than the Kodaira classification
allows on this locus at $m > 12$.

It is easy to verify that  for $\F_m$, the lattice $\Gamma =H_2 (B,\Z)$
spanned by $D_v, D_s$ can be written in an appropriate basis as
\begin{equation}
\Gamma=
\Gamma_0 = \left(\begin{array}{cc}
 0& 1\\
1 & 0\\
\end{array} \right), m\; {\rm even}, \;\;\;\;\;
\Gamma=
\Gamma_1 = \left(\begin{array}{cc}
 1& 0\\
0 & -1\\
\end{array} \right), m\; {\rm odd} \,.
\label{eq:gamma-01}
\end{equation}
In fact, all the even Hirzebruch surfaces $\F_{2m}$ are topologically
equivalent to one another, as are all the odd surfaces $\F_{2m +1}$.
The difference between two even or odd Hirzebruch surfaces with
distinct $m$ lies in the complex structure, which is associated with a
different set of effective irreducible divisors for each $m$.

Blowing up a generic point on a surface $B$ adds an additional divisor
class $E$ (the exceptional divisor), giving a new surface $\tilde{B}$ with
\begin{equation}
E^2 = -1, \;\;\;\;\; \tilde{K} \cdot E = -1 \,.
\end{equation}
In the new surface, $\tilde{K} = K + E$.  The dimension of the second
homology group increases by 1 with the blow-up so that $h_{1, 1}
(\tilde{B}) =h_{1, 1} (B) +
1$.  For example, if we blow up a generic point on $\F_1$ the
resulting surface is the {\em del Pezzo} surface $dP_2$.  The
intersection form on $dP_2$ is given by adding a new dimension
corresponding to the orthogonal generator $E$ to
$\Gamma_1$ from eq.\ (\ref{eq:gamma-01}), and  taking
$K_{dP_2} = K_{\F_1} + E$,
\begin{equation}
\Gamma_{dP_2} =
 \left(\begin{array}{ccc}
1 & 0 &0\\
0 & -1 & 0\\
0 &  0 & -1
\end{array} \right) \,,
\end{equation}
in a basis where $K_{dP_2} = (-3, 1, 1)$.
Blowing up a
generic point on $dP_k$ gives $dP_{k +1}$.
Blowing up non-generic points gives a more complicated tree of
resulting surfaces.  Each blow-up, however, has the effect of reducing
$K^2$ by 1 through the addition of the new exceptional divisor.   In
general, for any base we have
\begin{equation}
K^2 = 10-h_{1, 1} (B) \,.
\label{eq:k2-a}
\end{equation}

The number of tensor multiplets in the 6D supergravity theory
associated with F-theory on an elliptic fibration over the base $B$ is
given by 
\begin{equation}
T = h_{1, 1} (B) -1 \,.
\label{eq:t-h}
\end{equation}
The scalar fields in the $T$ tensor multiplets correspond to the
relative K\"ahler (volume) moduli for the $h_{1, 1}$ two-cycles in the
base, leaving out the overall volume modulus of $B$, which is
controlled by a scalar hypermultiplet.  Thus, fibrations over $\P^2$
correspond to theories with no tensor multiplets, fibrations over
$\F_m$ give theories with one tensor multiplet, and each additional
blow-up of the base gives an additional tensor multiplet.  Combining
\eq{eq:k2-a} and \eq{eq:t-h} gives
\begin{equation}
K^2 = 9-T =  10-h_{1, 1} (B) \,.
\label{eq:k2}
\end{equation}

The mathematical connection between a Calabi-Yau
elliptically fibered over a base $B$ and a Calabi-Yau fibered over a
base $\tilde{B}$ realized by blowing up $B$ at a point is mirrored in
the physics of the associated supergravity theories.  In the
supergravity theory, such a transition is realized by a phase
transition in the theory in which a single tensor multiplet is
exchanged for 29 scalar multiplets \cite{Seiberg-Witten-6D,
  Witten-phase, Morrison-Vafa-II}.  Starting from the theory on $\tilde{B}$
with the
larger value of $T$, this transition can be described by  a
limit in which a string becomes tensionless as $j \cdot x \rightarrow 0$ for
some $x \in \Gamma$.  From the F-theory point of view, the transition
from $B$ to $\tilde{B}$ can be seen as arising from a tuning of the
Weierstrass parameters for an elliptic fibration over $B$ so that a
singularity arises that is worse than any singularity in
the Kodaira Table.  For example, at a codimension two singularity
where deg $f$ = 4, deg $g$ = 6, and deg $\Delta$ = 12, the singularity
must be resolved by blowing up a point in the base.  These transitions
connect F-theory models with different bases $B$ and different numbers
of tensor multiplets $T$.  We discuss this connectivity in the space
of supergravity theories further below in Section
\ref{sec:connectivity}.

We now consider a simple example of an F-theory compactification on an
elliptically fibered threefold, focusing on the topological data.  
As additional examples,
many of the other theories constructed through other methods are
mapped into F-theory in Section \ref{sec:6D-map}.
\vspace*{0.05in}

\noindent
{\bf Example:} $SU(N)$ on $\F_2$

This is a 6D supergravity theory with $T = 1$.  If the $SU(N)$ is
realized by $N$ D7-branes on the divisor class 
\begin{equation}
\xi =D_v
\label{eq:example-1-6D}
\end{equation}
then we have
\begin{equation}
-12K = 24 D_v + 48 D_s = N D_v + Y \,,
\end{equation}
where $Y = (24 -N) D_v + 48 D_s$.  Matter in the fundamental
representation of the $SU(N)$ arises at intersections between $\xi =
D_v$ and $Y$.  The number of such intersections is
\begin{equation}
\xi \cdot Y = D_v \cdot \left[ (24 -N) D_v + 48 D_s \right] = 48-2(24
-N) = 2 N \,,
\end{equation}
so the theory has $2 N$ matter fields in the fundamental
representation of $SU(N)$.  It is easy to check that this is an
anomaly-free spectrum.  An explicit Weierstrass model for this F-theory
compactification for $N \leq 14$ has been identified \cite{KMT}.

To fit the set of models described through other constructions into
the context of F-theory, it is useful to exploit the close
correspondence between the anomaly structure of 6D supergravity
and the structure of F-theory compactifications, to which we now turn.

\subsection{Mapping 6D supergravities to F-theory}
\label{sec:6D-map}

As discussed in Section \ref{sec:6D-anomalies}, every 6D supergravity
theory contains a signature $(1, T)$ lattice $\Gamma$ of dyonic string
charges that couple to the (anti-)self-dual fields $B^\pm$ in the
theory.  From the F-theory point of view these dyonic strings are
realized by D3-branes wrapping cycles in $H_2 (B,\Z)$ of the base.
Thus, 
\begin{equation}
\Gamma =H_2 (B,\Z)
\end{equation}
with inner product given by the intersection form on $B$.
In F-theory the self-duality of the lattice $\Gamma$ follows
immediately from Poincar\'{e} duality.

We now explicitly describe how the anomaly lattice $\Lambda$ of a 6D
supergravity theory is mapped into the charge lattice $\Gamma$ for
theories with an F-theory realization \cite{KMT, KMT-II}.
The divisor classes associated with nonabelian gauge group factors in a 6D
F-theory compactification live in $\Gamma$ since
\begin{equation}
\xi_i \in H_2 (B,\Z)\,.
\end{equation}
The nonabelian gauge group factor $G_i$ comes from a stack of 7-branes
wrapped on the cycle $\xi_i$.  The corresponding element $b_i$ in the
anomaly lattice comes from a gauge dyonic string in space-time
associated with an instanton in $G_i$.  In F-theory this 
instanton gives a 3-brane wrapped on $\xi_i$ within the
7-brane world-volume.  Thus, we can associate the element $b_i$ of the
anomaly lattice with $\xi_i$
\begin{equation}
b_i \rightarrow \xi_i \,.
\label{eq:map-b}
\end{equation}
From the anomaly condition (\ref{eq:aa}) and the condition (\ref{eq:k2})
we have $a^2 = K^2 = 9-T$.  This suggests that
\begin{equation}
a \rightarrow K \,.
\label{eq:map-a}
\end{equation}
Indeed, both (\ref{eq:map-b}) and (\ref{eq:map-a}) can be checked by
computing the intersections in the F-theory picture and the
corresponding anomaly lattice  \cite{Sadov, Grassi-Morrison}
\begin{eqnarray}
-a \cdot b & = & -K \cdot \xi_i\\
b_i \cdot b_j & = &\xi_i \cdot \xi_j
\end{eqnarray}

This gives a clear picture of how the topological data needed for
an F-theory compactification can be identified from the structure of
the 6D supergravity spectrum and anomaly lattice.
Given a 6D theory, the number of tensor multiplets $T$ determines
$h_{1, 1} (B)$ through \eq{eq:t-h}.  There must then exist a lattice embedding
\begin{equation}
\Lambda \hookrightarrow \Gamma
\label{eq:map}
\end{equation}
that maps $a, b_i \rightarrow K, \xi_i$ as in \eq{eq:map-a},
\eq{eq:map-b}.  
This map takes the $(1, T)$
vector $j$ of scalars in the tensor multiplets to the K\"ahler moduli
of the F-theory compactification
\begin{equation}
j \rightarrow J \,.
\end{equation}
The mapping \eq{eq:map} is not necessarily uniquely defined given only
the nonabelian gauge group and matter content of the 6D theory.  There
may be multiple possible surfaces $B$ giving compatible F-theory
compactifications, and for a given surface the lattice map may not be
uniquely defined, although in many cases it can be shown that the
embedding is unique up to automorphisms of $\Gamma$ using theorems of
Nikulin.  To uniquely determine the F-theory model corresponding to a
given 6D supergravity, further information may be needed.  Knowing the
dyon charge lattice of the low-energy theory and having an explicit
description of the space of possible K\"ahler moduli encoded in $j$
({\it i.e.}, the K\"ahler cone), for example, is sufficient to
uniquely determine the intersection form on $B$, along with the set of
effective divisors.  This information uniquely determines the F-theory
realization.  Knowing the $U(1)$ content of the theory can also in
principle help in determining the topology of the F-theory geometry.
In many simple cases, however, as we see in examples below, the
nonabelian gauge group and matter content are already sufficient to uniquely
determine the corresponding F-theory realization.

In the case $T = 1$, the map \eq{eq:map} can be written
explicitly in terms of the divisors $D_v, D_s$ on $\F_m$ satisfying
\eq{eq:fm}.
The F-theory divisor corresponding to an anomaly vector $b$
is given by \cite{KMT}
\begin{equation}
b =\frac{1}{2}(\alpha,\talpha) \; \rightarrow
\;\xi =\frac{\alpha}{ 2}  (D_v + \frac{m}{2} D_s)
+ \frac{\talpha}{ 2}  D_s 
\label{eq:map-1}
\end{equation}
where we have expressed $b$ in terms of $\alpha, \talpha$ as in
\eq{eq:t1}.  This map is also compatible with the corresponding
expression for $a$
\begin{equation}
-a = (2, 2)\rightarrow -K= 2D_v + (m + 2)D_s\,.
\end{equation}
Note that  the anomaly formalism is in this case invariant under
exchange of $\alpha, \tilde{\alpha}$.  This can in some cases give
distinct realizations of a given 6D model (which are often related
through duality, such as for models on $\F_0 = \P^1 \times \P^1$,
where exchange of $\alpha, \tilde{\alpha}$ corresponds to an exchange
of $D_v$ and $D_s$.)

Dualities between heterotic and F-theory models have been worked out
in many special cases and classes of examples in various parts of the
literature; this connection is explored in many of the papers
referenced in the discussion of heterotic orbifold models, in
particular in the work of Friedman, Morgan and Witten \cite{fmw}, as
well as in much of the F-theory literature.  The
map \eq{eq:map} gives a simple unified way of identifying such
dualities in terms of the discrete data of the 6D supergravity theory
and the corresponding F-theory topology.  We now consider a number of
explicit examples of this map.
\vspace*{0.05in}

\noindent {\bf Example:}

Consider again the theory with gauge group
$G =SU(N)$  and $2 N$ fundamental matter fields, with
$T = 1$.
A straightforward calculation gives
\begin{equation}
b =\frac{1}{2} (\alpha, \tilde{\alpha}) = (1, -1)
\rightarrow \xi =D_v + (m/2 - 1)D_s \,.
\end{equation}
The map will only give an integral effective divisor for $m = 2$.
This gives $\xi = D_v$ on $\F_2$, as described in \eq{eq:example-1-6D}.  
\vspace*{0.05in}

\noindent {\bf Example:}

Consider the intersecting brane model example \eq{eq:IBM-example-1}
with gauge group $U(8)$, 8 matter fields in the antisymmetric
two-index representation and one adjoint matter field.
We saw from \eq{eq:IBM-example-matrix} that the anomaly lattice is
generated by a single vector $b = -a$.  Using \eq{eq:map}, we see that
this model maps to $\F_0$ with
\begin{equation}
b = (2, 2) \rightarrow \xi =  2D_v +2 D_s \,.
\end{equation}
The genus of this curve is given by
\begin{equation}
(K + \xi) \cdot \xi = (a + b) \cdot b = 0 = 2g-2 \,.
\end{equation}
So the $SU(8)$ group lives on a discriminant locus curve of genus one,
as expected given the one adjoint representation.

\vspace*{0.05in}

\noindent {\bf Example:}

Now consider the example of a type I/heterotic K3
compactification with gauge group \eq{eq:K3-example} and matter
content \eq{eq:K3-example-matter}.  From the form of the anomaly
vector $b_3 = (1, -2)$ associated with the gauge group factor $SO(16)$,
we see that the only possible value for $m$ that gives an effective
divisor for the image of $b_3$ under \eq{eq:map-1}
is $m = 4$.  The divisor classes associated with the different gauge
group factors are then
\begin{eqnarray}
b_1 = (1, 2) & \rightarrow & D_u = D_v + 4 D_s \\
b_2 = (1, 6) & \rightarrow & D_v + 8 D_s \nonumber\\
b_3 = (1, -2) & \rightarrow & D_v\nonumber
\end{eqnarray}
So we expect a $D_8$ singularity on the divisor $D_v$, and $A_3$
singularities on $D_v + 4 D_s$ and $D_v + 8 D_s$.  This general
structure is typical for smooth heterotic compactifications of the
$SO(32)$ theory on K3; the F-theory realization of such models is
always associated with the base $\F_4$, and the residual $SO(M)$
factor resides on $D_v$ \cite{Morrison-Vafa-II}. 
\vspace*{0.05in}

\noindent
{\bf Example:} $E_7 \times E_8$ on $\F_{12}$

We discussed above the model (\ref{eq:78-model}) with gauge group $E_7
\times E_8$.  From the vector $b_8 = (1, -6)$ in (\ref{eq:78-b}), we
see that (\ref{eq:map}) only gives an effective divisor for $m = 12$.
The $E_8$ and $E_7$ loci are then $D_v, D_u = D_v + 12 D_s$
respectively.  There is no bifundamental matter, as  expected, since $D_v
\cdot D_u = 0$.  This is an example of a heterotic $E_8 \times E_8$
vacuum with all instantons in one $E_8$.  For general heterotic $E_8
\times E_8$ compactifications, the resulting F-theory model has a base
$\F_m$ where $12 \pm m$ of the instantons are embedded in each $E_8$
factor \cite{Morrison-Vafa-II}.
\vspace*{0.05in}

\noindent
{\bf Example:} Gimon-Polchinski model

Consider now the Gimon-Polchinski model (\ref{eq:Gimon-Polchinski}),
with gauge group $G = SU(16) \times SU(16)$, two matter fields
transforming under the antisymmetric ${\bf 120}$ for each gauge group
factor, and one bifundamental matter field.  It is easy to compute
that the anomaly matrix for this model is
\begin{equation}
\Lambda= \left(\begin{array}{ccc}
8 &  -2&-2\\
-2 & 0 & 1\\
-2 &  1 &  0
\end{array} \right) \,.
\end{equation}
It follows that in the canonical $T = 1$ basis (\ref{eq:t1}),
\begin{equation}
b_1 = (1, 0), \;\;\;\;\; b_2 = (0, 1) \,.
\end{equation}
This theory must therefore be realized through an F-theory
compactification on $\F_0 =\P^1 \times \P^1$, with divisor classes
$b_1 \rightarrow D_v, b_2 \rightarrow D_s$.  This F-theory realization
of the Gimon-Polchinski model has been explored in some detail as a
useful example for studying the heterotic-F-theory duality, in
particular clarifying the role of instantons with Spin(32)$/\Z_2$
structure but not $SO(32)$ vector structure \cite{Berkooz-all,
  Aspinwall-point, Sen-gp}.
\vspace*{0.05in}

The reader may find it illuminating to work out a few further
examples.  As far as the author is aware all known 6D supergravity
theories that arise from any kind of string compactification admit an
embedding into F-theory through a map of the form \eq{eq:map}, at
least at the level of the discrete data characterizing the gauge group
and matter content of the theory.

\subsection{Global structure of the space of 6D ${\cal N} = 1$ supergravities}
\label{sec:6D-global}

\subsubsection{Consistency conditions, and matching ${\cal G}$ to
  ${\cal V}$}

In Section \ref{sec:6D-anomalies} we summarized the known constraints
on the space ${\cal G}^{6D, {\cal N} = 1}$ (which we refer to simply
as ${\cal G}$ for the remainder of this section).  For $T < 9$ there
are a finite set of possible combinations of nonabelian gauge groups
and matter content compatible with anomaly cancellation conditions and
proper-sign gauge kinetic terms.  For $T \geq 9$ there are infinite
families of models that satisfy the known consistency conditions.
Consistent supergravity theories must furthermore satisfy the
condition that the dyonic string charge lattice $\Gamma$ is
unimodular.

Given the finite constraint on the set of possible groups and
matter for $T < 9$, it is possible in principle to enumerate all
models consistent with the known consistency constraints.  This can be
done in practice more easily for smaller values of $T$.  For $T = 0$
the constraints are strongest, as the vectors in the anomaly lattice
are just integers $b \in\Z$, and a complete classification of models
is tractable \cite{0}.  For $T = 1$ the space of possible combinations
of groups and representations becomes larger; an exploration of part
of this space of theories with some restriction on matter
representations was initiated by Avramis and Kehagias
\cite{Avramis-Kehagias}.  Using the same methods that lead to the
finiteness bounds, a complete classification of models has been
carried out for gauge groups with $SU(N)$ and certain matter
representations \cite{KMT}, such a classification can be done for
arbitrary gauge groups and matter.  Incorporating the unimodularity
constraint on the dyon charge lattice, such a systematic
classification of models can in principle be continued up to $T = 8$.

Six-dimensional F-theory constructions, as described in Section
\ref{sec:6D-F-theory}, give a wide range of string vacua, and define a
set ${\cal V} ={\cal V}^{6D, {\cal N} = 1}$ of known string vacuum
constructions in 6D.  For any 6D supergravity theory admitting an
F-theory realization, the anomaly lattice of the 6D theory is mapped
into the topological structure of the F-theory construction through
the map \eq{eq:map}.  This provides a framework for identifying
general features of models lying in the set ${\cal G} \setminus{\cal
  V}$, which satisfy all known consistency conditions and yet have no
known string theory construction.  Basically, we just need to
determine which models in ${\cal G}$ do not lead to acceptable data
for an F-theory model.  There are a number of specific ways in which
the map \eq{eq:map} can fail to take the anomaly lattice of an
apparently-consistent theory to valid topological data for an F-theory
compactification.  We list a number of these possible failure modes
and comment on each.
\vspace*{0.05in}

\noindent $\bullet$ 
Unimodular embedding.

A necessary condition for the existence of an embedding in F-theory
through \eq{eq:map} is the existence of an embedding of $\Lambda$ in a
unimodular lattice.  As discussed at the end of Section
\ref{sec:6D-anomalies}, however, it has been shown
\cite{Seiberg-Taylor} that every consistent 6D theory has a unimodular
charge lattice, of which $\Lambda$ is a sublattice.  Thus, every
consistent theory has an anomaly lattice that can be embedded in some
unimodular lattice $\Gamma$.  $\Gamma$ can then be taken to be the
homology lattice $H_2 (B,\Z)$ of an F-theory base.
\vspace*{0.05in}

\noindent $\bullet$ 
Exotic matter fields.

There are a variety of 6D supergravity models that satisfy the known
consistency conditions and that have exotic matter representations
not realized through known codimension two singularity types in
F-theory \cite{KMT, KMT-II, 0}.  As examples, consider the following theories
(listing only charged matter for each theory)
\begin{eqnarray}
T = 1, G = SU(8), &  & {\rm matter} ={\tiny\yng(1)}+ 3 \;{\tiny\yng( 2)}+ 2 \;{\tiny\yng(1,1,1)}
+  {\tiny\yng(1,1,1,1)}\\
T = 0, G = SU(4), &  &  {\rm matter} =1 \times {\tiny\yng(2,2)}+
64 \times {\tiny\yng(1)}\\[0.1in]
T = 0, G = SU(8), & &  {\rm matter} =1 \times {\tiny\yng(2,2)} \label{eq:box-8}
\end{eqnarray}
Each of these theories contains a matter representation that cannot
be realized using a standard codimension two F-theory singularity.
The range of possible matter fields that may be realized using more
exotic codimension two singularities is, however, quite large
\cite{Morrison-Taylor}.  It may be possible to implement at least the
first two of these models using more exotic types of codimension two
singularity.  One clue to this connection is given by a relationship
between the group theory of matter representations and the
corresponding F-theory geometry \cite{0}.  For any representation we can define
the quantity
\begin{equation}
g_R = \frac{1}{12}\left(2 C_R + B_R -A_R \right)  \,.
\end{equation}
From the anomaly relations it is possible to show that for models with
$T = 0$ the set of representations transforming under a gauge factor
$G$ satisfies
\begin{equation}
\sum_{R}x_R \, g_R = \frac{(b-1) (b-2)}{2}.
\label{eq:genus-relation}
\end{equation}
In a $T = 0$ F-theory realization, $b$ is the degree of the polynomial
defining the curve $\xi$ on $\P^2$ carrying the singularity associated
with the gauge factor $G$.  The quantity \eq{eq:genus-relation} is
precisely the arithmetic genus of such a curve \cite{Perrin}.  The
arithmetic genus carries a contribution from the topological
(geometric) genus as well as a contribution from singular points on
the curve.  So it is natural to expect that $g_R$ encodes the
arithmetic genus contribution of a singularity on the curve $\xi$
carrying the group factor $G$.  This correspondence works, for
example, for the two-index symmetric representation of $SU(N)$, which
has $g_{\tiny\yng(2)} = 1$, and is produced by an ordinary double
point singularity contributing 1 to the arithmetic genus \cite{Sadov,
  Morrison-Taylor}.  A complete analysis of codimension two
singularities in F-theory is a  challenge for future research.
Note, however, that some representations, such as the ``box''
representation of $SU(8)$ appearing in (\ref{eq:box-8}) cannot appear
in any known kind of F-theory model for other reasons; in the case of the
$SU(8)$ box representation, the theory violates the bound associated
with the Kodaira condition (which is discussed further below).  There
are also constraints on  matter representations that can be realized from the
heterotic point of view \cite{Dienes-mr, Dienes}.  Understanding how
the heterotic, F-theory, and supergravity constraints on allowed
representations may be related is an interesting open problem.
\vspace*{0.05in}

\noindent $\bullet$ 
Weierstrass formulation

In the discussion here we have focused on the topological structure of
the F-theory realization, in particular on the divisor classes
supporting the singularities associated with a given nonabelian gauge
group.  For a complete F-theory realization, an explicit Weierstrass
model is needed, where $f, g$ are described explicitly as sections of
the appropriate line bundles ${\cal O} (-4K),{\cal O} ( -6K)$.  In a
variety of cases that have been studied, Weierstrass models can be
found whenever a map of the form (\ref{eq:map}) can be found whose
image gives divisor classes satisfying all the topological constraints
from F-theory (including the positivity and Kodaira constraints
formulated below).  In cases where the topological conditions are
satisfied, the number of degrees of freedom in the Weierstrass model
that remain unfixed when the coefficients are tuned to achieve the
desired gauge group precisely matches the number of uncharged scalar
fields in the low-energy theory \cite{KMT}. This led Kumar, Morrison,
and the author to conjecture that the degrees of freedom will match in
all cases, and that a Weierstrass model will exist whenever the
topological constraints are satisfied.  Finding a proof of this
conjecture is left as an open problem for further research.
\vspace*{0.05in}

\noindent $\bullet$ 
Positivity conditions and effective divisors

There is a sign condition on the divisors appearing in the image of
the map (\ref{eq:map}).  The divisors $\xi_i$ (associated with the curves
where the 7-branes are wrapped) must be effective
divisors; this is equivalent to the statement that there exists a
moduli vector $J$ such that 
\begin{equation}
J \cdot \xi_i > 0 \;\forall i \,.
\end{equation}
We understand this condition physically in the 6D supergravity theory
as the constraint that the gauge kinetic term for $G_i$ must have the
proper sign, {\it i.e.} $j \cdot b_i > 0$.

It is also the case for all F-theory models that $-K$ is an effective
divisor, so that $J \cdot  (-K) > 0$.  This condition in the
supergravity theory states that
\begin{equation}
j \cdot a < 0 \,.
\end{equation}
This fixes the sign of the curvature-squared term in the 6D theory of
the form $(a \cdot j) R^2$.  It is not clear whether this sign constraint
is necessary for consistency of the low-energy theory.
It is possible that the sign of this term
is fixed through a causality constraint analogous to those studied by
Adams {\it et al.} \cite{Adams-Nima}.  A complete explanation of this
constraint from the supergravity point of view is left as another open
question for future work.

\vspace*{0.05in}

\noindent $\bullet$ Kodaira constraint

Related to the positivity conditions just mentioned, there is also a
condition on the residual divisor locus $Y =  -12K-\sum_{i} \nu_i
\xi_i$, which states that this divisor is effective.  This condition
arises from the Kodaira constraint (\ref{eq:Kodaira}).  In the 6D
supergravity theory this constraint states that
\begin{equation}
j \cdot (-12a-\sum_{i} \nu_i b_i) > 0 \,.
\label{eq:Kodaira-inequality}
\end{equation}
All of the known infinite families of models that satisfy the other
macroscopic 6D constraints violate this ``Kodaira bound''.  Thus, proving
this bound would potentially reduce the space of possibly consistent
combinations of $T, G,$ and ${\cal M}$  to a finite set
${\cal G}$.  One possible route to proving
(\ref{eq:Kodaira-inequality}) would be to follow an approach along the
lines described in Section \ref{sec:8D-global}.  It may be that a
careful analysis of world-sheet anomalies on the dyonic strings of a
6D supergravity theory will constrain the set of allowed models in
such a way that (\ref{eq:Kodaira-inequality}) is necessary for
consistency of the theory.  Further study is needed, however, to
see whether this speculation is borne out in practice.

\subsubsection{Connectivity and finiteness of the space of theories}
\label{sec:connectivity}

Many, if not all, 6D supergravity theories with different gauge groups
and matter content are connected through continuous deformations in
the scalar moduli space.  Moving in the moduli space can lead to a
familiar Higgs type transition where a gauge group is broken, removing
scalar fields and vector fields from the theory in such a way that the
difference $H-V$ is unchanged and the anomaly relation (\ref{eq:hv})
remains valid.  From the F-theory point of view this kind of
transition arises from changing the coefficients in the Weierstrass
functions $f, g$ so that the structure of the discriminant locus
changes.  It is clear from the F-theory point of view that any two
models associated with the same base $B$ for the elliptic fibration
are continuously connected through a  deformation of
the Weierstrass equation.  

There are, however, further connections in the space of supergravity
theories.  As discussed in Section \ref{sec:6D-bases}, supergravity
theories with different numbers of tensor multiplets $T$ can be
connected through exotic phase transitions where a tensor multiplet
$T$ is exchanged for 29 hypermultiplets (with a possible additional
change in the vector multiplet structure) in such a way that the
anomaly relation (\ref{eq:hv}) is preserved.  Such a transition occurs
in a limit of the moduli space where $j \cdot x \rightarrow 0$ for
some $x \in \Gamma$ in the lattice of dyonic string charges.  Thus,
this transition is associated with a string becoming tensionless
\cite{Seiberg-Witten-6D, Witten-phase, Morrison-Vafa-II}.  From the
F-theory point of view this corresponds to a cycle in the base $B$
shrinking to a point, {\it i.e.} to a blow-down (the opposite of a
blow-up) of a rational curve in the surface.  From the heterotic point
of view, the transition from the theory with fewer tensor multiplets
occurs when an instanton shrinks to a point, as mentioned above,
producing a 5-brane carrying the additional tensor field.  Since all
smooth F-theory bases corresponding to theories with one 6D
supersymmetry can be connected by a sequence of blow-up and blow-down
operations, this suggests that the full space of 6D ${\cal N} = 1$
supergravities is a single connected moduli space \footnote{Extremal
  transitions associated with theories having the Enriques surface as
  base are identified in \cite{Aspinwall-Enriques}.}.  This would mean
that there is really a single unified quantum theory of 6D
supergravity since the entire moduli space is in principle visible in
the structure of fluctuations of massless scalar fields around any
given vacuum.  This gives a simple and appealing picture of
supergravity in six dimensions.

From the point of view of F-theory, the number of distinct bases $B$,
as well as the number of combinations of $T, G,$ and matter
representations that can be realized must be finite.
We summarize briefly the
argument for finiteness \cite{KMT-II};
the related fact that the number of topological classes of
elliptically fibered Calabi-Yau threefolds (up to birational
equivalence) is finite was proven by Gross \cite{Gross-finite}.  
From the minimal surface point
of view, any F-theory model can be viewed as a Weierstrass model over
either $\P^2$ or one of the Hirzebruch surfaces $\F_m$.  If the number
of tensor multiplets is greater than 1, then the Weierstrass model can
be thought of as a Weierstrass model on $\F_m$ with singularities
tuned to require one or more blow-ups of points on $B$, as discussed
in Section \ref{sec:6D-bases}.  Each different combination of base,
gauge group, and matter content, then, corresponds to a locally closed
set in the finite-dimensional space of Weierstrass models.  (A locally
closed set is an intersection of a closed set with an open set.  Each
condition on the discriminant locus gives a closed set, with each
further condition giving an enhanced gauge group or further blow-up
giving a smaller closed set within the original set.)  A fundamental
theorem from commutative algebra and algebraic geometry, the Hilbert
basis theorem, guarantees that there are only a finite number of
distinct such locally closed sets.  Heuristically this follows from
the observation that each gauge group enhancement or blow-up requires
tuning a finite number of parameters, and the total number of
parameters is finite and equal to the number of neutral scalar fields
in the generic model, 273 on $\P^2$ and 244 on $\F_m$.  Since the
known constraints on ${\cal G}$ give an infinite set of distinct
models, at present, the ``apparent swampland'' of 6D ${\cal N} = 1$
theories is infinite.  Finding an argument for the Kodaira constraint
as a consistency condition for any supergravity theory could eliminate
the infinite number of apparently consistent models that are not
realized in F-theory.

While the set of allowed F-theory models is provably finite, the
precise extent of the set of F-theory compactifications has not been
determined.  For example, the maximum number of tensor multiplets $T$
which can appear has not been definitively identified.  The largest
number known is $T = 192$, for the theory with gauge group
\eq{eq:big-group} \cite{Aspinwall-Morrison, Candelas-pr2}.  The
structure of the anomaly lattice provides insight into the the
extremal transitions where $T$ changes; this may help in charting the
connectivity and global structure of the space of F-theory models and
definitively identifying the upper bound on $T$.  Ideally such an
understanding would lead to a clear picture of how the space of models
arising from string/F-theory is bounded using considerations based
only on the macroscopic structure of the supergravity theory.

\section{Comments on Four Dimensions and Other\\ Concluding Remarks}

In these lectures we have described some global aspects of the space of
supergravity theories in dimensions 10, 8, and 6.  We have used
constraints on these theories to determine a set ${\cal G}$ of
``apparently consistent'' theories in each dimension.  We have also
described the set ${\cal V}$ of known string theory compactifications
in each dimension.  We have focused on the discrete data
characterizing each supergravity theory: the spectrum of fields and
the gauge symmetry group of the theory.  In 10D, ${\cal G} ={\cal V}$,
so that we have ``string universality''; any theory of supergravity in
ten dimensions that cannot be realized in string theory is
inconsistent (at least at the level of the field content and symmetry
of the theory).  In 8D and 6D, ${\cal G} \supset{\cal V}$, and the set
of possible models in the ``apparent swampland'' ${\cal G}
\setminus{\cal V}$ is infinite.  In each case, however, we have a set
of fairly well-understood criteria for determining which models in
${\cal G}$ admit a string realization (and hence are in ${\cal V}$).
It is possible that both in 8D and in 6D these criteria will
eventually be understood in terms of additional consistency
constraints on the low-energy theory; we have outlined possible
approaches that may lead to such an understanding.  It is also
possible that new approaches to string compactification may be found
that will expand the space ${\cal V}$.  Whether or not it can be
proven that string universality holds for supergravity theories in
eight and/or six dimensions, focusing on theories in the set ${\cal G}
\setminus{\cal V}$ is a promising direction for further progress.  By
searching for new quantum consistency conditions that rule out such
theories, or novel realizations of such theories in string theory, our
understanding of the global space of quantum gravity theories is
improved.

An important direction on which we have not focused in these lectures
is the investigation of the relationship between macroscopic and
stringy consistency conditions for more detailed aspects of
supergravity theories, beyond the field content and symmetry
structure.  For example, for 6D ${\cal N} = 1$ supergravity theories,
the hypermultiplets live in a quaternionic K\"ahler manifold.  To go
beyond the discrete version of string universality, it would be
necessary to demonstrate that the set of such manifolds, along with
their metrics, is determined uniquely from the low-energy theory to
match the set of possibilities arising from string compactifications.
Compact quaternionic K\"ahler manifolds have been classified
\cite{qk-1, qk-2, Dasgupta-hw}.  But a full analysis of which of these
manifolds are compatible with string theory remains as a challenge for
the future.  In most supergravity theories there are also terms that
might be added to the action of the theory, such as higher-derivative
terms.  For string realizations of these theories, the coefficients of
these additional terms in the low-energy theory are fixed by the
quantum physics of string theory.  For string universality to hold in
a complete fashion, there need to be consistency conditions that fix
these coefficients to the values determined by string theory.  It has
been shown that in some cases, supersymmetry places very strong
constraints on such coefficients \cite{Green-Sethi}.  But a systematic
understanding of the extent to which such constraints uniquely
constrain low-energy theories to those realized through string theory
is another direction open for future progress.

Extending the kind of global analysis done here to four dimensions
would obviously be an interesting and important endeavor.  In
principle this could be a very promising direction for future
research.  String theory constructions do impose strong constraints on
which low-energy 4D supergravity theories can be realized, just as in
higher dimensions.  In many constructions these consistency conditions
come from tadpole conditions closely analogous to the Kodaira
constraint and related tadpole conditions we have described here.
Anomalies have provided a clear window on these constraints in six and
ten dimensions.  Just as in eight dimensions, however, for
four-dimensional theories some principles beyond space-time anomalies
seem to be required to identify the string theoretic constraints from the
structure of the low-energy theory.  A number of new consistency
conditions on 4D low-energy theories have been identified in recent
years \cite{Vafa-swamp, Nima-weak, Adams-Nima}, though not directly related to
string constructions.  One natural approach that we have discussed
here for deriving further constraints on theories both in 8D and in 6D
is the analysis of anomalies in the world-volume of solitonic strings
in supergravity theories.  The same approach could be used for 4D
theories.  4D supersymmetric gravity theories generally contain axions
that play the same role as $B$ fields in higher dimensions.  (Axions
are dual to two-form fields in 4D.)  Most if not all supersymmetric
string vacuum constructions give rise to strings in 4D that couple to
these axions just as strings in higher dimensions couple to two-form
fields $B_{\mu \nu}$.  Thus, a careful analysis of the world-volume
theories of axionic strings in supersymmetric 4D gravity theories may
give new constraints on what structures are allowed in these theories.

One complication in approaching 4D theories with less than ${\cal N} =
4$ supersymmetry is that the set of string vacua is vast, and still
poorly understood.  Even for ${\cal N} = 2$ theories there are
fundamental mathematics problems that must be solved to get a global
picture of the space of string vacua.  Compactifications of type II
string theory on Calabi-Yau threefolds (not necessarily elliptically
fibered) give rise to such ${\cal N} = 2$ supergravity theories.
There is at this time no proof that the set of topological classes of
Calabi-Yau threefolds is finite, and no systematic classification.
While we were able to use the mathematical structure of lattices to
great efficacy in the description of higher-dimensional supergravity
theories, the analogous structure in 4D is more complex.  For example,
the 2-homology and intersection form on a complex surface is described
by a lattice that must be self-dual due to Poincar\'{e} duality.  For
a general Calabi-Yau threefold, the intersection structure is given by
a trilinear form $C: (H_2 (B,\Z))^3 \rightarrow\Z$.  The space of such
trilinear forms, and the constraints on such forms, are much less well
understood than the space of unimodular lattices.  Thus, even for
${\cal N} = 2$ theories in 4D, new mathematics probably must be
developed to attain a true global classification of theories.  

The situation becomes even more complicated for ${\cal N} = 1$
theories.  Studying different string approaches to ${\cal N} = 1$ 4D
supergravity theories has been a major industry for decades.  The
close correspondence identified in six dimensions between the anomaly
structure in supergravity theories and the topology of F-theory
compactifications, along with the observation that in 6D F-theory
provides a fairly universal framework for describing string vacua,
suggest that F-theory is a natural place to look for a global
characterization of the space of string vacua.  From the F-theory
point of view, however, the set of possible compactifications to 4D
depends upon elliptically fibered Calabi-Yau fourfolds.  Such
fourfolds are fibered over a three-complex dimensional base manifold
$B$.  The classification of threefolds is in a much more limited state
of development than that of surfaces.  For surfaces, the minimal model
approach leads to a systematic characterization of possible F-theory
bases, and provides a geometric understanding of the connection
between these bases.  For threefolds, the program of Mori \cite{Mori}
aspires to a similar classification.  But the mathematics is much more
complicated, and seems to need significantly more development to be
usable by physicists as an approach to a global characterization of
the set of bases that can be used for 4D F-theory constructions.  While
in recent years 4D F-theory models have provided a rich new
perspective on string phenomenology \cite{Donagi-Wijnholt,
  Beasley-hv1, Beasley-hv2, Marsano-F-theory, Blumenhagen-F-theory,
  Heckman-review, Weigand-review}, the full range of possible 4D vacua
is still poorly understood.  Some  geometric constraints on a class of 4D
F-theory models have been found by Cordova \cite{Cordova} based on the
assumption that gravity can decouple from the low-energy physics.
Inclusion of fluxes (``G-flux'') in F-theory should significantly
expand the range of possible 4D vacua, but the tools for understanding
this from a global/nonperturbative point of view are just being
developed \cite{Curio-Donagi, Hayashi-all, Marsano-ss}.  As for the 6D
models discussed here, U(1) factors also have some subtle features in
global constructions, and are a subject of current interest in 4D
F-theory models \cite{Grimm-u1}, where anomaly constraints also play a role
\cite{Marsano-u1, Dolan-u1}.

Other very general approaches to building new string vacua suggest
that we may have only seen the tip of the iceberg of 4D string
constructions.  Non-K\"ahler compactifications
\cite{Strominger-torsion, Fu-Yau, Becker-torsion, Adams-torsion, bty,
Becker-Sethi}, non-geometric fluxes \cite{Hellerman-mw,
Dabholkar-Hull, Kachru-stt, Hull, stw, Wecht-lectures}, $G_2$
compactifications of M-theory \cite{Duff-np, g2, Acharya-Witten},
asymmetric orbifolds \cite{Narain-sw}, and other constructions all suggest
that much work must be done to begin to get some global grasp of the
range of possibilities.

Finally, the discrete nature of the set of 4D ${\cal N} = 1$ vacua
makes it clear that attaining a global classification of models, as
well as determining macroscopic constraints, will require
qualitatively different insights from those used in the simpler
classes of theories in higher dimensions and/or with more
supersymmetry.  For the 8D and 6D ${\cal N} = 1$ supergravities we
have studied here,  the theories are connected in a single
continuous moduli space of Minkowski vacua.  In 4D, the inclusion of
fluxes and other discrete structure can stabilize many moduli, giving
a ``discretium'' of isolated regions of (generally AdS) supersymmetric
vacua \cite{Douglas-Kachru}.  The number of discrete families of
supersymmetric 4D vacua is infinite in some families of string
compactifications \cite{dgkt-IIA}.
Showing that certain apparently continuous
parameters in the description of the low-energy theory can only take
the specific values associated with string vacua presents an added
challenge for the program of identifying constraints that rule out
all models other than those realized in string theory.
It may be that statistical methods \cite{Douglas-statistics, Denef-Douglas} are
needed to make sense of this large set of discrete solutions.  Even in
4D, however, in certain cases such as intersecting brane models on a
toroidal orbifold it has been possible to place finite bounds on the
set of possible vacuum solutions \cite{Douglas-Taylor}, and to
systematically enumerate these models \cite{Gmeiner,
Rosenhaus-Taylor}.  Generalizing such results to more generic
backgrounds such as smooth Calabi-Yau manifolds, which may be possible
using the trilinear intersection form, would be a promising step
forward in attaining a more global picture of large classes of vacua.

Despite the preceding cautionary remarks regarding the challenges of
developing a global analysis of 4D ${\cal N} = 1$ vacua, a number of
the lessons learned from higher-dimensional supergravity theories
should be relevant for 4D physics.  Consideration of 6D theories
satisfying anomaly cancellation and other constraints has have
suggested some new structures that may arise in string theory, such as
novel types of matter from codimension two singularities in F-theory.
Understanding the stringy derivation of such phenomena in six
dimensions will also give new tools for describing 4D physics.  In 6D,
the close correspondence between the structures underlying
supergravity theories and F-theory has provided a ``bottom-up'' map
which can uniquely identify the F-theory geometry of any string vacuum
realizing a supergravity model with given symmetries and spectra.
Progress towards explicitly realizing such a correspondence in four
dimensions would help narrow the range of string constructions which
might correspond with observed physics.  This systematic
characterization of string vacua has also helped to identify new
supergravity constraints in six dimensions, such as the self-dual
nature of the dyonic charge lattice, which suggests that new
constraints on 4D gravity theories remain to be discovered.  Insights
into the duality relationships between different string theories in
higher dimensions has played an important role in our developing
understanding of string theory over the last 15 years.  Incorporating
these relationships into a global picture of the space of 6D string
vacua will help provide insight into how the different string vacuum
constructions are related in 4D.

\section*{Acknowledgments}

The perspective presented in these lectures incorporates the
contributions of many people.  I would particularly like to thank
Vijay Kumar and David Morrison, with whom many of the ideas developed
in these lectures have been explored.  Thanks also to my other
collaborators in the part of the work presented here in which I was
involved in the original research: Allan Adams, Oliver DeWolfe,
Michael Douglas, Satoshi Nagaoka, Daniel Park, Vladimir Rosenhaus, and
Nathan Seiberg.  Thanks to Oliver DeWolfe, Vijay Kumar, David Morrison
and Daniel Park for comments on a preliminary version of these notes.
I am grateful for helpful discussions and correspondence with Massimo
Bianchi, Ralph Blumenhagen, Mirjam Cvetic, Frederik Denef, Noam
Elkies, Dan Freedman, Michael Green, Vladimir Ivashchuk, Jim
Halverson, Gabriele Honecker, Shamit Kachru, John McGreevy, Greg
Moore, Joe Polchinski, Bert Schellekens, John Schwarz, Ashoke Sen,
Isadore Singer, Cumrun Vafa, Edward Witten, and Barton Zwiebach.  I
would like to thank the Simons Center for Geometry and Physics for
hospitality during the completion of these notes. And of course, many
thanks to the organizers of the TASI 2010 summer school for making
these lectures possible and to the students at the school for many
insightful and provocative questions and discussions.  This research
was supported by the DOE under contract \#DE-FC02-94ER40818.

\bibliographystyle{ws-procs9x6}
\bibliography{TASI}

\end{document}